\documentclass[aps,10pt,twocolumn,superscriptaddress,longbibliography]{revtex4-1}

\usepackage[utf8]{inputenc}
\usepackage{lmodern}
\usepackage{amsmath}
\usepackage{graphicx}
\usepackage{float}
\usepackage{amsfonts}
\usepackage{url}
\usepackage{epstopdf}
\usepackage{placeins}
\usepackage[usenames,dvipsnames]{xcolor} 
\usepackage{verbatim}
\usepackage{bbold}

\usepackage{times}
\usepackage{hyperref}
\hypersetup{
colorlinks=true,
linkcolor=blue,
citecolor=blue,
filecolor=magenta,
urlcolor=cyan
}
\usepackage[capitalize]{cleveref}

\begin{document}

\title{Stabilization of multi-mode Schr\" odinger cat states via normal-mode dissipation engineering}

\author{Petr Zapletal}
\affiliation{Friedrich-Alexander University Erlangen-N\"urnberg (FAU), Department of Physics, 91058 Erlangen, Germany}
\affiliation{Cavendish Laboratory, University of Cambridge, Cambridge CB3 0HE, United Kingdom}

\author{Andreas Nunnenkamp}
\affiliation{Faculty of Physics, University of Vienna, Boltzmanngasse 5, 1090 Vienna, Austria}
\affiliation{
School of Physics and Astronomy and Centre for the Mathematics and Theoretical Physics of Quantum Non-Equilibrium Systems, University of Nottingham, Nottingham, NG7 2RD, United Kingdom}
\affiliation{Cavendish Laboratory, University of Cambridge, Cambridge CB3 0HE, United Kingdom}

\author{Matteo Brunelli}
\affiliation{Cavendish Laboratory, University of Cambridge, Cambridge CB3 0HE, United Kingdom}
\affiliation{Department of Physics, University of Basel, Klingelbergstrasse 82, 4056 Basel, Switzerland}

\begin{abstract}

Non-Gaussian quantum states have been deterministically prepared and autonomously stabilized in single- and two-mode circuit quantum electrodynamics architectures via engineered dissipation. However, it is currently unknown how to scale up this technique to multi-mode non-Gaussian systems. Here, we upgrade dissipation engineering to collective (normal) modes of nonlinear resonator arrays and show how to stabilize multi-mode Schr\" odinger cat states. These states are multi-photon and multi-mode quantum superpositions of coherent states in a single normal mode delocalized over an arbitrary number of cavities.
We consider tailored dissipative coupling between resonators that are parametrically driven and feature an on-site nonlinearity, which is either a Kerr-type nonlinearity or an engineered two-photon loss.
For both types of nonlinearity, we find the same exact closed-form solutions for the two-dimensional steady-state manifold spanned by superpositions of multi-mode Schr\" odinger cat states.
We further show that, in the Zeno limit of strong dissipative coupling, the even parity multi-mode cat state can be deterministically prepared from the vacuum. Remarkably, engineered two-photon loss gives rise to a fast relaxation towards the steady state, protecting the state preparation against decoherence due to intrinsic single-photon losses and imperfections in tailored dissipative coupling, which sets in at longer times. The relaxation time is independent of system size making the state preparation scalable. 
Multi-mode cat states are naturally endowed with a noise bias that increases exponentially with system size and can thus be exploited for enhanced robust encoding of quantum information.

\end{abstract}

\maketitle

\section{Introduction}
Schr\" odinger cat states---quantum superpositions of macroscopically distinct (or ``classical") states---are a fundamental resource for quantum communication~\cite{Hacker2019}, quantum metrology \cite{Gilchrist2004,Giovannetti2011,Pezze2018} and quantum computation~\cite{vlastakis2013,chamberland2020,grimm2020}. 
Recently, the development of bosonic quantum error-correcting codes has led to a surge of interest in exploiting Schr\" odinger cat states for encoding and manipulating quantum information in an hardware-efficient manner \cite{leghtas2013,mirrahimi2014,leghtas2015,ofek2016,puri2017,puri2019,touzard2018,guillaud2019,grimm2020,Puri2020,Lescanne2020,chamberland2020}. The two-dimensional subspace spanned by superpositions of Schr\" odinger cat states can be used to encode quantum information, implementing a so-called cat qubit. Interestingly, such a qubit is naturally endowed with a large noise bias, i.e.~the qubit has a single dominant noise channel while all other types of noise are largely suppressed, which offers protection from errors in a quantum memory~\cite{leghtas2013,ofek2016}, and eases up the requirements of quantum error correction~\cite{puri2019} and quantum annealing \cite{Puri2017b}. The implementation of bias-preserving quantum gates \cite{touzard2018,guillaud2019,grimm2020,Puri2020} opens the door for universal fault-tolerant quantum computing based on cat qubits \cite{chamberland2020}. 
The potential for applications of Schr\" odinger cat states further increases when considering extensions to multiple modes. Two-mode Schr\" odinger cat states, also known as entangled coherent states~\cite{sanders1992}, are a resource for continuous-variable quantum information \cite{vanenk2001,sanders2012} and metrology \cite{joo2011,zhang2013}. They have been experimentally prepared using feedback control in superconducting circuits~\cite{wang2016}. A unique feature of these states is that they possess non-Gaussian entanglement, which allows circumventing many no-go theorems  imposed by Gaussian quantum resources~\cite{braunstein2005}. 
Extending the stabilization of non-Gaussian entangled states---and Schr\" odinger cat states in particular---to an arbitrary number of modes would render non-Gaussian quantum resources scalable~\cite{wang2001}.

The widespread application of Schr\" odinger cat states to photonic quantum technologies is also due to the unique control capabilities available in  circuit quantum electrodynamics (circuit QED) architectures, which guarantee unprecedented control  of nonlinear interactions and dissipation \cite{Blais2020b,blais2020,ma2021}. In this context, a particularly convenient approach is dissipation engineering~\cite{poyatos1996,metelmann2015}, which exploits tailored interactions with a dissipative environment to attain deterministic and robust preparation of desired target states or operations. The dissipative preparation of single-mode Schr\" odinger cat states using engineered two-photon loss and two-photon (parametric) drive have been first envisioned in Ref.~\cite{gilles1994}. For a circuit QED setup, the autonomous stabilization of a two-dimensional subspace spanned by single-mode cat states has been recently proposed in Ref.~\cite{mirrahimi2014} and realized in Ref.~\cite{leghtas2015}. It has also been proposed how to generate approximate two-mode Schr\" odinger cat states via dissipation engineering \cite{mamaev2018}. However, engineering \emph{multi-mode} Schr\" odinger cat states and the stabilization of steady-state manifolds thereof in driven-dissipative parametrically coupled systems have remained an unexplored avenue.

In this work we propose how to dissipatively generate multi-mode cat states in a scalable fashion and show how to autonomously stabilize a decoherence-free subspace (DFS), which is spanned by multi-mode Schr\"odinger  cat states. Our approach consists in upgrading dissipation engineering to collective modes (normal modes) of \emph{dissipatively} coupled arrays of nonlinear resonators. 
We present our ideas in terms of two closely related models that are readily realizable with parametrically driven superconducting circuits. The first model is, as sketched in Fig.~\ref{scheme}a, based on an array of parametrically driven Kerr resonators, also known as Kerr parametric resonators (KPRs). In the second model, the Kerr nonlinearity is replaced by engineered two-photon dissipation, depicted in Fig.~\ref{scheme}b. In both cases, the only source of coupling is provided by engineered nonlocal dissipation connecting neighboring pairs of modes.  
This kind of non-local dissipation has recently been considered in the proposal for the generation of two-mode Schrödinger cat states \cite{mamaev2018} and in several other contexts, e.g.~to realize non-reciprocal photon transport in a pair of modes~\cite{metelmann2015} or in cavity arrays~\cite{Porras2019,Wanjura2020}. In contrast, the more conventional scenario of arrays of coherently (tunnel-) coupled KPRs have been studied in Refs.~\cite{Savona2017,Rota2019}, in connection with the emergence of critical behavior. Fully connected networks of tunnel-coupled KPRs have also been proposed for solving combinatorial optimization problems~\cite{Goto2016,Nigg2017}. 

As we shall see, the choice of the pairwise dissipative coupling is responsible for the unique features of our model.
The idea is simple and yet extremely powerful, as it allows to select a single normal mode and completely suppress its decay, while all the other normal modes can be heavily damped. The interaction between the (driven) non-dissipative normal mode and the other dissipative modes leads to an asymptotically stable Schr\"odinger  cat manifold for the non-dissipative mode, which we characterize analytically. Since these cat states are stabilized in a normal mode, they are indeed multi-mode superpositions delocalized over the entire array. We further show that the stabilization of multi-mode cat states is robust against intrinsic single-photon loss and imperfections in the dissipative coupling. In particular, the model featuring local two-photon loss [cf.~Fig.~\ref{scheme}b] guarantees a fast relaxation towards the asymptotic manifold, which allows for an efficient steady-state preparation of the even parity cat state from the vacuum even in the presence of intrinsic single-photon loss. Multi-mode cat states can be stabilized and prepared in the Zeno limit even in the presence of imperfections in non-local dissipation, albeit with decreased fidelity as an additional decoherence channel sets in.  The relaxation time towards the Schr\" odinger cat manifold is independent of system size, making the state preparation scalable. Crucially, the noise bias of multi-mode cat states exponentially increases with system size, which can be exploited for realizing a quantum memory with enhanced protection with respect to single-mode implementations. 

The remainder of this work is structured as follows. In Sec.~\ref{s:Model} we introduce our first model, namely the Kerr-resonator array of Fig.~\ref{scheme}a. In Sec.~\ref{s:DFS} we derive an analytic expression for the steady states of the model, which take the form of a two-dimensional DFS spanned by even and odd multi-mode Schr\"odinger cat states. 
In Sec.~\ref{s:SteadyCat} we  discuss how to target such states, and we show that, for sufficiently strong non-local dissipation, the even multi-mode cat state can be faithfully prepared by initializing the array in the vacuum. In Sec.~\ref{s:Zeno} we explain this behavior by developing an effective single-mode theory, valid in the Zeno limit of strong nonlocal dissipation. We describe the long-time dynamics of the non-dissipative mode and extract an analytical expression of the dissipative gap, dictating the slowest convergence time to the steady state. The expression shows that the requirements of preparing a cat state from the vacuum with near-unit fidelity and guaranteeing a fast convergence cannot be simultaneously satisfied. In Sec.~\ref{s:TwoPhotonLoss} we then show how this drawback can be remedied by replacing the Kerr non-linearity with local two-photon loss as shown in Fig.~\ref{scheme}b. In Sec.~\ref{s:Loss} we account for the effect of unwanted decoherence mechanisms. We include in both models intrinsic single-photon loss, which causes decoherence within the steady-state manifold. In Sec.~\ref{s:Imp}, we discuss effects of decoherence due to imperfections in non-local dissipation.
In Sec.~\ref{bias}, we discuss the noise bias of the multi-mode cat manifold and how it can be exploited for a robust encoding of quantum information.
In Sec.~\ref{s:Exp}, we estimate experimentally relevant parameters of our model based on state-of-the-art circuit QED technology. Sec.~\ref{s:Conclusions} contains concluding remarks and possible directions for future investigation.

\section{Model}\label{s:Model}

\begin{figure}[t]
\centering
\includegraphics[width=0.9\linewidth]{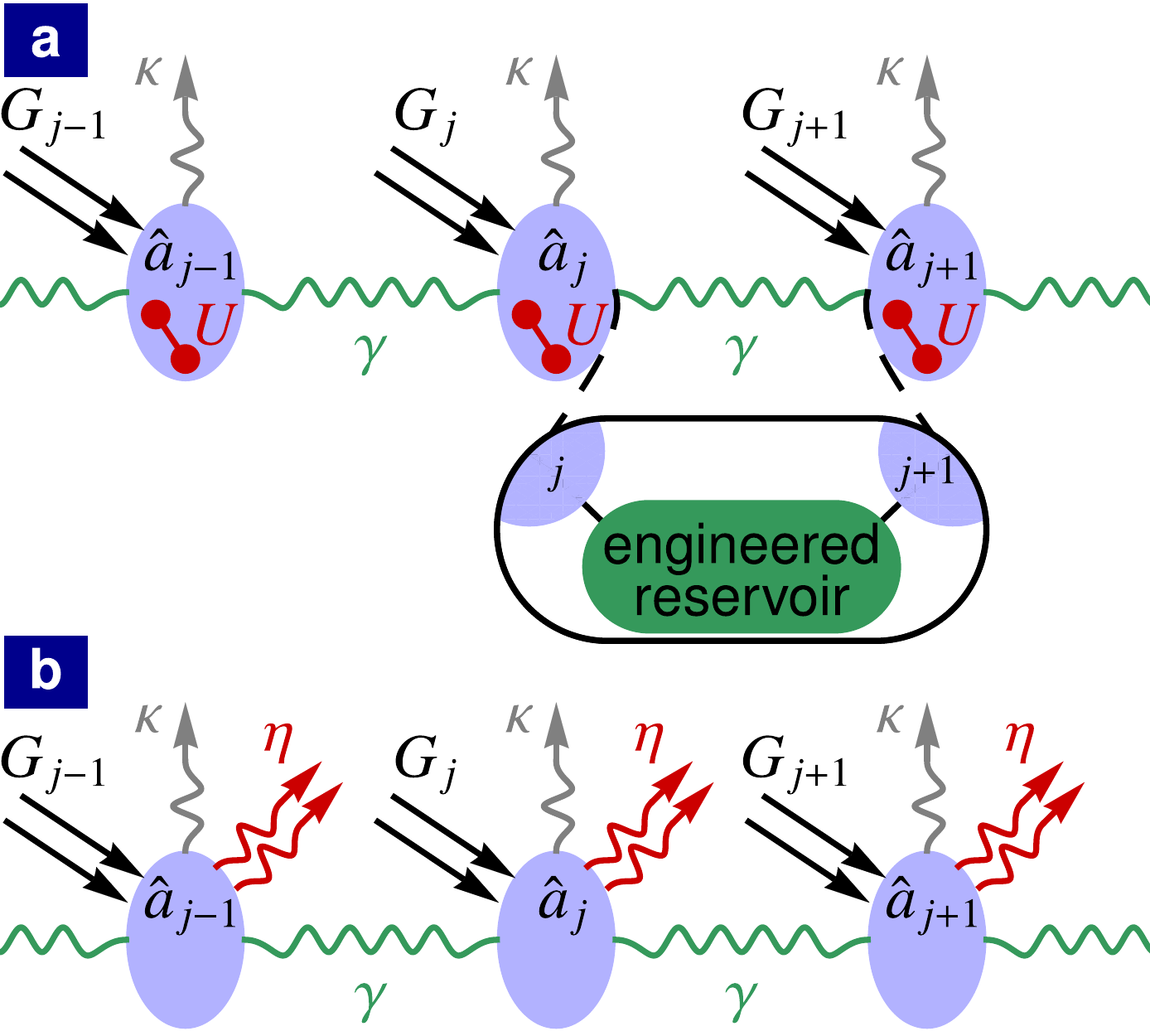}
\caption{Two cavity arrays considered for the stabilization of multi-mode Schr\" odinger cat states; straight (wiggly) lines represent Hamiltonian (dissipative) contributions. (a) One-dimensional array of $N$ resonators $\hat{a}_j$ with on-site Kerr nonlinearity $U$, non-local dissipation at rate $\gamma$, parametric two-photon drive with amplitudes $G_j = Ge^{-i\theta j}$, intrinsic photon loss at rate $\kappa$, and periodic boundary conditions $\hat{a}_{N+1}=\hat{a}_1$. The inset shows that non-local dissipation is implemented via coupling to an engineered reservoir. Note that each dissipative process (wiggly lines) is due to coupling to an independent reservoir. (b) Alternative model where the Kerr nonlinearity is replaced by local two-photon loss at rate $\eta$.}
\label{scheme}
\end{figure}

We consider a one-dimensional array of $N$ identical Kerr resonators, as depicted in Fig.~\ref{scheme}a, each subject to a parametric two-photon drive. In a frame rotating at the resonator frequency and for the pump frequency matching twice the resonator frequency, the system is described by the Hamiltonian $\hat{H} = \hat{H}_U + \hat{H}_G$ ($\hbar=1$)
\begin{gather}\label{kerr ham}
\hat{H}_U = \sum_{j=1}^{N} U\hat{a}_j^{\dagger}\hat{a}_j^{\dagger}\hat{a}_j\hat{a}_j,\\
\hat{H}_G = \sum_{j=1}^{N} \left[G_j\hat{a}_j^{\dagger}\hat{a}_j^{\dagger} + \textrm{h.c.}\right],
\end{gather}
where $U$ is the strength of the Kerr nonlinearity and  $G_j = G\,e^{-i\theta j}$ is the amplitude of the two-photon drive at site $j$ with a phase difference $\theta$ between neighboring resonators; without loss of generality we assume $G>0$.

The distinctive feature of our model lies in the coupling between resonators, which is not described by a Hamiltonian term, but has instead a \emph{dissipative} nature, inasmuch as it correlates photon emission events from neighboring resonators. This is formally described by a non-local dissipation term in the Lindblad master equation
\begin{equation}\label{master}
\dot{\hat{\rho}} = - i\left[ \hat{H},\hat{\rho}\right] + \gamma \sum_{j=1}^{N}\mathcal{D}\left[\hat{a}_j - e^{i\phi}\hat{a}_{j+1} \right]\hat{\rho} = \mathcal{L} \hat{\rho},
\end{equation}
where $\gamma$ is the rate of non-local dissipation, $\mathcal{L}$ is the Liouvillian superoperator and $\nobreak{\mathcal{D}[\hat{\mathcal{O}}]\hat{\rho} = \hat{\mathcal{O}}\hat{\rho}\hat{\mathcal{O}}^{\dagger} -\frac{1}{2}\left(\hat{\mathcal{O}}^{\dagger}\hat{\mathcal{O}}\hat{\rho}  + \hat{\rho}\hat{\mathcal{O}}^{\dagger}\hat{\mathcal{O}} \right) }$. Loosely speaking, this unconventional dissipator can be interpreted as a dissipative analogue of a hopping term. In fact, it corresponds to indirect hopping via an auxiliary, fast decaying, mode, which acts as an engineered reservoir, see the inset of Fig.~\ref{scheme}a. It can be implemented in different ways, e.g.~via coupling the neighboring resonators $\hat{a}_j$ and $\hat{a}_{j+1}$ to a transmission line with respective coupling amplitudes having a relative phase difference $\phi$~\cite{metelmann2015,mamaev2018}.  

We assume periodic boundary conditions $\hat{a}_{N+1} = \hat{a}_1$. In this case, non-local dissipation can be expressed in the plane-wave basis $\hat{b}_k = \frac{1}{\sqrt{N}}\sum_{j=1}^N e^{ijk}\hat{a}_j$ as 
\begin{equation}\label{dissipator}
\gamma\sum_{j=1}^{N}\mathcal{D}\left[\hat{a}_j - e^{i\phi}\hat{a}_{j+1} \right]\hat{\rho} =  \sum_{k=1}^{N}\gamma_k\mathcal{D}\left[\hat{b}_k \right]\hat{\rho},
\end{equation}
where the rate of dissipation $\gamma_k = 2\gamma\left[ 1-\cos\left(k-\phi\right)\right]$ depends on the quasi-momentum $k$ which takes the values $k = \frac{2\pi}{N}, 2 \frac{2\pi}{N},...,2\pi$. From Eq.~\eqref{dissipator} we see that the effect of engineered nonlocal dissipation is best understood in momentum space, where it takes the form of a non-uniform damping for the normal modes. In particular, the normal mode with $k=\phi$ is a dark mode of the non-local dissipator, which does not experience any dissipation as $\gamma_{\phi}=0$.  Therefore the relative phase $\phi$ allows to select a single \emph{non-dissipative normal mode}. This behavior is to be contrasted with intrinsic, i.e.~non-engineered, single-photon loss, which instead leads to homogeneous decay of all normal modes. For now we assume that~\eqref{dissipator} is the only dissipation channel acting on the system; effects of intrinsic photon loss will be accounted for in Sec.~\ref{s:Loss}.

Writing the Hamiltonian in the plane-wave basis we get 
\begin{gather}\label{Hmomentum}
 \hat{H} _U=   \frac{U}{N} \sum_{k_1,k_2,k_3,k_4}\delta_{k_1+k_2,k_3+k_4}\hat{b}_{k_1}^{\dagger} \hat{b}_{k_2}^{\dagger} \hat{b}_{k_3} \hat{b}_{k_4},\\
\hat{H}_G =  \sum_k \left( G \,\hat{b}^{\dagger}_{k}  \hat{b}^{\dagger}_{\theta - k} + \textrm{h.c.}\right),
\end{gather}
where, $\delta_{k_1+k_2,k_3+k_4}$ is Kronecker delta and the arguments $k_1+k_2$ and $k_3+k_4$ are defined modulo $2\pi$. We can see that the Kerr nonlinearity $\hat{H}_U$ corresponds to a four-wave mixing process, which scatters a photon pair $k_1$ and $k_2$ to a photon pair $k_3$ and $k_4$ conserving the total quasi-momentum. 
The two-photon drive $\hat{H}_G$ corresponds to the creation of a photon pair with a total quasi-momentum $\theta$.

\begin{figure*}[t]
\centering
\includegraphics[width=1\linewidth]{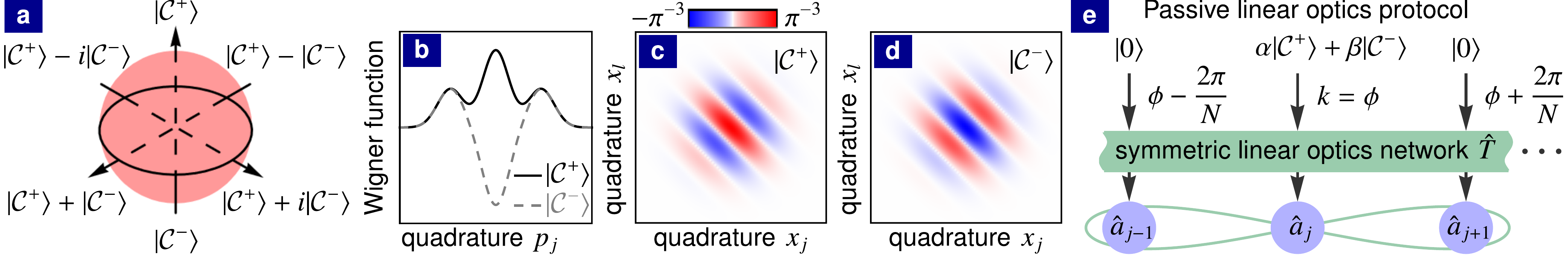}
\caption{Multi-mode decoherence-free subspace. (a) Two-dimensional subspace consisting of dark states of the Kerr-resonator array corresponding to all superpositions of cat states $|\mathcal{C}^{\pm}\rangle$. (b) $N$-mode Wigner function $W\left(x_1,...,x_N,p_1,...,p_N\right)$ along $p_j =  p_N$ with $x_j = 0$ (for all $j$) for the even parity cat state $|\mathcal{C}^{+}\rangle$ (black solid lines) and for the odd parity cat state $|\mathcal{C}^{-}\rangle$ (gray dashed lines) stabilized in normal mode $\phi=0$, where $\hat{x}_j\propto \hat{a}_j^{\dagger} + \hat{a}_j$ and $\hat{p}_j\propto i(\hat{a}_j^{\dagger} - \hat{a}_j)$ are quadratures of resonator $j$. Similarly to single-mode cat states, we can see two satellite peaks corresponding to the coherent states $\bigotimes_{j=1}^N |\pm\zeta_j\rangle_j$. (c) and (d) Interference fringes of the $N$-mode Wigner function in the $x_j$-$x_l$ plane (for any $j\neq l$) for $|\mathcal{C}^{+}\rangle$ and $|\mathcal{C}^{-}\rangle$, respectively, a unique feature of multi-mode cat states absent for single-mode cat states. (e) Mixing of a single-mode cat state with $N-1$ vacuum states on a symmetric linear optics network that corresponds to expressing the multi-mode cat state, which is stabilized in the normal mode $\phi$, in the basis of local modes.}
\label{multi DFS}
\end{figure*}

\section{Multi-mode Decoherence free subspace}\label{s:DFS}

Our first goal is to characterize the steady states of the master equation (\ref{master}). In particular, we look for steady states in the form of pure states $|\Psi\rangle$, which are referred to as \textit{dark states} \cite{kraus2008}. Combining Eqs.~\eqref{master} and~\eqref{dissipator} it readily follows that dark states must satisfy the following two conditions 
\begin{gather}
\hat{b}_k|\Psi\rangle = 0,\,\,\textrm{for all}\,\,k\neq\phi,\label{cond 1}\\
\hat{H}|\Psi\rangle = \epsilon |\Psi\rangle.\label{cond 2}
\end{gather}
The condition (\ref{cond 1}) is satisfied if all modes $k\neq\phi$ are in the vacuum, i.e.~$|\Psi \rangle = \left(\bigotimes_{k\neq\phi}|0\rangle_k\right)\otimes|\psi\rangle_{\phi}$, from which it is clear that the role of engineered dissipation is to select a single normal mode that is (possibly) populated in the infinite-time limit. The dark state condition (\ref{cond 2}) can then be written as
\begin{equation}\label{cond zeta}
\sum_k\hat{b}_k^{\dagger}\hat{b}_{2\phi - k}^{\dagger}\left(\hat{b}_{\phi}^2 - \zeta^2\right)|\Psi\rangle = \zeta^{*\,2} \left(\hat{b}^2_{\phi} - \frac{\epsilon}{G}\right) |\Psi\rangle,
\end{equation}
where $\zeta = i\sqrt{\frac{NG}{U}}$ and we hereafter set $\theta = 2\phi$ \footnote{Phases $\phi$ and $\theta$ are connected by the local gauge transformation $\hat{a}_j\rightarrow\hat{a}_je^{ij\nu}$ yielding $\phi\rightarrow\phi+\nu$ and $\theta\rightarrow \theta + 2\nu$. While these phases are gauge dependent, the phase difference $2\phi - \theta$ is gauge invariant. For  $2\phi - \theta = 0$ (corresponding to $\theta = 2\phi $ set in the main text), we are able to express the dark state condition in the form of Eq.~\eqref{cond zeta} allowing us to write exact closed-form solutions for the dark states.}. 
For $\epsilon = G\zeta^2$, overall destructive interference is guaranteed if the degenerate photon pairs annihilated in mode $\phi$ satisfy the following condition
\begin{equation}\label{cond factor}
\left(\hat{b}_{\phi}^2 - \zeta^2\right)|\psi\rangle_{\phi} = \left(\hat{b}_{\phi} - \zeta\right) \left(\hat{b}_{\phi} + \zeta\right)|\psi\rangle_{\phi} = 0,
\end{equation}
in which case both sides of Eq.~(\ref{cond zeta}) vanish.

Eq.~(\ref{cond factor}) represents a simplified dark state condition. From its factorized form we can deduce that the two coherent states $\hat{b}_{\phi}|\pm\zeta\rangle_{\phi} = \pm\zeta|\pm\zeta\rangle_{\phi}$ and their superpositions are dark states. All possible dark states span a two-dimensional subspace shown in Fig.~\ref{multi DFS}a, whose basis states are superpositions of coherent states, known as Schr\" odinger cat states $\nobreak{|\mathcal{C}^{\pm}\rangle_{\phi} = \mathcal{N}_{\pm}\left(|\zeta\rangle_{\phi}\pm|-\zeta\rangle_{\phi} \right)}$, where $\mathcal{N}_{\pm} = \left[2\left(1\pm e^{-2|\zeta|^2}\right)\right]^{-1/2}$. Note that the cat states $|\mathcal{C}^{\pm}\rangle_{\phi}$ are exactly orthogonal in contrast to the coherent states $|\pm\zeta\rangle_{\phi}$, which have a finite overlap. For convenience, from now on we drop the subscript $\phi$ from the state of the non-dissipative mode. Due to the linearity of the master equation, incoherent mixtures of dark states are also steady states. All dark states and their incoherent mixtures span a decoherence-free subspace (DFS)
\begin{align}\label{DFS}
\hat{\rho}_\mathrm{ss} =& c_{++}|\mathcal{C}^+\rangle\langle\mathcal{C}^+| + c_{--}|\mathcal{C}^-\rangle\langle\mathcal{C}^-|\nonumber\\
& + \left(c_{+-}|\mathcal{C}^+\rangle\langle\mathcal{C}^-| + \textrm{h.c.}  \right),
\end{align}
where the coefficients $c_{\mu}$ depend on the initial state. 
Even though steady states which are neither dark states nor their incoherent mixtures can in principle exist \cite{kraus2008}, we numerically verified that the master equation (\ref{master}) has no steady states lying outside of the DFS (\ref{DFS}) (see Appendix~\ref{app:conserved} for more details).

Notice that DFSs with a similar structure have recently been realized in circuit QED~\cite{leghtas2015,touzard2018,Lescanne2020,Wang2019,grimm2020}. However, compared to these cases, here the crucial difference is that the DFS is spanned by \emph{multi-mode} Schr\"odinger cat states.
In fact, the ground state of a single non-dissipative KPR supports a two-dimensional manifold  spanned by a \emph{single-mode} cat states of amplitude $\zeta = i\sqrt{G/U}$~\cite{puri2017,Wielinga1993}. The steady state of a parametrically driven resonator with engineered two-photon dissipation (hereafter simply referred to  as two-photon driven dissipation)~\cite{mirrahimi2014} also supports a similar manifold, in this case with amplitude $\zeta = \sqrt{-iG/\eta}$, where $\eta$ is the rate of two-photon dissipation (see Sec.~\ref{s:TwoPhotonLoss} for more details). We now show that in our case the states in the DFS~\eqref{DFS} are multi-mode.

The cat states are encoded in an arbitrary normal mode $k=\phi$. As the normal modes are delocalized over the entire array, the cat states that we discuss here are highly nonlocal. This can be seen explicitly by moving into the basis of local modes
\begin{equation}\label{many cats}
|\mathcal{C^{\pm}}\rangle = \mathcal{N}_{\pm} \left(\bigotimes_{j=1}^N |\zeta_j\rangle_j\pm\bigotimes_{j=1}^N |-\zeta_j\rangle_j\right),
\end{equation}
where $\zeta_j = \frac{\zeta}{\sqrt{N}}e^{-ij\phi}$ \footnote{Eq.~\eqref{many cats} can be derived using \unexpanded{$|\pm\zeta\rangle = \mathcal{D}_{\hat{b}_{\phi}}(\pm\zeta)|0\rangle$} and \unexpanded{$ \mathcal{D}_{\hat{b}_{\phi}}(\pm\zeta) = \prod_{j=1}^N\mathcal{D}_{\hat{a}_{j}}(\pm\zeta_j)$} which follows from \unexpanded{$\hat{b}_k = \frac{1}{\sqrt{N}}\sum_{j=1}^N e^{ijk}\hat{a}_j$ and $[\hat{a}_j,\hat{a}^{\dagger}_m] = \delta_{jm}$}, where \unexpanded{$\mathcal{D}_{\hat{c}}(\zeta)=\exp(\zeta\hat{c}^{\dagger} - \zeta^*\hat{c})$ is the displacement operator and $|0\rangle = \bigotimes_{j=1}^N|0\rangle_j= \bigotimes_{k}|0\rangle_k$} is the $N$-mode vacuum state.}. Controlling the phase $\phi$ (and that of the drive $\theta = 2\phi$) it is possible to prepare different instances of these cat states. In particular, if $\phi$ is an even multiple of $\pi$, Eq.~\eqref{many cats} resembles a GHZ-like state states, with real and opposite amplitudes; if $\phi$ is an odd multiple of $\pi$, the amplitudes being superimposed are staggered.

These pure states are the multi-mode version of standard Schr\"odinger cat states~\cite{Ansari1994}. They have the remarkable property of being
both multi-photon and multi-mode quantum superpositions, and possess genuine multipartite entanglement~\cite{sanders2012,wang2001,Vogel2014}. Moreover, such entanglement is of the most useful kind, namely non-Gaussian, which makes~\eqref{many cats} a resource for quantum metrology~\cite{joo2011,zhang2013,Gilchrist2004}, quantum information based on  continuous variables~\cite{vanenk2001,braunstein2005,sanders2012} and continuous-variable quantum computation~\cite{Jeong2002}. 
We also stress the potential of multi-mode cat states  is still largely unexplored. In fact, theoretical studies of multi-mode cat states has been mostly limited to $N=2$ (also known as entangled coherent states~\cite{sanders2012}) due to the current lack of preparation protocols that are at the same time scalable and robust to errors.

To contrast the multi-mode cat states with standard single-mode cat states, we plot in Fig.~\ref{multi DFS}b, \ref{multi DFS}c and \ref{multi DFS}d the $N$-mode Wigner function $W\left(x_1,...,x_N,p_1,...,p_N\right)$ of the cat states $|\mathcal{C}^{\pm}\rangle$ stabilized in the normal mode $\phi=0$ \footnote{The joint Wigner function of the $N$-mode state $\hat{\rho}$ is \unexpanded{$W(\vec{x},\vec{p})=\pi^{-N}\int_{\mathbb{R}^{N}}\textrm{d}\vec{y}\,\exp(-2\,i\,\vec{y}\cdot\vec{p})\langle \vec{x}+\vec{y}|\hat{\rho}|\vec{x}-\vec{y}\rangle$}, where $\vec{x}$, $\vec{p}$ and $\vec{y}$ are real-valued $N$-dimensional vectors, and $x_j$ and $p_j $ are quadratures of resonator $j$. For cat states stabilized in mode $\phi\neq0$, the amplitudes $\zeta_j$ and  $\zeta_{j+1}$ at neighboring resonators have a relative phase difference $\phi$ [see Eq.~\eqref{many cats}]. This results in a relative rotation between the local coherent states \unexpanded{$|\zeta_j\rangle_j$} and \unexpanded{$|\zeta_{j+1}\rangle_{j+1}$} in phase space by an angle $\phi$.}. The unique feature of multi-mode cat states is the quantum interference in multi-mode phase space, which manifests itself in interference fringes of the Wigner function in the $x_j$-$x_l$ plane (for any $j\neq l$) depicted in \ref{multi DFS}c and \ref{multi DFS}d for $|\mathcal{C}^{+}\rangle$ and $|\mathcal{C}^{-}\rangle$, respectively. This feature is absent for single-mode cat states.

An interesting observation comes from comparing our protocol with known techniques for the generation of multi-mode cat states in linear optics~\cite{Gilchrist2004}. In a linear optics network, multi-mode cat states can be prepared by mixing a \emph{single-mode cat state}  $|\mathcal{C}^{\pm}\rangle$ with $N-1$ vacua. The optics network is represented by a unitary $\hat{T}$ performing the linear transformation $\nobreak{\hat{a}_j = \hat{T}\hat{b}_{2\pi j/N}\hat{T}^{\dagger} = \sum_k T_{jk}\hat{b}_k}$ of the input modes $\hat{b}_k$, where $T_{jk}= \frac{1}{\sqrt{N}} e^{-ijk}$.
In our case, as depicted in Fig.~\ref{multi DFS}e, this linear transformation is automatically implemented by engineering the state of a normal mode. The obvious differences are that our method allows to stabilize a cat state in stationary (cavity) modes rather than itinerant modes, and that it does not rely on an external resource (i.e., the input single-mode cat state). Besides that, our strategy possesses yet another  fundamental advantage. For a single-mode cat state  of amplitude $\zeta$ at the input of an interferometer, the corresponding multi-mode cat state at the output has rescaled local amplitudes $\zeta/\sqrt{N}$; this feature reflects the passive character of the optical network, which does not pump energy into the system. On the other hand, in the driven-dissipative process that we exploit to stabilize the multi-mode cat state, the total power $\propto NG$ injected into the array increases with the number of resonators $N$.
This in turn  allows for the amplitude of the state $|\zeta|\propto\sqrt{N}$ to increase with $N$. 
As a result, the multi-mode cat states can be distributed over an arbitrary number of local modes $N$ with the local amplitudes $|\zeta_j| = |\zeta|/\sqrt{N} = \sqrt{G/U}$ independent of $N$.
This scaling of amplitudes allows for increasing noise bias by using multi-mode cat states instead of single-mode cat states, which we discuss in detail in Sec.~\ref{bias}.

\section{Steady-state preparation of multi-mode cat states}\label{s:SteadyCat}

In the last section we showed that the steady states of Eq.~\eqref{master} belong to a DFS  spanned by multi-mode cat states. Therefore, depending on the initial state $\hat{\rho}_{\rm in}$, the state of the array asymptotically converges towards a pure dark state $\alpha|\mathcal{C}^{+}\rangle + \beta|\mathcal{C}^-\rangle$ or their incoherent mixtures. Given the unique properties displayed by these cat states [cf.~the discussion following Eq.~\eqref{many cats}], we now focus on the stabilization of $|\mathcal{C}^{\pm}\rangle$, i.e., we wish to find initial states $\hat{\rho}_{\rm in}$ for which the array asymptotically approaches these states. At the same time, we should keep in mind that the choice of the initial state $\hat{\rho}_{\rm in}$ must be experimentally feasible. 
 
\begin{figure}[t]
\centering
\includegraphics[width=0.9\linewidth]{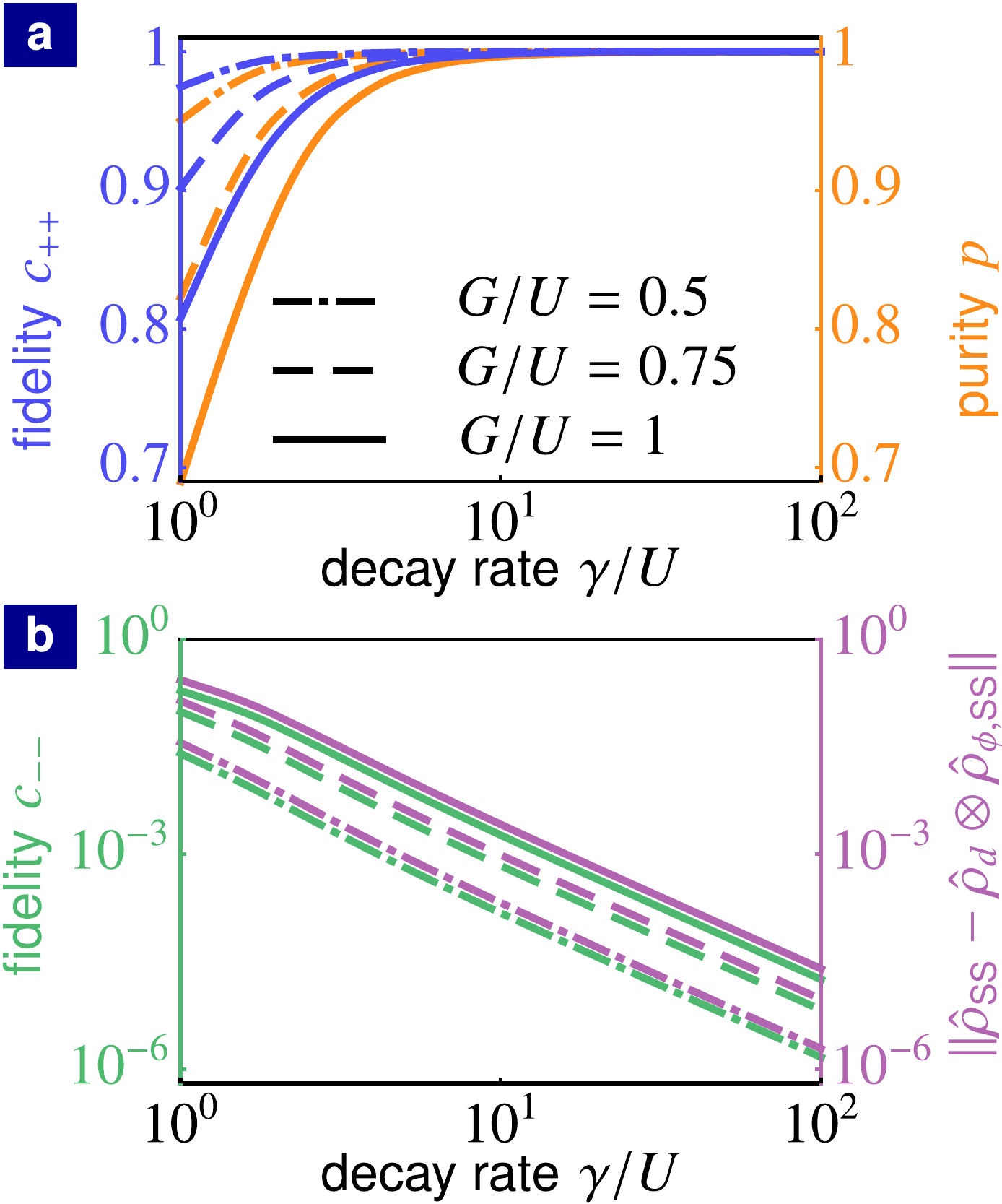}
\caption{Steady state for the array initially in the vacuum state. (a) Fidelity $c_{++}$ with $|\mathcal{C}^+\rangle$ (blue lines) and purity $p$ (orange lines) as a function of the decay rate $\gamma$ for different values of the two-photon-pump strength $G$. 
(b) Fidelity $c_{--}$ with $|\mathcal{C}^-\rangle$ (green lines) as well as difference $\lVert\hat{\rho}_\mathrm{ss} - \hat{\rho}_d \otimes \hat{\rho}_{\phi,\mathrm{ss}}\rVert$ (purple lines) between the steady state $\hat{\rho}_{\mathrm{ss}}$ of the full master equation (\ref{master}) and the steady state $\hat{\rho}_{\phi,\mathrm{ss}}$ of the effective master equation (\ref{master zeno}) in the Zeno limit as a function of the decay rate $\gamma$ for different two-photon-pump strengths $G$ [identical to panel (a)]. (Parameters: $N=3$.)}
\label{fidelity}
\end{figure}

In analogy with the preparation of a single-mode even parity cat state, which can be stabilized starting from the vacuum~\cite{mirrahimi2014,puri2017}, we initialize the array in the $N$-mode vacuum state $\nobreak{\hat{\rho}_{\rm in}=|0\rangle\langle0|}$. In Fig.~\ref{fidelity} we plot the fidelity $c_{++}$ $(c_{--})$ of the steady state with the multi-mode cat state $|\mathcal{C}^+\rangle$ ($|\mathcal{C}^-\rangle$) as a function of the non-local decay rate for different two-photon pump strengths (see Appendix~\ref{app:conserved} for more details about the numerical simulations). We can see that $c_{++}$ (blue lines in Fig.~\ref{fidelity}a) is in general smaller than unity and $c_{--}$ (green lines in Fig.~\ref{fidelity}b) is non-zero. Therefore, at variance with the single-mode case, starting from the vacuum does not guarantee to reach the target steady state $|\mathcal{C}^+\rangle$. Nonetheless, Fig.~\ref{fidelity} provides a clear indication that, for sufficiently strong non-local dissipation, fidelities with $|\mathcal{C}^+\rangle$ arbitrarily close to unity can be achieved, which renders the preparation of multi-mode cat states from vacuum feasible in the strong dissipation regime.

The cat state $|\mathcal{C}^+\rangle$ can be transformed into a superposition $\cos\alpha|\mathcal{C}^+\rangle+i\sin\alpha|\mathcal{C}^-\rangle$ of cat states $|\mathcal{C}^+\rangle$ and $|\mathcal{C}^-\rangle$ by a rotation within the stabilized DFS induced by a weak single-photon drive as demonstrated for single-mode cat states \cite{touzard2018} (see Appendix~\ref{app:Rot} for more details).

The target steady state $|\mathcal{C}^+\rangle$ is not deterministically approached from the initial vacuum for all $\gamma$ since photon parity is not conserved during the time evolution \footnote{Note that the dynamics of an array with only two resonators, i.e. $N=2$, is fundamentally different from that of larger arrays with $N\geq3$ as the photon parity $\hat{\mathcal{P}}_{\phi} = \exp{\bigl(i\pi \hat{b}_{\phi}^{\dagger}\hat{b}_{\phi}\bigr)}$ of mode $\phi$ is conserved, where the quasi-momentum can take only values $\phi = 0,\pi$. As a result, the even-parity cat state \unexpanded{$|\mathcal{C}^{+}\rangle$} is deterministically approached from the initial vacuum state for any $\gamma$. However, in this manuscript we focus on arrays $N\geq3$ for which the photon parities $\hat{\mathcal{P}}$ and $\hat{\mathcal{P}}_{\phi}$ are not conserved.}.
The multi-mode cat states $|\mathcal{C^{\pm}}\rangle$ spanning the DFS have a well-defined parity as they are $\pm 1$ eigenstates of the generalized parity operator $\hat{\mathcal{P}} = \exp{\bigl(i\pi\sum_{j=1}^N \hat{a}_j^{\dagger}\hat{a}_j\bigr)}$. However, the photon parity is not a conserved quantity of the master equation \eqref{master} --- a conserved quantity is an operator $\hat{J}$ such that  $\mathcal{L}^{\dagger}\hat{J} = 0$, while $\mathcal{L}^{\dagger}\hat{\mathcal{P}} = -2\sum_{k\neq\phi}\hat{b}_k^{\dagger}\hat{b}_k\hat{\mathcal{P}}\neq 0$, where $\mathcal{L}^{\dagger}$ is the adjoint Liouvillian~\cite{albert2014}.

For the initial vacuum, the system starts off in the even-parity subspace but during the time evolution it leaks to the odd-parity subspace leading to non-unit (non-zero) values of the fidelity $c_{++}$ $(c_{--})$ shown in Fig.~\ref{fidelity}.
  For $\gamma\sim G,U$ this leakage leads to a decreased purity $p$ of the steady state (orange lines in Fig.~\ref{fidelity}a), while for sufficiently strong dissipation the leakage is suppressed and the generalized parity is approximatively  conserved (to an arbitrary degree of accuracy).
 This further confirms that the regime of strong non-local dissipation is relevant for steady-state preparation of the even cat state $|\mathcal{C}^{+}\rangle$.
 In this regime, the dynamics at all times can be captured by an effective single-mode theory as we will show in the next section.
 On the other hand, due to non-conservation of parity and the many-body interaction $\hat{H}_U$, for values of the dissipation rate comparable to the Kerr nonlinearity we expect the \emph{transient dynamics} to exhibit genuine many-body features. 

\section{The Zeno limit}\label{s:Zeno}
We now have numerical evidence that, in the limit of dominant non-local dissipation, the even multi-mode cat state $|\mathcal{C}^{+}\rangle$ can be deterministically prepared from the $N$-mode vacuum. In order to better explain this behavior, in this section we develop an analytical insight into the regime of large non-local decay rate $\gamma$.
For $\gamma\gg G,U$, the many-body interaction $\hat{H}_U$ between normal modes  is suppressed: all decaying modes $k\neq\phi$ are strongly damped and only the single normal mode $\phi$ is populated. Transitions from the vacuum state to excited states of the decaying modes $k\neq\phi$ are inhibited, i.e.~they are in the vacuum \emph{at all times}. This is analogous to the quantum Zeno effect \cite{misra1977,koshino2005}, i.e.~the inhibition of quantum transitions due to a frequent projective measurement of a quantum system. In this analogy, the measurement is enacted by the engineered environment via the non-local dissipation $\mathcal{L}_{\gamma}=\sum_{k}\gamma_k\mathcal{D}[\hat{b}_k]$, which continuously  projects all decaying modes onto the vacuum state. In the following, we refer to the limit of a large non-local decay rate $\gamma\gg G,U$ as the Zeno limit.

In Appendix~\ref{app:Zeno} we show that in this limit the reduced density matrix  $\hat{\rho}_{\phi}$ of the non-decaying mode $\phi$ is described by the effective master equation
\begin{equation}\label{master zeno}
\dot{\hat{\rho}}_{\phi}=  -i\left[ \hat{H}_{\phi},\hat{\rho}_{\phi}\right] + \Gamma \mathcal{D}\left[\hat{b}_{\phi}^2 - \zeta^2 \right]\hat{\rho}_{\phi} = \mathcal{L}_{\phi} \hat{\rho}_{\phi},
\end{equation}
where $\mathcal{L}_{\phi} $ is the effective single-mode Liouvillian. This effective master equation is derived by treating the Hamiltonian $\hat{H}$ in~\eqref{master} as a perturbation to the dominant dissipation $\mathcal{L}_{\gamma} $ and projecting all decaying modes $k\neq\phi$ onto the vacuum state. Perturbation theory to first order in $\hat{H}$ describes a \emph{unitary Zeno dynamics} governed by the effective Hamiltonian $\hat{H}_{\phi} =    \frac{U}{N}\left(\hat{b}_{\phi}^{\dagger \,2} - \zeta^{*\,2} \right)\left(\hat{b}_{\phi}^2 - \zeta^2 \right)$, namely that of a single-mode KPR~\cite{puri2017}. 
On the other hand, processes which are of second order in  $\hat{H}$ lead to an \emph{effective two-photon driven dissipation} term described by the dissipator $\mathcal{D}\left[\hat{b}_{\phi}^2 - \zeta^2 \right] $ and occurring at a rate $\Gamma = 4\frac{U^2}{N^2}\sum_{{k\neq\phi}}\frac{1}{\gamma_k}$; this effective dissipation is facilitated by virtual excitations of the decaying modes $k\neq\phi$. We thus come to an important conclusion: the emergence of  effective two-photon dissipation is the mechanism responsible for the relaxation towards the steady state of the non-decaying mode.

We note that, since it involves second-order processes, dissipation sets in only at long times $t\sim\frac{1}{\Gamma} \sim\frac{\gamma N^2}{U^2}$. At short times $t\sim\frac{N}{U}$, the mode $\phi$ undergoes a unitary Zeno dynamics governed by the effective Hamiltonian $\hat{H}_{\phi}$. A general (model-independent) treatment of Lindblad master equations with strong dissipation acting on a subset of the degrees of freedom, is discussed in detail in Ref.~\cite{popkov2018}, including the derivation of the unitary Zeno dynamics and effective weak dissipation of non-decaying modes. 

In Fig.~\ref{fidelity}b, we plot the difference $\lVert\hat{\rho}_\mathrm{ss} - \hat{\rho}_d \otimes \hat{\rho}_{\phi,\mathrm{ss}}\rVert$ (purple lines) between the steady state $\hat{\rho}_\mathrm{ss}$ of the full master equation (\ref{master}) and the steady state $\hat{\rho}_d\otimes\hat{\rho}_{\phi,\mathrm{ss}}$ of the effective master equation (\ref{master zeno}) as a function of the decay rate $\gamma$ for different values of $\frac{G}{U}$, where $\hat{\rho}_{d} = \bigotimes_{k\neq\phi} |0\rangle_k\langle0|$ is the $\left(N-1\right)$-mode vacuum state and $\lVert \hat{A} \rVert^2 = {\rm Tr }\left[ \hat{A}^{\dagger}\hat{A} \right]$ is the Hilbert-Schmidt norm. We can see that the difference vanishes as $\mathcal{O}\left(\frac{U^2}{\gamma^2}\right)$ with the increasing decay rate $\gamma$. This confirms that, for large enough $\gamma$, the effective master equation (\ref{master zeno}) accurately describes the dynamics of the Kerr-resonator array at long times.

In the Zeno limit, the photon parity $\mathcal{P}_{\phi} = (-1)^{\hat{b}_{\phi}^{\dagger}\hat{b}_{\phi}}$ is a \emph{conserved quantity} of the effective Liouvillian, i.e., $\mathcal{L}_{\phi}^{\dagger}\mathcal{P}_{\phi}=0$, as it commutes with both the Hamiltonian $\hat{H}_{\phi}$ and the jump operator $\hat{b}_{\phi}^{2} - \zeta^2$~\cite{albert2014}. 
Therefore, in this limit there is no leakage to the odd-parity subspace. This provides a neat explanation for the behavior observed in Fig.~\ref{fidelity}a for large $\gamma$ and proves that the dissipative dynamics of the master equation (\ref{master}) allows for the steady-state preparation of the even cat state $|\mathcal{C}^{+}\rangle$ from the vacuum state $|0\rangle$ with a fidelity arbitrarily close to unity.

The rate of convergence towards the steady state is given by the dissipative gap $\Delta_d$, which is determined from the spectrum of the Liouvillian as the smallest non-vanishing real part of the eigenvalues \cite{albert2014}. Since the rate of two-photon dissipation is inversely proportional to the non-local decay rate, i.e., $\Gamma\propto\gamma^{-1}$, the dissipative gap $\Delta_d \propto \gamma^{-1}$ retains the same dependence (see Appendix~\ref{app:gap} for more details). For $|\zeta|\gtrsim1$, the dissipative gap can be approximated as 
\begin{equation}\label{eq gap}
\Delta_d \approx 8\frac{U^2}{N^2}\left(\sum_{{k\neq\phi}}\frac{1}{\gamma_k}\right)|\zeta|^2\propto \frac{U G}{N\gamma}.
\end{equation}
Eq.~\eqref{eq gap} is an important result as it expresses a trade-off between the suppression of the leakage to the odd-parity subspace and the time required to reach the steady-state manifold. The price of an accurate (i.e., near-unit-fidelity) preparation of the even multi-mode cat state $|\mathcal{C}^{+}\rangle$ (from the vacuum) is a slow convergence. Although this does not \emph{per se} preclude the stabilization of $|\mathcal{C}^{+}\rangle$, it becomes problematic in the presence of competing decoherence mechanisms, such as intrinsic photon loss. In Sec.~\ref{s:Loss}, we will discuss the impact of this source of decoherence in the Kerr-resonator array and how it limits the steady-state preparation. Moreover, increasing the number of modes $N$ in the superposition leads to further increase of the convergence time, which poses a theoretical limit to the scalability of our approach.

\section{Alternative model including two-photon loss}\label{s:TwoPhotonLoss}
To avoid the trade-off between the suppression of the leakage to the odd-parity subspace and the rate of convergence towards the steady state, we consider an alternative model which allows the stabilization of multi-mode cat states $|\mathcal{C}^{\pm}\rangle$ while at the same time featuring \emph{a large dissipative gap}.
This model is based on an array of dissipatively coupled linear resonators ($U=0$), subject to local two-photon loss at rate $\eta$ and two-photon pump with the amplitude $G$, as sketched in Fig.~\ref{scheme}b. The model only requires a small adjustment with respect to our previous model, namely replacing the on-site Kerr nonlinearity with engineered two-photon loss. This alternative model is described by the master equation 
\begin{equation}\label{master loss}
\dot{\hat{\rho}} = 2\eta\sum_{j=1}^{N}\mathcal{D}\left[a_j^2 - e^{-2ij\phi}\frac{\tilde{\zeta}^2}{N}\right]\hat{\rho} +\sum_{k}\gamma_k\mathcal{D}\bigl[\hat{b}_k\bigr]\hat{\rho} = \mathcal{L}\hat{\rho},
\end{equation}
where $\tilde{\zeta} = \sqrt{-iNG/\eta}$. The local two-photon driven dissipation described by the term $\mathcal{D}[a_j^2 - e^{-2ij\phi}\frac{\tilde{\zeta}^2}{N}]$ is known to stabilize a DFS whose basis states are even and odd single-mode cat states \cite{gilles1994,mirrahimi2014}. The local dissipator has been realized experimentally in Refs.~\cite{leghtas2015,touzard2018,Lescanne2020}.

\begin{figure}[t]
\centering
\includegraphics[width=0.9\linewidth]{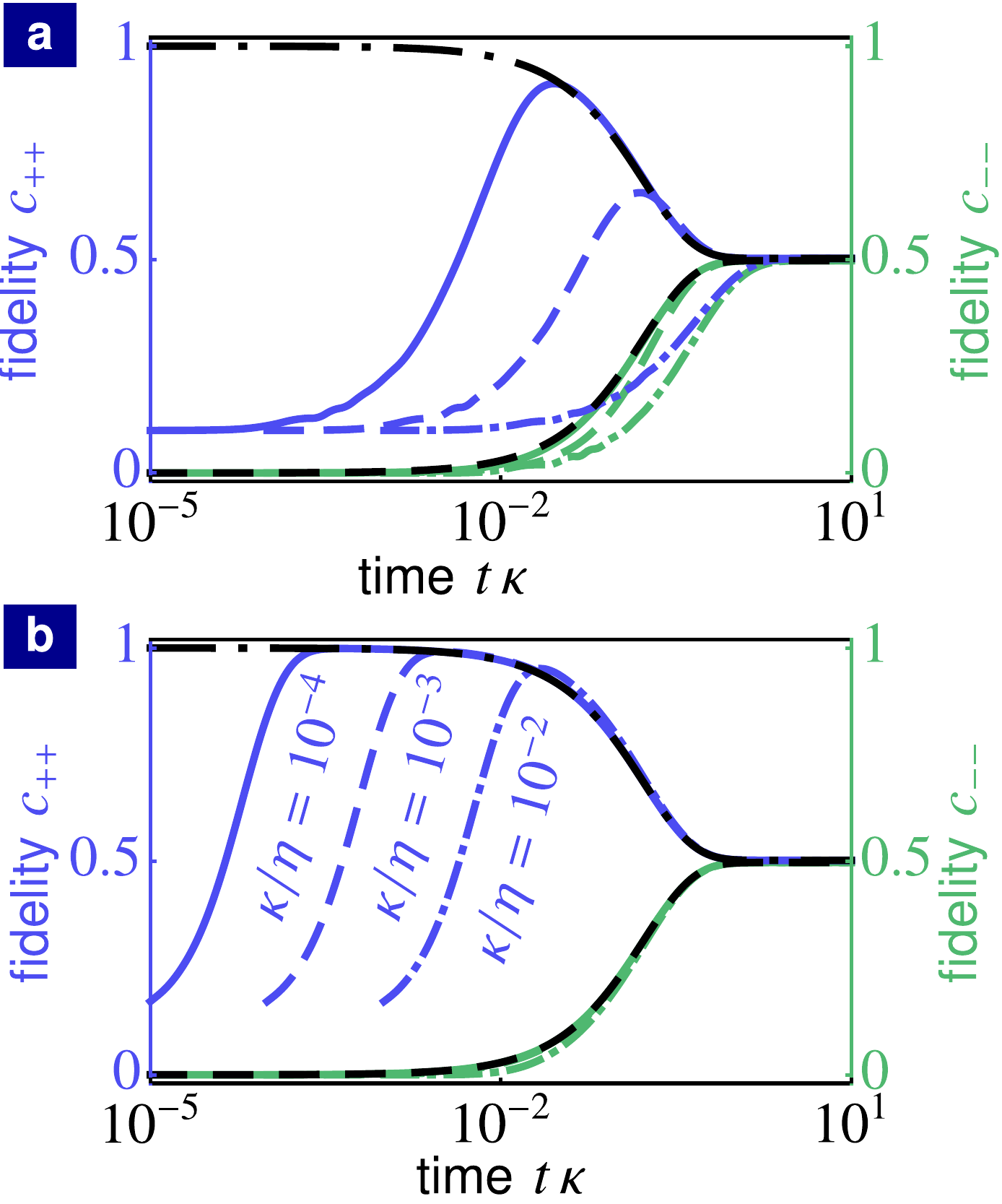}
\caption{Dissipative preparation of the even cat state $|\mathcal{C}^+\rangle$ from the vacuum state in the presence of intrinsic photon loss in the Zeno limit. Fidelity of the instantaneous state $\hat{\rho}_d\otimes\hat{\rho}_{\phi}(t)$ with $|\mathcal{C}^+\rangle$ (blue line) and with $|\mathcal{C}^-\rangle$ (green line)  as a function of time $t$ for (a) the Kerr-resonator array and (b) the alternative model with local two-photon loss as well as for intrinsic single-photon loss $\kappa/U=10^{-4}$ (solid lines), $\kappa/U=10^{-3}$ (dashed lines) and $\kappa/U=10^{-2}$ (dot-dashed lines). 
Black dot-dashed line and black dashed line show the fidelity $c_{++}$ and the fidelity $c_{--}$, respectively, for the array initially in the state $|\mathcal{C}^+\rangle$. 
[Parameters: (a) $N=3$, $G/U = 1$ and $\gamma/U = 100$, (b) $N=3$, $G/\eta = 1$.
Truncation: $M_{\phi}=20$].}
\label{fidelity 2p}
\end{figure}

The dissipatively coupled array of resonators described by the master equation (\ref{master loss}) exhibits the same DFS as the Kerr-resonator array [see Eq.~(\ref{DFS})] with a modified amplitude of the cat states $\zeta\rightarrow\tilde{\zeta} = e^{-i\frac{3\pi}{4}} \zeta$ (see Appendix~\ref{app:SS al} for a detailed derivation of the DFS). In the Zeno limit of strong non-local dissipation $\gamma\gg G,\eta$, the time evolution of mode $\phi$ is described by an effective master equation 
\begin{equation}
\dot{\hat{\rho}}_{\phi} = 2 \frac{\eta}{N} \mathcal{D}[\hat{b}_{\phi}^2 - \tilde{\zeta}^2]\hat{\rho}_{\phi}\,,\label{zeno al}
\end{equation} 
which is derived by employing the first-order perturbation theory where $\mathcal{L}_{\eta} = 2\eta\sum_{j=1}^{N}\mathcal{D}[a_j^2 - e^{-2ij\phi}\frac{\tilde{\zeta}^2}{N}]$ is treated as a perturbation to the dominant dissipation $\mathcal{L}_{\gamma}$ (see Appendix~\ref{app:Zeno al} for more details). In comparison to the Kerr-resonator array [cf.~Eq.~(\ref{master zeno})], \emph{the unitary Zeno dynamics is absent} and processes that are of first order in the perturbation $\mathcal{L}_{\eta}$ give rise to the dissipation $2 \frac{\eta}{N} \mathcal{D}[b_{\phi}^2 - \tilde{\zeta}^2] $ leading to the convergence towards the DFS. This is in contrast to the Kerr-resonator array, where the convergence towards the DFS is due to second-order processes involving virtual excitations of modes $k\neq\phi$.

For this second model we can completely suppress the leakage to the odd-parity subspace in the Zeno limit (the photon parity $\mathcal{P}_{\phi}$ is a conserved quantity) and at the same time maintain a large dissipative gap 
\begin{equation}
\Delta_d\approx2\eta|\tilde{\zeta}|^2/N = 2G,
\end{equation}
which increases with the strength $G$ of the two-photon drive (see Appendix~\ref{app:gap} for more details). Note that, in contrast to the Kerr-resonator array~\eqref{eq gap}, the dissipative gap $\Delta_d$ is independent of $\gamma$ in the Zeno limit. Importantly, the gap is also independent of system size $N$, making the preparation of cat states in large arrays feasible.

\section{Effects of intrinsic single-photon loss}\label{s:Loss}

In all current circuit QED implementations, unwanted intrinsic photon loss is the dominant source of decoherence, both for the case of the Kerr-resonator array~\cite{kirchmair2013, grimm2020} and for the model with two-photon driven dissipation~\cite{leghtas2015,touzard2018,Lescanne2020}. Photon loss at each resonator is described by the Liouvillian $\mathcal{L}_{\kappa}\hat{\rho} = \kappa\sum_{j=1}^{N} \mathcal{D}\left[\hat{a}_j \right]\hat{\rho} $. We take intrinsic loss into account by including it in the master equation~\eqref{master}, thus obtaining $\dot{\hat{\rho}} = \left(\mathcal{L} + \mathcal{L}_{\kappa}\right)\hat{\rho}$. In contrast to the non-local dissipation $\mathcal{L}_\gamma$, which leads to non-uniform dissipation in momentum space, intrinsic photon loss induces uniform dissipation of all normal modes, i.e., $\mathcal{L}_{\kappa}\hat{\rho} = \kappa\sum_{k} \mathcal{D}[\hat{b}_k ]\hat{\rho}$. For modes $k\neq\phi$, intrinsic photon loss only increases the rate of single-photon dissipation $\gamma_k \rightarrow \gamma_k +\kappa$ which results in their faster decay.

Crucially, intrinsic photon loss gives rise to the single-photon dissipation $\mathcal{D}[\hat{b}_{\phi}]$ of mode $\phi$ in addition to the two-photon driven dissipation $\mathcal{D}[\hat{b}_{\phi}^2-\zeta^2]$. We remind the reader that the latter is due to local two-photon loss in the model of Sec.~\ref{s:TwoPhotonLoss} and arises due to the coupling to modes $k\neq\phi$ in the Kerr-resonator array, see Sec.~\ref{s:Zeno}. The single-photon dissipation opens a decoherence channel $\hat{b}_{\phi}|\mathcal{C}^{\pm} \rangle = \zeta |\mathcal{C}^{\mp} \rangle$ within the DFS (\ref{DFS}) at rate $|\zeta|^2 \kappa$ \cite{mirrahimi2014}. This results in quantum jumps between even and odd cat states, leading to decoherence but not to leakage out of the DFS; even in the presence of single-photon loss, the steady state is confined within the cat manifold. In addition, intrinsic photon loss also causes a decrease in the amplitude $|\zeta|$ of the cat states. However, this decrease is negligible provided that the intrinsic loss rate $\kappa\ll G$ is small (which is the case for state-of-the-art superconducting cavities \cite{kirchmair2013}) as the two-photon drive quickly repumps depleted photons. 

Due to the decoherence, the array (for both models) converges at times $t\sim 1/|\zeta|^2 \kappa$ towards the fully mixed state $\hat{\rho}_\mathrm{ss} \approx \frac{1}{2}\left(|\mathcal{C}^{+}\rangle\langle\mathcal{C}^{+}| + |\mathcal{C}^{-}\rangle\langle\mathcal{C}^{-}|\right)$. Therefore, for the deterministic  preparation of  $|\mathcal{C}^+\rangle$, it is important how the rate of decoherence $|\zeta|^2 \kappa$ compares to the rate of convergence towards this state, which corresponds to the dissipative gap $\Delta_d$. 

We start by investigating the effects of intrinsic photon loss on the Kerr-resonator array in the Zeno limit, i.e., we consider  $\dot{\hat{\rho}}_{\phi} =-i\left[ \hat{H}_{\phi},\hat{\rho}_{\phi}\right] + \Gamma \mathcal{D}\left[\hat{b}_{\phi}^2 - \zeta^2 \right]\hat{\rho}_{\phi} +\kappa\mathcal{D}[\hat{b}_{\phi}]\hat{\rho}_{\phi} $. We plot the fidelity $c_{++}$ of the instantaneous state $\hat{\rho}_d\otimes\hat{\rho}_{\phi}(t)$ with the even cat state $|\mathcal{C}^+\rangle$ (blue lines) and the fidelity $c_{--}$ with the odd cat state $|\mathcal{C}^-\rangle$ (green lines) in Fig.~\ref{fidelity 2p}a as a function of time for different values of the intrinsic loss rate $\kappa$ and starting from the vacuum state. We can see that, small intrinsic loss $\kappa/U = 10^{-4}$ is required to reach a large fidelity $c_{++}\approx0.91$ (blue solid line) before
 both fidelities $c_{++}$ and $c_{--}$ converge towards the stationary values $c_{++} = c_{--} \approx 0.5$  and the fully mixed state is approached. 
On the other hand, for a larger loss rate $\kappa/U = 10^{-2}$, the fidelity $c_{++}$ (blue dot-dashed line) does not exceed the value $c_{++}=0.5$ at any time as the rate of decoherence $|\zeta|^2\kappa$ is larger than the rate $\Delta_d$ of convergence towards the even cat state. 
Since the dissipative gap $\Delta_d$ (\ref{eq gap}) is small in the Zeno limit $\gamma\gg G,U$, the steady-state preparation in the Kerr-resonator array is largely limited by decoherence. 

We now consider the impact of photon loss on our second model in the Zeno limit described by the master equation $\dot{\hat{\rho}}_{\phi} = 2 \frac{\eta}{N} \mathcal{D}[b_{\phi}^2 - \tilde{\zeta}^2]\hat{\rho}_{\phi}  +\kappa\mathcal{D}[\hat{b}_{\phi}]\hat{\rho}_{\phi} $. 
From Fig.~\ref{fidelity 2p}b we can see that a large fidelity $c_{++}$ builds up at times orders of magnitude shorter than the characteristic decoherence time $\frac{1}{\kappa}$ ($|\zeta|^2\sim1$ here) as the array quickly converges towards the even cat state. The quick convergence is guaranteed by a large dissipative gap $\Delta_d$ compared to the decoherence rate $|\zeta|^2\kappa$ with the ratio $\frac{|\zeta|^2\kappa}{\Delta_d} = \frac{N}{2}\frac{\kappa}{\eta}$.
For all values of the intrinsic loss rate $\kappa$, the maximal fidelity is orders of magnitude closer to unity than that for the Kerr-resonator array and is reached at orders of magnitude shorter times. 
 
We conclude that local two-photon loss leads to a fast convergence towards the DFS, which allows for an efficient steady-state preparation 
in the presence of intrinsic photon loss. 
This is in contrast to the steady-state preparation in the Kerr-resonator array, which is severely limited by decoherence due to the slow convergence towards the DFS.

\section{Robustness against imperfections in non-local dissipation}\label{s:Imp}

Leveraging strong and tunable non-local dissipation plays a crucial role in our preparation protocol, allowing to select a single non-dissipative normal mode and at the same time strongly damping the remaining modes. We now study the robustness of our protocol against imperfections in the non-local dissipation, which can appear due to the imperfect tuning of the couplings to the engineered reservoirs. 

As the first example of imperfection, we consider a finite accuracy in tuning the phase of the non-local dissipators. Imagine we select a plane-wave mode $\phi'\equiv k'$ (whose dissipation we aim to suppress) but due to finite precision, the externally imposed phase $\phi = \phi' + \delta\phi$ of the non-local jump operators differs from the target quasi-momentum $\phi'$ by a small offset $\delta\phi$, where we assume $\delta\phi>0$ for simplicity. 
Due to the quasi-momentum mismatch $\phi\neq\phi'$, mode $\phi'$ is no longer a dark mode of the non-local dissipator $\mathcal{L}_{\gamma}$ and, as a consequence, it exhibits non-vanishing single-photon loss at the rate $\gamma_{\phi'} =2\gamma[1- \cos(\delta\phi)]$.
As we discussed in Sec.~\ref{s:Loss}, single-photon loss of the selected mode leads to decoherence within the DFS, which limits the fidelity of prepared cat states.
In contrast to the intrinsic photon loss discussed in Sec.~\ref{s:Loss}, which is independent of $\gamma$, the single-photon loss rate $\gamma_{\phi'}$ is proportional to $\gamma$ and thus it can  reach significant values in the Zeno limit of large $\gamma$. 
The dependence of the unwanted single-photon loss on $\gamma$ leads to a trade-off between suppressing the excitations of remaining normal modes in the Zeno limit and keeping the unwanted single-photon loss rate small. As both of these competing processes decrease the fidelity of prepared cat states, $\gamma$ has to be optimized to attain efficient cat-state preparation.

We also notice that the second smallest single-photon loss rate $\gamma_{\phi' +2\pi/N } =2\gamma[1-\cos(2\pi/N-\delta\phi)]$ decreases due to the offset $\delta\phi$. Therefore, in order to achieve strong damping of the remaining normal modes, i.e. $\gamma_{\phi' +2\pi/N }\gg\eta,G $, we need not only a large rate of non-local dissipation $\gamma$ but also a small offset $\delta\phi\ll2\pi/N$, i.e., the non-local phase should be tunable with a sufficiently high resolution. The second condition becomes increasingly challenging for large system sizes $N$.

So far we have considered a common offset  $\delta\phi$ for all the cavities, so that the overall translational invariance of the model is preserved. We now move to address the effects of disorder. To this aim, we consider a general form of imperfections in the engineered non-local dissipation $\mathcal{L}_{\gamma}=\sum_{j}\gamma_j\mathcal{D}[\hat{a}_j - e^{\phi_j}\hat{a}_{j+1}]$ leading to rates $\gamma_j$ and phases $\phi_j$ of jump operators that vary at different bonds between resonators $j$ and $j+1$. These imperfections break the translational symmetry of the array. As a result, normal modes $\hat{b}_q= \sum_{j=1}^N T^*_{qj}\hat{a}_j$ of the non-local dissipator $\mathcal{L}_{\gamma}$ are no longer plane  waves, where $T_{jq}$ is a transformation matrix and $q=1,2,...N$ \footnote{The dissipator $\mathcal{L}_{\gamma}\hat{\rho}= \sum_{j,l=1}^N D_{jl} \left[\hat{a}_j\hat{\rho}\hat{a}_l^{\dagger} -\frac{1}{2}\left\{\hat{a}_l^{\dagger}\hat{a}_j,\hat{\rho} \right\}\right] = \sum_{q=1}^N  \bar{\gamma}_q \left[\hat{b}_q\hat{\rho}\hat{b}_q^{\dagger} -\frac{1}{2}\left\{\hat{b}_q^{\dagger}\hat{b}_q,\hat{\rho} \right\}\right]$ can be always diagonalized by the linear transformation $T_{jq}$ of annihilation operators, where columns of the transformation matrix $T_{jq}$ and single-photon loss rates $\bar{\gamma}_q$ are eigenvectors and eigenvalues, respectively, of the dynamical matrix $D_{jl}$.}. As a consequence, pure steady states of the master equations \eqref{master} and \eqref{master loss} in the form of dark states do, in general, not exist.

However, it is still possible to show that if all but one normal modes $q\neq 1$ experience large single-photon loss $\bar{\gamma}_{q\neq1}$, cat states can be stabilized and prepared in the selected mode $q=1$ in the Zeno limit. Here we ordered normal modes $q$ by the loss rate $\bar{\gamma}_q$ from smallest to largest, so that $q=1$ is now the closest mode to achieve the ideal non-dissipative condition. For concreteness, we focus here on the alternative model \eqref{master loss} with local two-photon loss and with a vanishing two-photon pump phase $\theta = 0 $ \footnote{Jump-operator phases $\phi_j\rightarrow\phi_j+\nu_j-\nu_{j+1}$ and two-photon pump phases $\theta_j\rightarrow\theta_j+2\nu_j$ transform under the local gauge transformation $\hat{a}_j\rightarrow\hat{a}_j e^{i\nu_j}$. The master equation \eqref{master loss} with vanishing two-photon pump phases $\theta_j=0$, for all $j$, describes a general situation with any two-photon pump phases $\theta_j\neq 0$ (that can also vary at different resonators $j$) for the particular gauge choice $\nu_j = - \theta_j/2$.}. In the Zeno limit $\bar{\gamma}_{q\neq1}\gg\eta,G,\bar{\gamma}_1$, the time evolution of mode $q=1$ is described by an effective master equation 
\begin{equation}
\dot{\hat{\rho}}_{1} = 2 \frac{\bar{\eta}}{N} \mathcal{D}[\hat{b}_{1}^2 - \bar{\zeta}^2]\hat{\rho}_{1} + \bar{\gamma}_1 \mathcal{D}[\hat{b}_{1}]\hat{\rho}_{1},\label{zeno dis}
\end{equation}
where $\bar{\eta} = \eta N\sum_{j=1}^{N}|T_{j1}|^4$, $\bar{G}= G\sum_{j=1}^{N}T_{j1}^{*\,2}$ and $\bar{\zeta} = \sqrt{-iN\bar{G}/{\bar{\eta}}}$ (see Appendix~\ref{app:Zeno dis} for more details). Despite the broken translational invariance,  we still recover two-photon driven dissipation at rate $\bar{\eta}$, which stabilizes a DFS spanned by cat states with amplitudes $\bar{\zeta}$. We then get to a remarkable conclusion: in the Zeno limit, disorder affects only quantitatively the nature of two-photon driven dissipation, leading to a modified amplitude of cat states $\bar{\zeta} = \sqrt{-iN\bar{G}/{\bar{\eta}}}$ and of the dissipative gap $\Delta_d\approx 2|\bar{G}|$. This provides an analytical insight into the robustness of the Zeno limit to disorder.
However, from Eq.~\eqref{zeno dis} we also see a qualitative change with respect to the disorderless case~\eqref{zeno al}, namely the appearance of additive single photon loss term.
Due to imperfections, non-local dissipation can give rise to single-photon loss of the selected mode $\nobreak{q=1}$. 
In particular, for a non-trivial overall phase $\nobreak{\Phi =\sum_{j=1}^N \phi_j\neq2n\pi}$, where $n$ is any integer, the dark mode condition, $\nobreak{T_{(j+1)1}  = e^{\phi_j}T_{j1}}$ for all $j$, is not compatible with periodic boundary conditions and, as a consequence, the selected mode exhibits a non-vanishing single-photon loss rate $\bar{\gamma}_1\neq0$.

\begin{figure}[t]
\centering
\includegraphics[width=0.9\linewidth]{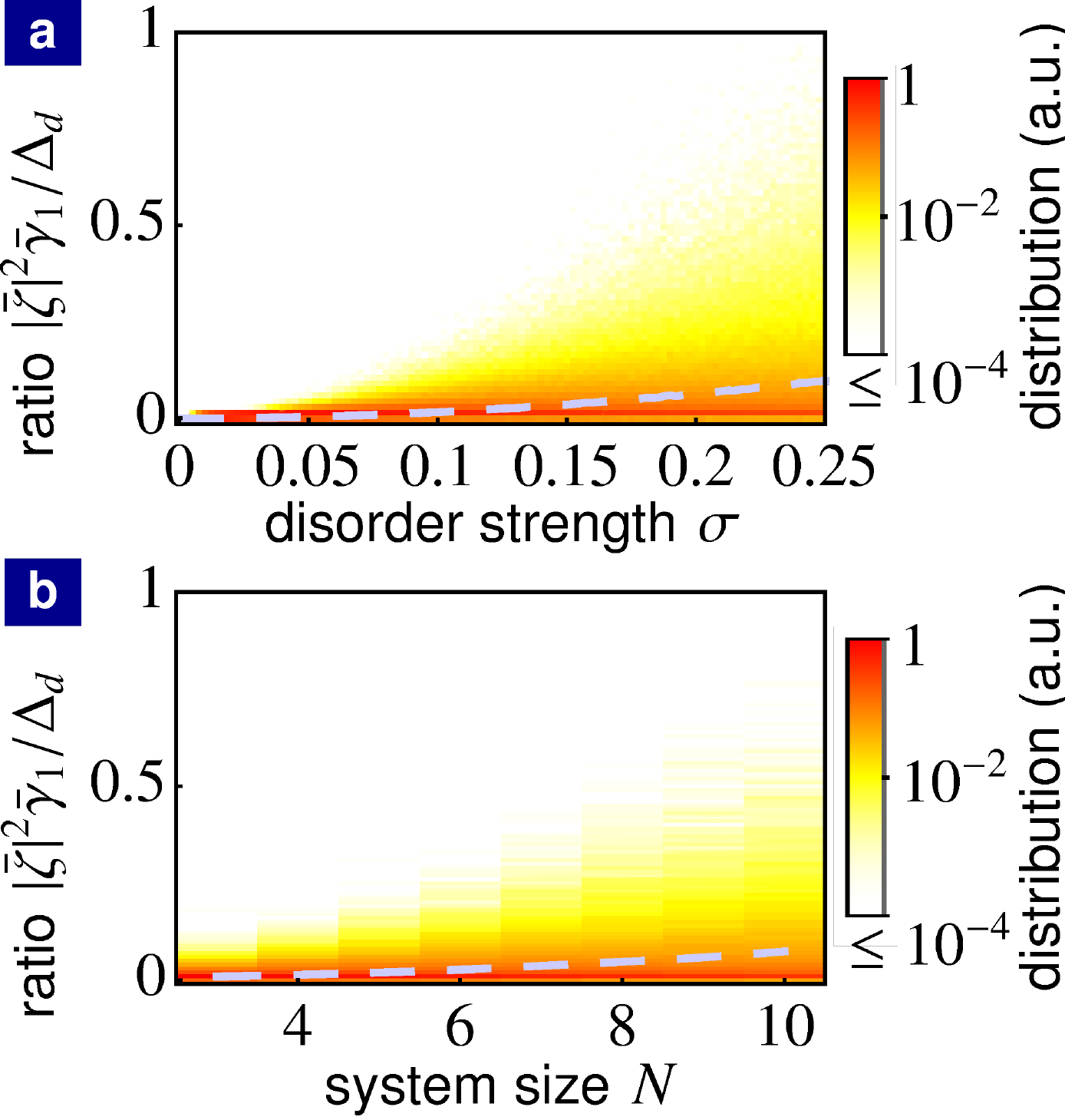}
\caption{Effects of disorder in non-local dissipation in the alternative model. Distribution (color gradient) of the ratio between the rate of decoherence $|\bar{\zeta}|^2\bar{\gamma}_1$ and the dissipative gap $\Delta_d$ for disorder in nonlocal decay rates $\gamma_j$ and jump-operator phases $\phi_j$ as a function (a) of the disorder strength $\sigma$ and (b) of system size $N$. Blue dashed lines show the average value of the ratio. [Parameters: (a) $N=3$, $ \gamma/\eta=3$, $G/\eta=1$, (b) $\sigma=0.075$, $ \gamma/\eta=4.5/(1-\cos(\frac{2\pi}{N}))$, $G/\eta=1$]}
\label{gapdis}
\end{figure}

We now consider disorder in non-local dissipation rates $\gamma_j = \gamma(1 + \sigma\delta\gamma_j)$ and jump-operator phases $\phi_j = \frac{\pi}{3}\sigma\delta\phi_j$, where $\delta\gamma_j$ and $\delta\phi_j$ are normally distributed random variables and $\sigma$ is the strength of disorder. The cat-state amplitude $\bar{\zeta}$ and the dissipative gap $\Delta_d$  are only marginally modified in the presence of moderate disorder $\sigma \lesssim 0.25$ (see Appendix~\ref{app:Zeno dis} for more details). On the other hand, the single-photon loss rate can reach considerable values even for moderate disorder, leading to significant decoherence within the DFS and, as a result, to a decreased fidelity of prepared cat states. The fidelity of prepared state is determined by the ratio of the decoherence rate $|\bar{\zeta}|^2\bar{\gamma}_1$ and the dissipative gap $\Delta_d$, see Sec.~\ref{s:Loss}. We plot the distribution of the ratio in a disordered array as a function of the disorder strength in Fig.~\ref{gapdis}a. We can see that with the increasing strength of disorder it is more likely to obtain a decoherence rate  $|\bar{\zeta}|^2\bar{\gamma}_1\sim \Delta_d$ comparable to the dissipative gap $\Delta_d$, which severely limits the fidelity of prepared cat states. On the other hand, the decoherence rate $|\bar{\zeta}|^2\bar{\gamma}_1$ remains, on average (blue dashed line), order of magnitude smaller than the dissipative gap $\Delta_d$ even for moderate disorder $\sigma\sim0.25$, allowing the preparation of cat states with a large fidelity.

We now study the effects of disorder in arrays with increasing system size $N$. 
In order to keep the smallest loss rate of the remaining modes $\bar{\gamma}_2\approx\gamma'$ approximately constant with increasing systems size, we increase the non-local dissipation rate $\nobreak{ \gamma=\gamma'/2[1-\cos(\frac{2\pi}{N})]}$.
We plot the distribution of the ratio between the decoherence rate $|\bar{\zeta}|^2\bar{\gamma}_1$ and the dissipative gap $\Delta_d$ for a constant disorder strength $\sigma = 0.075$ as a function of system size $N$ in Fig.~\ref{gapdis}b. We can see that, with increasing system size, the ratio increases on average (blue dashed line) and it is more likely to obtain a decoherence rate  $|\bar{\zeta}|^2\bar{\gamma}_1\sim \Delta_d$ comparable to the dissipative gap $\Delta_d$. The decoherence rate $|\bar{\zeta}|^2\bar{\gamma}_1\propto N$ increases due to the increasing cat-state amplitude $|\bar{\zeta}|\propto\sqrt{N}$ while the single-photon loss rate $\bar{\gamma}_1$ and the dissipative gap $\Delta_d$ stay approximately constant with increasing system size (see Appendix~\ref{app:Zeno dis}).

We note that for some disorder realizations the overall phase $\Phi$ is vanishingly small and, as a consequence, the single photon loss rate $\bar{\gamma}_1$ is negligible. For these disorder realizations, the dominant source of decoherence within the DFS is single-photon amplification $\mathcal{D}[\hat{b}_1^{\dagger}]$. Single-photon amplification is a second-order perturbation process in the two-photon pump $\hat{H}_G$ and it is facilitated by virtual excitations of strongly decaying normal modes $q\neq1$. 
However, single-photon amplification does not severely limit the fidelity of prepared cat states since it is suppressed in the Zeno limit as its rate is inversely proportional to $\gamma$ (see Appendix~\ref{app:amp} for more details).

In conclusion, due to imperfections in non-local dissipation, the selected normal mode is not a dark mode of the non-local dissipator. Remarkably,  cat states can still be stabilized and prepared in the selected mode in the Zeno limit, albeit with decreased fidelity as decoherence due to unwanted single-photon loss sets in.
Since the single-photon loss rate is proportional to the non-local dissipation rate $\gamma$, we face a trade-off between suppressing the excitations of remaining normal modes and keeping the decoherence rate small. The effects of decoherence are more pronounced for large system sizes due to the increasing cat-state amplitude. However, we can conclude that for moderate disorder strengths and moderate system sizes, the decoherence rate remains, on average, order of magnitude smaller that the dissipative gap, thus allowing for the preparation of cat states with a large fidelity even in the presence of imperfections.

\section{Noise bias of multi-mode cat states}\label{bias}

A  manifold spanned by Schr\"odinger cat states is advantageous for encoding and protecting quantum information because it experiences biased noise~\cite{mirrahimi2014}, which means that states in the manifold display an asymmetric response to different kinds of noise channels. This can considerably simplify the quantum-error-correcting protocols required to protect quantum information from decoherence. In this Section we first recall the most important features of biased noise and then show that multi-mode cat states offer an enhanced noise bias over single-mode cats considered so far. 
 
 For concreteness, let us assume we encode the logical quantum states into the following superposition of single-mode cat states, $\vert0/1\rangle_L\equiv (|\mathcal{C}^{+}\rangle\pm|\mathcal{C}^{-}\rangle)/\sqrt2\approx|\pm\zeta\rangle$ (the last approximation is excellent for $\vert\zeta\vert\gtrsim 2$), as for instance realized with a  single KPR cat state of amplitude $\zeta= i\sqrt{G/U}$~\cite{puri2017}. The overlap between the coherent states $\vert\langle +\zeta|-\zeta\rangle\vert=\exp(-2\vert \zeta\vert^2)$ is exponentially suppressed with the number of photons $\vert\zeta\vert^2$. This simple fact endows the encoding with a natural protection against noise. Indeed, by increasing the amplitude, the coherent states move further apart and it becomes extremely unlikely for any noise process to cause a jump between $\vert0\rangle_L$ and $\vert1\rangle_L$, i.e., to generate a so-called bit-flip error. Note that increasing the amplitude comes at the price of an increased phase-flip error rate, i.e., noise-induced flips between $\vert\pm\rangle_L\equiv |\mathcal{C}^{\pm}\rangle$ eigenstates. However, it can be shown that the phase-flip error rate increases only linearly with respect to the number of photons in the mode. As a result, there is a net bias $\exp(-2\vert \zeta\vert^2)/\vert \zeta\vert^2$ between the error rates  experienced by a single resonator~\cite{mirrahimi2014}. Such noise bias has been recently experimentally observed and tuned in Ref.~\cite{Lescanne2020}. 

Let us now consider a similar encoding, but replacing the single mode cat states with multi-mode cat states  $|\mathcal{C}^{\pm}\rangle_\phi$. This choice allows for a further exponential improvement in the noise bias, achieving $\exp(-2N\vert \zeta\vert^2)/(N\vert \zeta\vert^2)$; the expression is written in terms of  single-mode amplitude, for instance for the case of Kerr nonlinearity  $\zeta_{N\text{-mode}} \equiv i\sqrt{\frac{NG}{U}}=\sqrt{N} \zeta$.
The bias exponentially increases with respect to both the number of coherent photons in each resonator (set by the ratio between the two-photon pump and either the Kerr non-linearity or the two-photon loss rate) and the number of resonators in the array. While the first feature leads to `standard noise bias' and is already exploited for qubit encoding in a single-mode cat manifold, the second feature is unique to our model. The physical reason for this is that the total number of photons in the multi-mode cat state stabilized in the resonator array increases with the number of resonators as discussed in Sec.~\ref{s:DFS}.
Therefore, the multi-mode DFS~\eqref{DFS} can encode a qubit with bit-flip errors exponentially suppressed with system size $N$, gaining an exponential improvement in the noise bias compared to single-mode cat states. This can be exploited to realize a \emph{protected quantum memory} which only suffers from phase-flip errors, which can be in turn corrected by simple classical error correction techniques (e.g.~a simple repetition code)~\cite{guillaud2019}. 

\section{Experimental parameters}\label{s:Exp}

For both models of cat-state preparation, the regime $\gamma\gg G,U,\eta\gg\kappa$ is required to suppress excitations of remaining normal modes  ($\gamma\gg G,U,\eta$, as discussed in Sec.~\ref{s:Zeno}) while at the same time reducing the effects of decoherence due to intrinsic single-photon loss ($U,\eta\gg\kappa$, as discussed in Sec.~\ref{s:Loss}).
Strong Kerr nonlinearities can be effectively induced both in three-dimensional microwave cavities~\cite{kirchmair2013} and on-chip resonators~\cite{Wang2019}, via coupling to a Josephson junction, resulting in  photon-photon interactions far exceeding the photon loss rate.
The induced Kerr nonlinearity $U\sim100-1000\,{\rm kHz}$ can be larger than intrinsic loss rates $\kappa\sim10-100\,{\rm Hz}$ of state-of-the-art three-dimensional cavities \cite{reagor2013,reagor2016} by orders of magnitude, with feasible ratios $\kappa/U\sim10^{-5}-10^{-3}$. The stabilization of Schr\" odinger cat states in a single Kerr resonator using a parametric two-photon drive has been demonstrated in Ref.~\cite{grimm2020} reporting an amplitude $|\zeta|\sim1$, which corresponds to $G/U \sim 1$. In an array, the amplitude $|\zeta|\propto\sqrt{N}$ of the cat states increases with the number of Kerr resonators. Local two-photon loss can be realized in a similar three-dimensional cQED architecture \cite{mirrahimi2014} at rate $\eta\sim100\,{\rm kHz}$ \cite{leghtas2015} with a feasible ratio $\kappa/\eta=10^{-2}$ \cite{touzard2018}. It has been employed for the stabilization of a single-mode Schr\" odinger cat state with an amplitude $|\zeta|\sim1$  \cite{leghtas2015}.  

In addition to coupling neighboring resonators to a transmission line as discussed in \cite{metelmann2015}, nonlocal dissipation can be also implemented by the coupling to a strongly decaying microwave mode with a feasible decay rate $\kappa_A \sim 100\,{\rm MHz}$ \cite{leghtas2015}. Microwave modes can be coupled using parametrically driven elements consisting of several Josephson junctions reaching coupling strengths $g\sim100\,{\rm MHz}$ as demonstrated in planar superconducting circuits \cite{sliwa2015,lecocq2017}. Using a weaker parametric drive such that $g\ll\kappa_A$, one can achieve effective nonlocal dissipation at rate $\gamma \sim g^2/\kappa_A $ by adiabatically eliminating the strongly decaying mode. It is also feasible to achieve the Zeno limit of strong non-local dissipation $\gamma\gg G,U,\eta$ as the parametric coupling $g$ and the decay rate $\kappa_A$ can be sufficiently large.

Finally, we stress that, while multi-mode cat states can be stabilized and prepared in resonator arrays with a large system size $N$, all characteristic features of the multi-mode cat-state preparation discussed in this manuscript---and especially the exponentially enhanced noise bias---can be observed already for a small system size $N=3$.

\section{Conclusions}\label{s:Conclusions}

We showed that a two-dimensional manifold spanned by superpositions of multi-mode Schr\" odinger cat states can be stabilized in an array of resonators coupled via non-local dissipation. The required non-linearity, which is either of the Kerr type or an engineered two-photon loss, acts locally on each resonator while the dissipative coupling is linear, which makes our proposal particularly convenient  for experimental implementations.
The two models we put forward are readily realizable with state-of-the-art circuit QED architectures.

In the Zeno limit of strong non-local dissipation, we showed that the even-parity multi-mode cat state can be prepared from the initial vacuum state with a fidelity approaching unity. 
In the Kerr-resonator array, the steady-state preparation is limited by decoherence due to intrinsic photon loss, which sets in at long times, as the relaxation towards the steady state is due to a weak effective two-photon dissipation resulting in a slow rate of convergence.
On the other hand, local two-photon loss gives rise to a quick convergence towards the multi-mode cat states allowing for their efficient preparation. Multi-mode cat states can be stabilized and prepared in the Zeno limit even in the presence of imperfections in non-local dissipation, albeit with decreased fidelity as an additional decoherence channel sets in.  Importantly, the rate of convergence towards the cat states is independent of system size making the steady-state preparation scalable.
Scaling towards large system sizes requires a good suppression of imperfections in non-local dissipation as they lead to decoherence which is more pronounced in large systems.
Our protocol exploits $N$ local two-photon pumps allowing  for the amplitude $|\zeta|\propto\sqrt{N}$ of the multi-mode cat states to increase with the number of resonators in the array $N$.
This in turn leads to an exponentially increasing noise bias of the cat manifold, which can be exploited for the implementation of a quantum memory with enhanced protection.

Being able to prepare multi-mode cat states in a deterministic, robust and scalable fashion is of direct relevance for applications in quantum metrology, quantum computation and quantum information. In  quantum computation, especially in the context of bosonic codes based on cat qubits, multi-mode cat states correspond to GHZ states, which are a required resource for universal fault-tolerant quantum computing, e.g. by enabling Toffoli state preparation~\cite{chamberland2020}. In quantum metrology, multi-mode cat states can provide a source of non-Gaussian multimode light needed for attaining Heisenberg scaling~\cite{joo2011,zhang2013}; they are also a source of non-Gaussian multipartite entanglement to generate non-Gaussian cluster states~\cite{Ra2020}.
Moreover, our proposal may also foster multimode extensions of existing protocols currently limited by the lack of non-Gaussian resources beyond the single- or two-mode case. Multi-mode cat states are also relevant to fundamental aspects of quantum theory, e.g. to test predictions of decoherence theory and explore the quantum-to-classical transition~\cite{Zurek2003}. 

Our treatment of dissipatively coupled cavity arrays opens new avenues in quantum reservoir engineering~\cite{poyatos1996} as it allows for tailored dissipation in momentum space. Interesting future directions are the stabilization of other non-Gaussian multi-mode entangled states using different on-site models, reservoir engineering with multiple normal modes with different quasi-momenta by employing different non-local reservoirs, and tailoring dissipation in momentum space of higher dimensions. Finally, for the specific models considered in this work, even though the steady states can be described analytically for any system parameters and the transient dynamics can be well described in the Zeno limit by an effective master equation for a single normal mode, much less has been understood about the driven-dissipative many-body dynamics for the non-local dissipation rate comparable to the Kerr nonlinearity, which will be subject of future investigation. 

\section*{acknowledgements}
This work was supported by the European Union’s Horizon 2020 research and innovation programme under grant agreement No 732894 (FET Proactive HOT). P.Z. acknowledges funding from the European Union’s Horizon 2020 research and innovation programme under grant agreement No 828826 (Quromorphic) and the EPSRC (grant No EP/N509620/1). A.N. holds a University Research Fellowship from the Royal Society. 

\section*{Note added}
During the final stage of this project, a related paper appeared, investigating mixed steady states of dissipatively coupled Kerr resonator arrays, which correspond to the ground states of a frustrated antiferromagnet \cite{li2020}.
After the completion of this project, another work appeared, studying the generation and detection of two-mode cat states in dissipatively-coupled parametric oscillators \cite{zhou2021a}, however only treating the particular case $N=2$.
\appendix

\section{Numerical simulations of the master equation  \eqref{master}}\label{app:conserved}

In this appendix we provide details about numerical simulations of the master equation \eqref{master} and its conserved quantities.

In general, steady states which are neither dark states nor their incoherent mixtures can exist \cite{kraus2008}. However, we verified that the master equation (\ref{master}) has no steady states lying outside of the DFS (\ref{DFS}) by numerically solving for $\mathcal{L} \hat{\rho}_\mathrm{ss} = 0$ for $N=3,4$ and various values of the non-local decay rate $\gamma/U$ and two-photon drive strength $G/U$. We truncate the infinitely dimensional Hilbert space considering only $M_k$ lowest energy levels of each mode $k$.

Each coefficient $c_{\mu} = \textrm{Tr}\left[\hat{\rho}_{\rm in}\hat{J}_{\mu} \right]$ of the DFS~\eqref{DFS} is associated with a conserved quantity $\hat{J}_{\mu}$
 and encodes the information about $\hat{\rho}_{\rm in}$ that is preserved during the time evolution.
As a result, conserved quantities can be used to find the particular steady state for a given initial state. Conserved quantities $\hat{J}_{\mu}$ are formally defined as solutions of $\mathcal{L}^{\dagger}\hat{J}_{\mu}=0$, where  $\mathcal{L}^{\dagger}$ is the adjoint Liouvillian and 
\begin{equation}\label{ad master}
\mathcal{L}^{\dagger} \hat{J}_{\mu} = i\left[ \hat{H},\hat{J}_{\mu} \right] +  \sum_{k\neq\phi}\gamma_k\left[\hat{b}_k^{\dagger}\hat{J}_{\mu} \hat{b}_k  - \frac{1}{2}\left\{\hat{J}_{\mu},\hat{b}_k^{\dagger}\hat{b}_k\right\}\right],
\end{equation}
and $\left\{\cdot,\cdot \right\}$ is the anticommutator \cite{albert2014}. We solve numerically for the conserved quantities. Finding conserved quantities, which are bi-orthogonal to the basis operators $\hat{\xi}_{\pm\pm}=|\mathcal{C}^{\pm}\rangle\langle\mathcal{C}^{\pm}|$ of the DFS,~i.e. ${\rm Tr[\hat{J}_{\mu}^{\dagger}\hat{\xi}_{\beta}]} = D_{\mu}\delta_{\mu\beta}$, we can determine coefficients $c_{\mu} = {\rm Tr[\hat{J}_{\mu}^{\dagger}\hat{\rho}_{\rm in}]}$ of the particular steady state for a given initial state $\hat{\rho}_{\rm in}$. Normalization $D_{\mu}=1$, for all $\mu$, guarantees that $\hat{\rho}$ is a density matrix. We plot in Fig.~\ref{fidelity} the fidelities with the even cat state and the odd cat state as well as the purity of the particular steady state for the initial vacuum state, where we used the following truncation of the Hilbert space: $M_{k=\phi}=12$ and $M_{k\neq\phi}=3$ for $G/U=0.5$, $M_{k=\phi}=16$ and  $M_{k\neq\phi}=3$ for $G/U=0.75$, and $M_{k=\phi}=18$ and $M_{k\neq\phi}=3$ for $G/U=1$.

\section{Effective master equation in the Zeno limit}\label{app:Zeno}
In this appendix, we describe the effective Zeno dynamics of mode $k=\phi$ in the regime of strong non-local dissipation, which acts as a continuous measurement projecting all modes $k\neq\phi$ onto the vacuum state. We derive an effective master equation for mode $\phi$ employing the Dyson series of the Liouvillian dynamics, which was described in detail in Ref.~\cite{popkov2018}.

We start by rescaling time $\tau = \gamma\,t$  in the master equation (\ref{master}) obtaining
\begin{equation}\label{master per}
\frac{\partial\hat{\rho}(\tau)}{\partial \tau} = \sum_{k\neq\phi}\frac{\gamma_k}{\gamma}\mathcal{D}[\hat{b}_k]\hat{\rho}(\tau) - \frac{i}{\gamma}\left[ \hat{H},\hat{\rho}(\tau)\right] = \left(\bar{\mathcal{L}}_d + \mathcal{K}\right)\hat{\rho}(\tau),
\end{equation}
where $\bar{\mathcal{L}}_d =   \sum_{k\neq\phi}\frac{\gamma_k}{\gamma}\mathcal{D}[\hat{b}_k] $ and $\mathcal{K} = -\frac{i}{\gamma}[\hat{H},\,\cdot\,]$.
In the limit of strong dissipation $\gamma_k \gg U,G$, for all $k\neq \phi$, we can treat $\mathcal{K}$ as a perturbative term. A general solution of equation (\ref{master per}) can be written as
\begin{equation}
\hat{\rho}(\tau) = \mathcal{U}(\tau)\rho(0), 
\end{equation}
where $\mathcal{U}(\tau)$ is the propagator satisfying
\begin{equation}\label{propagator}
\mathcal{U}(\tau) = e^{\bar{\mathcal{L}}_d \tau}\left( 1 + \int_{0}^{\tau}{\rm d}s \,e^{-\bar{\mathcal{L}}_d s}\,\mathcal{K}\,\mathcal{U}(s)\right).
\end{equation}
By iterating this equation, we obtain the Dyson series for the propagator
\begin{align}
&\mathcal{U}(\tau) = e^{\bar{\mathcal{L}}_d \tau}\biggl( 1 + \int_{0}^{\tau}{\rm d}\tau_1 \,e^{-\bar{\mathcal{L}}_d \tau_1}\,\mathcal{K}\,e^{\bar{\mathcal{L}}_d \tau_1}\nonumber\\ 
&+  \int_{0}^{\tau}{\rm d}\tau_1 \,e^{-\bar{\mathcal{L}}_d \tau_1}\,\mathcal{K}\,e^{\bar{\mathcal{L}}_d \tau_1} \int_{0}^{\tau_1}{\rm d}\tau_2 \,e^{-\bar{\mathcal{L}}_d \tau_2}\,\mathcal{K}\,e^{\bar{\mathcal{L}}_d \tau_2} + ...\biggr),\label{dyson}
\end{align}
where the ellipsis denotes terms of third and higher order in $\mathcal{K}$.

Mode $\phi$ does not experience any dissipation $({\rm Tr}_d \bar{\mathcal{L}}_d)\hat{\rho}_{\phi} =\hat{\rho}_{\phi}$, where $\hat{\rho}_{\phi} = {\rm Tr}_{d}\hat{\rho}$ is the reduced density matrix of mode $\phi$ and ${\rm Tr}_{d}$ denotes the trace over all decaying modes $k \neq \phi$. The dissipator $\bar{\mathcal{L}}_d$ targets a unique state $\hat{\rho}_d$ in the reduced Hilbert space of all decaying modes $k\neq\phi$,~i.e.
\begin{equation}
({\rm Tr}_{\phi} \bar{\mathcal{L}}_d)\hat{\rho}_{d} = 0,
\end{equation}
where ${\rm Tr}_{\phi}$ is the trace over mode $\phi$. The unique steady state is the vacuum state $\hat{\rho}_{d} = \bigotimes_{k\neq\phi} |0\rangle_k\langle0| $. The projection onto the kernel of $ \bar{\mathcal{L}}_d$ is $ \mathcal{P}_d = \lim_{\tau\rightarrow\infty}\exp(\bar{\mathcal{L}}_d\tau)$ obeying following relations
\begin{equation}\label{rel proj}
\bar{\mathcal{L}}_d \mathcal{P}_d =  \mathcal{P}_d \bar{\mathcal{L}}_d =  \mathcal{P}_d,
\end{equation}
and
\begin{equation}\label{act proj}
 \mathcal{P}_d \hat{X} =  \hat{\rho}_d \otimes  {\rm Tr}_{d} \hat{X},
\end{equation}
for an arbitrary density matrix or operator $\hat{X}$.

For strong dissipation $\gamma_k \gg U,G$, the dynamics of the system is constrained to the dissipation-free subspace $\hat{\rho} (\tau) = \hat{\rho}_d \otimes \hat{\rho}_{\phi}(\tau)$ at all times $\tau\gg1$ \cite{popkov2018}. The dynamics of the reduced density matrix $\hat{\rho}_{\phi}(\tau)$ is described by the propagator $\mathcal{P}_d \,\mathcal{U}(\tau) \,\mathcal{P}_d$. Writing the Dyson series for the propagator up to the second order in $\mathcal{K}$, we obtain
\begin{align}
& \mathcal{P}_d \,\mathcal{U}(\tau) \,\mathcal{P}_d = \mathcal{P}_d  + \tau\, \mathcal{P}_d \,\mathcal{K}\,\mathcal{P}_d  \nonumber\\
&+ \mathcal{P}_d \,\mathcal{K} \int_{0}^{\tau}{\rm d}\tau_1\int_{0}^{\tau_1}{\rm d}\tau_2 \,e^{\bar{\mathcal{L}}_d \left( \tau_1 - \tau_2\right)}\,\mathcal{K}\,\mathcal{P}_d  + \mathcal{O}\left(\frac{U^3}{\gamma^3} \right),\label{dyson proj}
\end{align}
where we used equation (\ref{rel proj}). We explicitly evaluate the second-order term in the Dyson series in the Supplementary Material to obtain
\begin{align}
 \mathcal{P}_d \,\mathcal{U}(\tau) \,\mathcal{P}_d = &\mathcal{P}_d  + \tau\, \mathcal{P}_d \,\mathcal{K}\,\mathcal{P}_d + \frac{\tau^2}{2}\, \left(\mathcal{P}_d \,\mathcal{K}\,\mathcal{P}_d \right)^2  \nonumber\\
 &+ \tau\,\frac{\Gamma}{\gamma} \mathcal{D}[\hat{b}_{\phi}^2 - \zeta^2 ]\mathcal{P}_d + \mathcal{O}\left(\frac{U^3}{\gamma^3} \right).\label{dyson proj 2}
\end{align}

We now consider that we initially start with a state in the dissipation-free subspace,~i.e.~$\hat{\rho}(0) = \hat{\rho}_d\otimes\hat{\rho}_{\phi}(0)$. The time evolution within the dissipation-free subspace is described by $\hat{\rho}_{\phi}(\tau) = {\rm Tr}_d\left[\mathcal{P}_d \,\mathcal{U}(\tau) \,\mathcal{P}_d\left(\hat{\rho}_d\otimes\hat{\rho}_{\phi}(0)\right)\right]$. To obtain a master equation in a differential form, we use the equivalence $\partial \hat{\rho}_{\phi}/\partial \tau = \lim_{\tau\rightarrow0}\left\{ \hat{\rho}_{\phi}(\tau) - \hat{\rho}_{\phi}(0)\right\}/\tau$. Using the Dyson series (\ref{dyson proj 2}) and rescaling back time $t = \tau/\gamma$, we obtain the effective master equation (\ref{master zeno}) in Lindblad form. The effective Hamiltonian $\hat{H}_{\phi} = {\rm Tr}_d\left[\left(\hat{\rho}_d\otimes\hat{I}_{\phi}\right) \hat{H}\right]$ is the projection of the full Hamiltonian $\hat{H}$ onto the dissipation-free subspace, where $\hat{I}_{\phi}$ is the identity operator in the reduced Hilbert space of mode $\phi$. 

\section{Rotation within the stabilized DFS}\label{app:Rot}

We consider a weak single-photon drive $\hat{H}_{\lambda} = \lambda(e^{ij\phi}\hat{a}_j+e^{-ij\phi}\hat{a}_j^{\dagger})$ applied to an arbitrary single resonator $j$. 
In the Zeno limit $\gamma\gg U,G,\lambda$, the dynamics of mode $\phi$ is described by the effective master equation
\begin{equation}\label{1p zeno}
\dot{\hat{\rho}}_{\phi}=  -i\left[ \hat{H}_{\phi}+\hat{H}_{\phi,\,\lambda},\hat{\rho}_{\phi}\right] + \Gamma \mathcal{D}\left[\hat{b}_{\phi}^2 - \zeta^2 \right]\hat{\rho}_{\phi}.
\end{equation}
To derive the effective master equation, we followed the same steps as in Appendix~\ref{app:Zeno} for the Kerr resonator array without the single-photon drive. The single-photon drive $\hat{H}_{\lambda}$ of resonator $j$ gives rise to the single-photon drive $\hat{H}_{\phi,\,\lambda} = \frac{\lambda}{\sqrt{N}}(\hat{b}_{\phi}+\hat{b}_{\phi}^{\dagger})$ of mode $\phi$, which is a first order process in $\hat{H}_{\lambda}$. Note that the single-photon drive $\hat{H}_{\lambda}$ in combination with the Kerr nonlinearity $\hat{H}_{U}$ and the two-photon drive $\hat{H}_{G}$ does not lead to any second-order effect due to the strong suppression of exactions of decaying modes and the conservation of total quasi-momentum.

It has been shown in Ref.~\cite{mirrahimi2014} that the single-photon drive $\hat{H}_{\phi,\,\lambda}$ induces the rotation 
\begin{align}
\hat{R}(\alpha) =&  \cos\alpha \left(|\mathcal{C}^+\rangle\langle\mathcal{C}^+| + |\mathcal{C}^-\rangle\langle\mathcal{C}^-|\right)\nonumber\\
 &+ i\sin\alpha \left(|\mathcal{C}^+\rangle\langle\mathcal{C}^-| + |\mathcal{C}^-\rangle\langle\mathcal{C}^+|\right)
\end{align}
within the DFS stabilized by two-photon driven dissipation $\mathcal{D}[\hat{b}_{\phi}^2-\zeta^2]$, provided that the drive strength $\lambda/\sqrt{N}$ is small compared to the rate $\Gamma$ of two-photon driven dissipation. 
The even cat state $|\mathcal{C}^+\rangle$, which is dissipatively prepared from the initial vacuum state, can be transformed into a superposition $\hat{R}(\alpha)|\mathcal{C}^+\rangle = \cos\alpha|\mathcal{C}^+\rangle+i\sin\alpha|\mathcal{C}^-\rangle$ of cat states $|\mathcal{C}^+\rangle$ and $|\mathcal{C}^-\rangle$ by the rotation $\hat{R}(\alpha)$.

\section{Decoherence-free subspace of the alternative model \eqref{master loss}}\label{app:SS al}
In this appendix, we discuss the steady states of the alternative model with local two-photon los described by the master equation (\ref{master loss}).

We start by writing the dissipator
\begin{equation}
2\eta\sum_{j=1}^{N}\mathcal{D}[\hat{a}_j^2 - e^{-2ij\phi}\frac{\tilde{\zeta}^2}{N}] = - i \left(\hat{H}_{\rm eff}\hat{\rho} - \hat{\rho}\hat{H}_{\rm eff}^{\dagger}\right) +  \mathcal{J}\hat{\rho}
\end{equation}
in a form of the non-Hermitian Hamiltonian $\hat{H}_{\rm eff}$ and the jump term $\mathcal{J}\hat{\rho}$ which can be expressed in the plane wave basis as
\begin{align}
 &\hat{H}_{\rm eff} =   -i\frac{\eta}{N} \sum_{k_1,k_2,k_3,k_4}\delta_{k_1+k_2,k_3+k_4} \nonumber\\
 & \times\left(\hat{b}_{k_1}^{\dagger}\hat{b}_{k_2}^{\dagger} - \delta_{k_1\phi} \delta_{k_2\phi}\zeta^{*\,2}\right) \left(\hat{b}_{k_3}\hat{b}_{k_4} - \delta_{k_3\phi} \delta_{k_4\phi}\tilde{\zeta}^{2}\right),\\
 &\mathcal{J}\hat{\rho} =  2\frac{\eta}{N} \sum_{k_1,k_2,k_3,k_4}\delta_{k_1+k_2,k_3+k_4} \nonumber \\ 
 &\times\left(b_{k_1}b_{k_2} - \delta_{k_1\phi} \delta_{k_2\phi}\tilde{\zeta}^{2}\right)\hat{\rho} \left(\hat{b}_{k_3}^{\dagger}\hat{b}_{k_4}^{\dagger} - \delta_{k_3\phi} \delta_{k_4\phi}\tilde{\zeta}^{*\,2}\right),
\end{align}
where the arguments $k_1+k_2$ and $k_3+k_4$ are defined modulo $2\pi$.
Similarly to the Kerr-resonator array, we look for pure steady states $\hat{\rho}_{\rm SS} = |\Psi\rangle\langle\Psi|$ with all modes $k\neq\phi$ being in the vacuum state, in which case the non-local dissipators $\sum_{k}\gamma_k\mathcal{D}[\hat{b}_k ]\hat{\rho}_{\rm SS}  = 0$ in Eq.~(\ref{master loss}) vanish. Using $\hat{b}_k\hat{\rho}_{\rm SS}  = \hat{\rho}_{\rm SS} \hat{b}_k^{\dagger} = 0$ for all $k\neq\phi$, we evaluate all remaining terms in the master equation (\ref{master loss}) obtaining $\hat{H}_{\rm eff}\hat{\rho}_{\rm SS}  =   -i\frac{\eta}{N} \sum_{k} \left(\hat{b}_{k}^{\dagger}\hat{b}_{2\phi - k}^{\dagger} - \delta_{k\phi} \tilde{\zeta}^{*\,2}\right) \left(\hat{b}_{\phi}^2 - \tilde{\zeta}^{2}\right)\hat{\rho}_{\rm SS} $,  $\hat{\rho}_{\rm SS}  \hat{H}_{\rm eff}^{\dagger} =   i\frac{\eta}{N} \sum_{k} \hat{\rho}_{\rm SS}  \left(\hat{b}_{\phi}^{\dagger\,2} - \tilde{\zeta}^{*\,2}\right) \left(\hat{b}_{k}\hat{b}_{2\phi - k} - \delta_{k\phi} \tilde{\zeta}^{2}\right)$ and $\mathcal{J}\hat{\rho}_{\rm SS}  =  2\frac{\eta}{N}  \left(\hat{b}_{\phi}^{2} - \tilde{\zeta}^{2}\right)\hat{\rho}_{\rm SS}  \left(\hat{b}^{\dagger\,2}_{\phi} - \tilde{\zeta}^{*\,2}\right)$.
All three remaining terms vanish if the pure steady state is one of the coherent states $ \hat{b}_{\phi}|\psi\rangle =  \hat{b}_{\phi}|\pm\tilde{\zeta}\rangle = \pm\tilde{\zeta} |\tilde{\zeta}\rangle$ or an arbitrary superposition of these coherent states. We conclude that an arbitrary superposition of coherent states $|\pm\tilde{\zeta}\rangle $ is a steady state of the master equation satisfying $\mathcal{L}\hat{\rho}_{\rm SS}  = 0$. Similarly as for the array of Kerr resonators, all possible superpositions of coherent states $|\pm\tilde{\zeta}\rangle $ and their incoherent admixtures form the DFS (\ref{DFS}). We numerically confirm that there are no steady states lying outside the DFS  (\ref{DFS}).

\section{Effective master equation in the Zeno limit for the alternative model \eqref{master loss}}\label{app:Zeno al}
In this appendix, we derive the effective master equation for mode $\phi$ for the alternative model \eqref{master loss} with local two-photon loss in the Zeno limit of strong non-local dissipation. We follow the same steps in the derivation as for the Kerr-resonator array in Appendix~\ref{app:Zeno}.

We start by rescaling time $\tau = \gamma\,t$ in the master equation (\ref{master loss}) obtaining
\begin{align}
\frac{\partial\hat{\rho}(\tau)}{\partial \tau} =& \sum_{k\neq\phi}\frac{\gamma_k}{\gamma}\mathcal{D}[\hat{b}_k]\hat{\rho}(\tau) + \frac{1}{\gamma}\left[\mathcal{J}\hat{\rho} - i \left(\hat{H}_{\rm eff}\hat{\rho} - \hat{\rho}\hat{H}_{\rm eff}^{\dagger}\right) \right]\nonumber\\
= &\left(\bar{\mathcal{L}}_d + \mathcal{K}\right)\hat{\rho}(\tau),
\end{align}
where $\mathcal{K} \hat{\rho}= \frac{1}{\gamma}\left[\mathcal{J}\hat{\rho}- i \left(\hat{H}_{\rm eff}\hat{\rho} - \hat{\rho}\hat{H}_{\rm eff}^{\dagger}\right) \right]$ and $\nobreak{\bar{\mathcal{L}}_d =  \sum_{k\neq\phi}\frac{\gamma_k}{\gamma}\mathcal{D}[\hat{b}_k]}$.
In the limit of strong non-local dissipation $\gamma_k \gg \eta, G$, for all $k\neq \phi$, we can treat $\mathcal{K}$ as a perturbative term. A general solution of the master equation is $\hat{\rho}(\tau) = \mathcal{U}(\tau)\rho(0)$, where the propagator $\mathcal{U}(\tau)$ can be expanded in the Dyson series (\ref{dyson}). 

The dissipator  $\bar{\mathcal{L}}_d$ is identical to the dominant dissipator for the Kerr-resonator array in the Zeno limit. It targets a unique state $\hat{\rho}_d$ in the reduced Hilbert space of modes $k\neq\phi$. For strong non-local dissipation $\gamma_k \gg \eta, G$, the dynamics of the system is constrained to the subspace $\hat{\rho} (\tau) = \hat{\rho}_d \otimes \hat{\rho}_{\phi}(\tau)$ at all times $\tau\gg1$ \cite{popkov2018}. The dynamics of the reduced density matrix $\hat{\rho}_{\phi}(\tau)$ is described by the propagator
\begin{align}
\mathcal{P}_d \,\mathcal{U}(\tau) \,\mathcal{P}_d &= \mathcal{P}_d  + \tau\, \mathcal{P}_d \,\mathcal{K}\,\mathcal{P}_d  + \mathcal{O}\left(\frac{\eta^2}{\gamma^2} \right) \nonumber\\
&=  \mathcal{P}_d  + \tau\frac{2}{N}\frac{\eta}{\gamma} \mathcal{D}[\hat{Z}_{\phi}]\,\mathcal{P}_d  + \mathcal{O}\left(\frac{\eta^2}{\gamma^2} \right),\label{dyson proj loss}
\end{align}
where $\hat{Z}_{\phi} = \hat{b}_{\phi}^2 - \tilde{\zeta}^{2}$. We used the Dyson series (\ref{dyson}) as well as the relation (\ref{rel proj}) in the first equality and, in the second equality, we explicitly evaluated the first-order term $\mathcal{P}_d \,\mathcal{K}\,\mathcal{P}_d$ using
$\mathcal{P}_d\, \hat{H}_{\rm eff}\mathcal{P}_d\,\hat{\rho} =   -i\frac{\eta}{N}\hat{Z}_{\phi} ^{\dagger} \hat{Z}_{\phi}\mathcal{P}_d\,\hat{\rho}$, $\nobreak{\mathcal{P}_d\,\left[\left(\mathcal{P}_d \,\hat{\rho} \right)\hat{H}_{\rm eff}^{\dagger}\right] =   i\frac{\eta}{N} \left(\mathcal{P}_d\, \hat{\rho}\right) \hat{Z}_{\phi}^{\dagger}\hat{Z}_{\phi}}$ as well as $\nobreak{\mathcal{P}_d\,\mathcal{J}\mathcal{P}_d\,\hat{\rho} =  2\frac{\eta}{N}  \hat{Z}_{\phi}\left(\mathcal{P}_d\,\hat{\rho}\right) \hat{Z}_{\phi}^{\dagger}}$. Note that the first-order term in the Dyson series leads to decay towards the DFS (\ref{DFS}) at times $t \sim \frac{N}{\eta}$ \cite{gilles1994}. As a result, it is sufficient to consider the Dyson series only up to the first order in $\mathcal{K}$ as higher-order terms set in at longer times $t \sim\frac{\gamma}{\eta^2}$. This is in contrast to the Zeno dynamics of the Kerr-resonator array, where the first-order term in the Dyson series leads to the unitary dynamics and decay towards the DFS is due to the second-order term.

We now consider that we initially start with a state in the subspace $ \hat{\rho}_d\otimes\hat{\rho}_{\phi}$. The time evolution within this subspace is described by $\hat{\rho}_{\phi}(\tau) = {\rm Tr}_d\left[\mathcal{P}_d \,\mathcal{U}(\tau) \,\mathcal{P}_d\left(\hat{\rho}_d\otimes\hat{\rho}_{\phi}(0)\right)\right]$. We obtain the effective master equation (\ref{zeno al}) in Lindblad form
where we used the Dyson series (\ref{dyson proj loss}) and rescaled back time  $t = \tau/\gamma$.

\section{Dissipative gap in the Zeno limit}\label{app:gap}
\begin{figure}[t]
\centering
\includegraphics[width=0.666\linewidth]{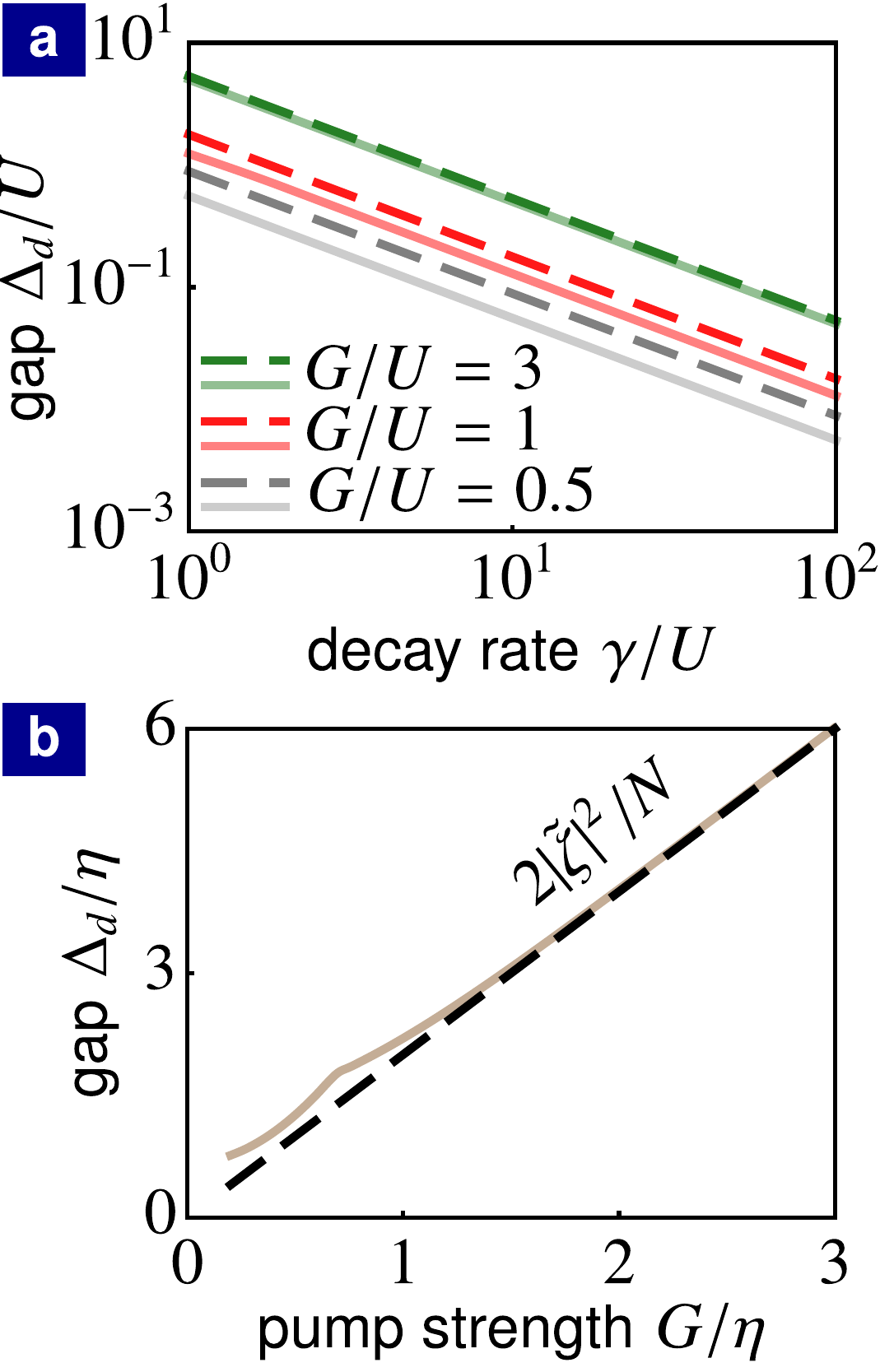}
\caption{Dissipative gap $\Delta_d$ of the effective Liouvillian $\mathcal{L}_{\phi}$. (a) The dissipative gap of the Kerr-resonator array as a function of the decay rate $\gamma$ for different values of $\frac{G}{U}$. The exact values of the dissipative gap obtained by diagonalizing the Liouvillian superoperator (solid lines) and the estimated values of the dissipative gap (dashed lines) according to Eq.~(\ref{eq gap}). (b) The dissipative gap of the alternative model \eqref{master loss} with local two-photon loss as a function of the two-photon pump strength $G$. (Parameters: $N=3$. Truncation: $M_{k=\phi}=40$)}
\label{fig gap}
\end{figure}

In this appendix we study the dissipative gap $\Delta_d$ in the Zeno limit, which is determined from the spectrum of the effective Liouvillian $\mathcal{L}_{\phi}$ as the smallest non-vanishing real part of the eigenvalues \cite{albert2014}. We investigate the dissipative gap first for the Kerr-resonator array and then for the alternative model with local two-photon loss.

For the Kerr-resonator array, we numerically determine the spectrum of the Liouvillian, and from that we extract the dissipative gap. We plot the dissipative gap in Fig.~\ref{fig gap}a as a function of the non-local decay rate $\gamma$ for several values of the two-photon pump strength $G$. From the numerical data, we infer that the dissipative gap is $\Delta_d = \frac{\Gamma}{2}\frac{N}{U}\epsilon_1$, where $\epsilon_1$ is the energy of the first excited state of the effective Hamiltonian $\hat{H}_{\phi}$. For large $|\zeta| \gtrsim 1$, the energy of the first excited state approaches $\epsilon_1 \approx 4\frac{U}{N}|\zeta|^2$ and, as a result, the dissipative gap can be approximated by Eq.~(\ref{eq gap}). In Fig.~\ref{fig gap}a, we can see that the exact values (solid lines) approach the values given by Eq.~(\ref{eq gap}) (dashed lines) as the strength $G$ of the two-photon drive and, as a consequence, $|\zeta|$ increase.

We now study the dissipative gap $\Delta_d$ for the alternative model, which is extracted from the numerically calculated spectrum of the Liouvillian $\mathcal{L}_{\phi}$. Note that, in contrast to the Kerr-resonator array, the Liouvillian and, as a consequence, the dissipative gap are independent of $\gamma$ in the Zeno limit for large $\gamma$. We plot the dissipative gap $\Delta_d$ in Fig.~\ref{fig gap}b as a function of the pump power $G$. It shows that the dissipative gap grows as $\Delta_d \approx2\eta|\zeta|^2/N$ with the amplitude $\zeta$.

\section{Effective master equation \eqref{zeno dis} in the presence of imperfections}\label{app:Zeno dis}

\begin{figure*}[t]
\centering
\includegraphics[width=1\linewidth]{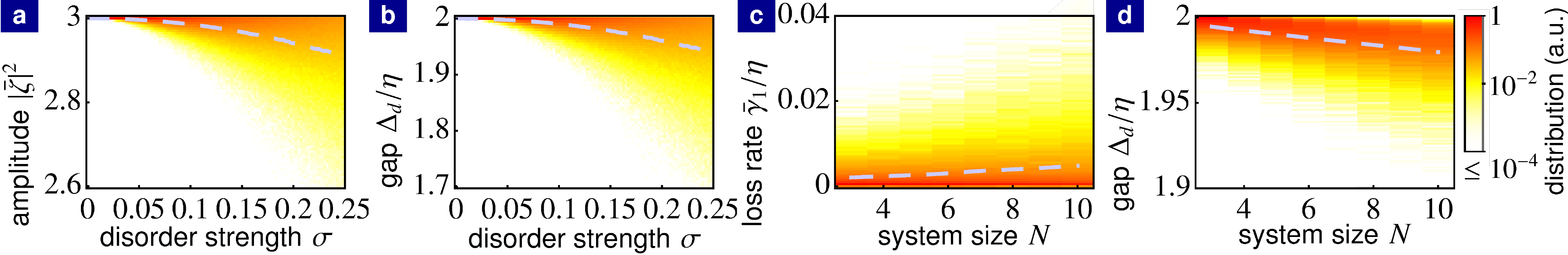}
\caption{Effects of disorder in non-local dissipation in the alternative model. Distribution (color gradient) (a) of the cat-state amplitude $|\zeta|^2$  and (b) of the dissipative gap $\Delta_d$ for disorder in nonlocal decay rates $\gamma_j$ and jump-operator phases $\phi_j$ as a function of the disorder strength $\sigma$. Distribution (c) of the singe-photon loss rate $\bar{\gamma}_1$ and (d) of the dissipative gap as a function of system size $N$. Blue dashed lines show average values of the corresponding quantities. [Parameters: (a) and b $N=3$, $ \gamma/\eta=3$, $G/\eta=1$; (c) and (d) $\sigma=0.075$, $ \gamma/\eta=4.5/(1-\cos(\frac{2\pi}{N}))$, $G/\eta=1$]}
\label{fig:dis app}
\end{figure*}

In this appendix, we discuss the effective master equation for mode $q=1$ for the alternative model \eqref{master loss} with local two-photon loss in the Zeno limit of strong non-local dissipation in the presence of imperfections.

To derive the effective master equation in the Zeno limit $\bar{\gamma}_{q\neq1}\gg\eta,G,\bar{\gamma}_1$, we treat terms $\mathcal{K} \hat{\rho}= \frac{1}{\gamma}\left[\mathcal{L}_{\eta} +\bar{\gamma}_1\mathcal{D}[\hat{b}_1]\hat{\rho}\right]$ as a perturbation to the dominant dissipation $\nobreak{\bar{\mathcal{L}}_d =  \sum_{q\neq1}\frac{\bar{\gamma}_q}{\gamma}\mathcal{D}[\hat{b}_q]}$. Starting initially in the subspace $ \hat{\rho}_d\otimes\hat{\rho}_{1}$, the dynamics of the reduced density matrix $\hat{\rho}_{1}$ is described by the effective master equation (\ref{zeno dis}) in Lindblad form within first-order perturbation theory. In the derivation of the effective master equation, we followed the same steps as in Appendix~\ref{app:Zeno al} for the alternative model without imperfections.

The amplitude of cat states $\bar{\zeta} $ and the dissipative gap $\Delta_d$ are modified due to disorder in non-local decay rates $\gamma_j$ and jump-operator phase $\phi_j$. We plot the distribution of the cat-state amplitude and the dissipative gap in Figs.~\ref{fig:dis app}a and \ref{fig:dis app}b, respectively, in the presence of random disorder as a function of the disorder strength $\sigma$. We can see that for moderate disorder $\sigma\lesssim0.25$, the cat-state amplitude and the dissipative gap retain values that are comparable to those of the disorder-free array for $\sigma=0$. This is in contrast to the single-photon loss rate $\bar{\gamma}_1$, which reaches considerable values due to even moderate disorder $\sigma\sim0.25$ and leads to a significant decoherence rate $|\zeta|^2\bar{\gamma}_1$ in comparison to the dissipative gap as shown in Fig.~\ref{gapdis} in the main text.

We plot the distribution of the single-photon loss rate $\bar{\gamma}_1$ and the dissipative gap $\Delta_d$ in Figs.~\ref{fig:dis app}c and \ref{fig:dis app}d, respectively, for the constant disorder strength $\sigma=0.075$ as a function of system size $N$. We can see that the dissipative gap moderately decreases on average (blue dashed lines) with increasing system size but it retains values that are comparable to the value $\Delta_d\approx2G$ of the disorder-free array. The single-photon loss rate moderately increases on average (blue dashed lines) with increasing system size. This is in contrast to the cat-state amplitude $|\zeta|^2\propto N$ which increases proportionally to system size $N$ leading to a rapidly increasing decoherence rate $|\bar{\zeta}|^2\bar{\gamma}_1$ as discussed in the main text.

\section{Single-photon amplification in the presence of imperfections}\label{app:amp}

 \begin{figure}[t!]
\centering
\includegraphics[width=0.8\linewidth]{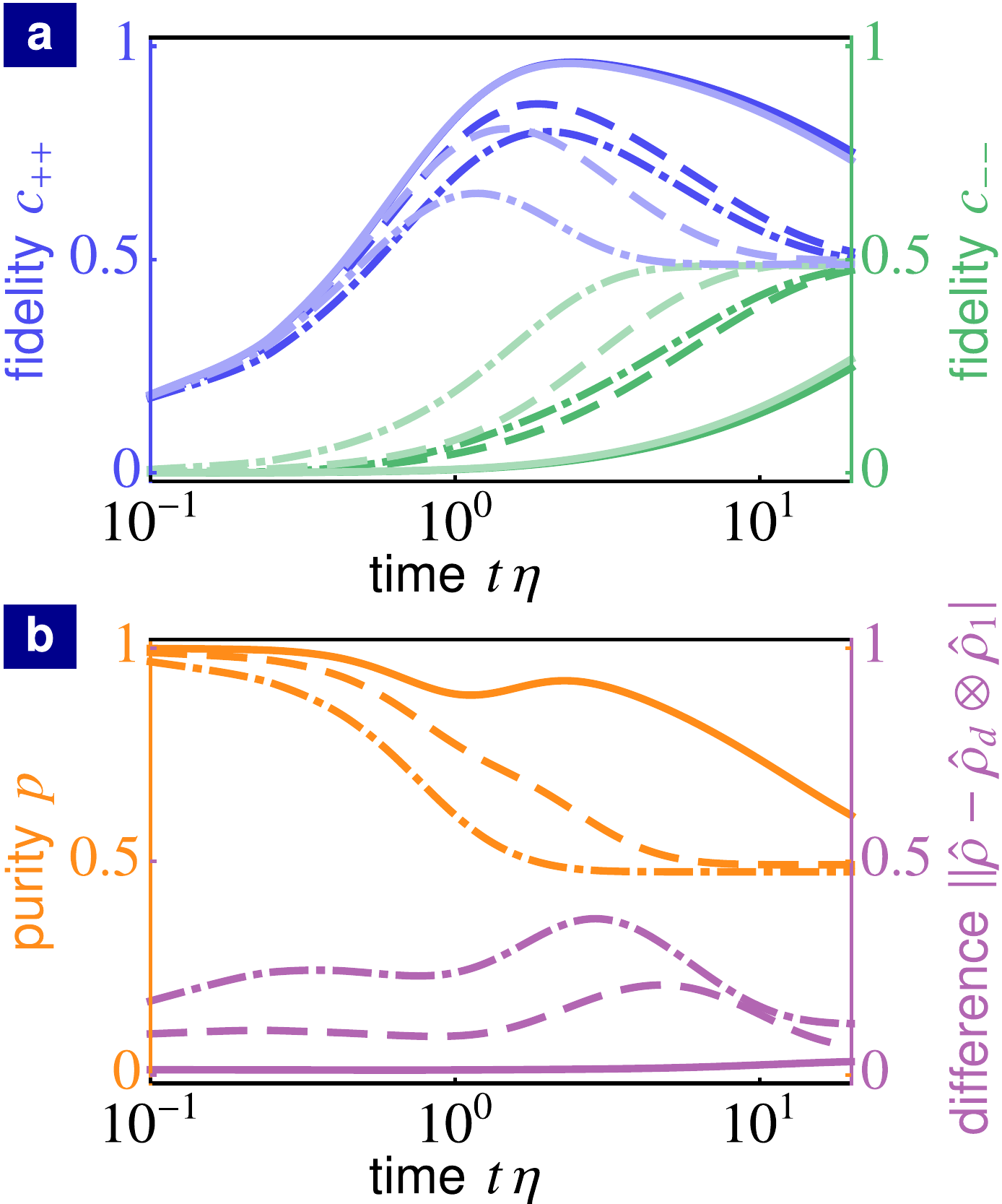}
\caption{Effects of single-photon amplification. (a) The fidelity with the even cat state $c_{++}$ (blue lines) and the odd cat state $c_{--}$ (green lines) as a function of time starting initially from the vacuum state for a particular disorder realization with a vanishing overall phase $\Phi = 0$, and for the non-local dissipation rate $\gamma/\eta=\sqrt{10}$ (dot-dashed lines), $\gamma/\eta=10$ (dashed lines) and $\gamma/\eta=10^2$ (solid lines). Full master equation \eqref{master loss} simulations (dark blue and dark green lines) are compared to the simulations of the effective master equation \eqref{zeno amp} in the Zeno limit  (light blue and light green lines). (b) Purity of the instantaneous state (orange lines) and difference $\lVert\hat{\rho} - \hat{\rho}_d \otimes \hat{\rho}_{1}\rVert$ (purple lines) between the instantaneous state $\hat{\rho}$ according to the full master equation and the instantaneous state $\hat{\rho}_{1}$ according to the master equation (\ref{zeno amp}) for the effective Zeno dynamics. (Parameters: $N=3$, $G/\eta=1$, $\sigma=0.1$, $\delta\gamma_{1} = 1.34$,  $\delta\gamma_{2} =4.59$,  $\delta\gamma_{3} = -5.65$, $\delta\phi_{1} = 2.18$, $\delta\phi_{2} = 0.83$, $\delta\phi_{3} = -3.00$)}
\label{fig:amp}
\end{figure}

In this appendix, we discuss single-phonon amplification which is a second order process in $\hat{H}_G$ in the Zeno limit. Crucially, single-photon amplification is the dominant source of decoherence within the DFS for some realizations of disorder in non-local dissipation. As a result, single-photon amplification has to be taken into account to accurately describe the dynamics of the selected mode in the Zeno limit.

In particular, we consider the alternative model with vanishing two-photon pump phase $\theta=0$ \cite{Note6} and with disordered non-local dissipation rates $\gamma_j$ and jump-operator phases $\phi_j$. Within second-order perturbation theory, the dynamics of the selected mode $q=1$ in the Zeno limit is described by the effective master equation
\begin{equation}
\dot{\hat{\rho}}_{1} = 2 \frac{\bar{\eta}}{N} \mathcal{D}[\hat{b}_{1}^2 - \bar{\zeta}^2]\hat{\rho}_{1} + \bar{\gamma}_1 \mathcal{D}[\hat{b}_{1}]\hat{\rho}_{1}+\mu \mathcal{D}[\hat{b}_{1}^{\dagger}]\hat{\rho}_{1}.\label{zeno amp}
\end{equation}
Single-photon amplification at rate 
\begin{equation}\label{eq:mu}
\mu = 16G^2\sum_{j,k=1}^{N}\sum_{q=2}^{N}\frac{T^*_{j1}T^*_{jq}T_{k1}T_{kq}}{\bar{\gamma}_q },
\end{equation}
 is a second order process in the two-photon pump $\hat{H}_G$ and is absent within the first order perturbation theory, see Eq.~\eqref{zeno dis}.

To derive the master equation \eqref{zeno amp}, we evaluate the second order term in the Dyson series \eqref{dyson proj}
\begin{gather}
 \mathcal{P}_d \,\mathcal{K} \int_{0}^{\tau}{\rm d}\tau_1\int_{0}^{\tau_1}{\rm d}\tau_2 \,e^{\bar{\mathcal{L}}_d \left( \tau_1 - \tau_2\right)}\,\mathcal{K}\,\mathcal{P}_d \approx \frac{\tau}{\gamma}{\mu} \mathcal{D}[\hat{b}_{1}^{\dagger}]\,\mathcal{P}_d
\end{gather}
where we consider only the perturbation in a form of the two-photon pump $\mathcal{K} = -\frac{i}{\gamma}[\hat{H}_G,\,\cdot\,]$. We neglect the remaining perturbations as they give rise to only marginal second-order corrections.

Single-photon amplification is facilitated by virtual excitations of strongly decaying modes $q\neq1$. For an ideal array with vanishing jump-operator phases $\phi_j = 0$, the simultaneous excitation of the selected mode $q=1$ and another mode $q\neq1$ via the two-photon pump is not allowed due to the conservation of quasi-momentum. As a result, the single-photon amplification rate $\mu$ vanishes. On the other hand, imperfections in the form of non-vanishing jump-operator phases $\phi_j \neq 0$ lead to the breaking of the quasi-momentum conservation. As a consequence, these imperfections enable the simultaneous excitation of the selected mode $q=1$ and another mode $q\neq1$ giving rise to non-vanishing single-photon amplification.

We will now show that for imperfections leading to vanishingly small overall phase $\Phi = \sum_{j=1}^N\phi_j$, single-photon amplification is a dominant source of decoherence as single-photon loss rate $\bar{\gamma}_1$ is negligible. We consider a disordered array with the vanishing overall phase $\Phi=0$ and, as a consequence, $\bar{\gamma}_1=0$. In Fig.~\ref{fig:amp}a we plot the fidelity $c_{++}$ of the instantaneous state $\hat{\rho}(t)$ with the even cat state $|\mathcal{C}^+\rangle$ (blue lines) and the fidelity $c_{--}$ with the odd cat state $|\mathcal{C}^-\rangle$ (green lines) as a function of time for different values of the non-local dissipation rare $\gamma$ and starting initially from the vacuum state. In Fig.~\ref{fig:amp}b, we plot the purity $p$ of the instantaneous state. We can see that at long times the array approaches the fully mixed state $\hat{\rho}_{1} \approx \frac{1}{2}\left(|\mathcal{C}_{+}\rangle\langle\mathcal{C}_{+}| + |\mathcal{C}_{-}\rangle\langle\mathcal{C}_{-}|\right)$ with $c_{++}\approx c_{--}\approx p \approx 0.5$ as decoherence due to single-photon amplification sets in. We compare full master equation \eqref{master loss} simulations (dark blue and dark green lines)  to the simulations of the effective master equation \eqref{zeno amp} in the Zeno limit  (light blue and light green lines) and we plot in Fig.~\ref{fig:amp}b difference $\lVert\hat{\rho} - \hat{\rho}_d \otimes \hat{\rho}_{1}\rVert$ (purple lines) between the instantaneous state $\hat{\rho}$ according to the full master equation and the instantaneous state $\hat{\rho}_{1}$ according to the master equation (\ref{zeno amp}) for the effective Zeno dynamics. We can see that in the Zeno limit for large non-local dissipation rate $\gamma$, the effective master equation (\ref{zeno amp}) describes well the time-evolution of the array as the difference $\lVert\hat{\rho} - \hat{\rho}_d \otimes \hat{\rho}_{1}\rVert$ is small throughout the entire time evolution.

Single-photon amplification excites the array out of the DFS. As single-photon amplification $\mu$ is a second-order processes, it is, in the Zeno limit, week compared to the two-photon driven dissipation $\bar{\eta}$, which stabilizes the DFS. As a result, upon excitation out of the DFS the array quickly converges back towards the DFS. However, the single-photon amplification process transforms odd-parity eigenstates to even-parity eigenstates and vice versa as it injects a single photon into the array. As a result, single-photon amplification leads to leaking between parity subspaces and convergence towards the fully mixed state.

Even though, single-photon amplification leads to decoherence within DFS, it does not severely limit the fidelity of prepared cat states in the Zeno limit. Since single-photon amplification is a second-order process facilitated by virtual excitations of strongly decaying modes, its rate is inversely proportional to $\gamma$, see Eq.~\eqref{eq:mu}. As a result, single-photon amplification is suppressed in the Zeno limit. This is in contrast to single-photon loss whose rate is proportional to $\gamma$ leading to strong decoherence in the Zeno limit.

\bibliography{Cat_paper}

\begin{thebibliography}{64}%
\makeatletter
\providecommand \@ifxundefined [1]{%
 \@ifx{#1\undefined}
}%
\providecommand \@ifnum [1]{%
 \ifnum #1\expandafter \@firstoftwo
 \else \expandafter \@secondoftwo
 \fi
}%
\providecommand \@ifx [1]{%
 \ifx #1\expandafter \@firstoftwo
 \else \expandafter \@secondoftwo
 \fi
}%
\providecommand \natexlab [1]{#1}%
\providecommand \enquote  [1]{``#1''}%
\providecommand \bibnamefont  [1]{#1}%
\providecommand \bibfnamefont [1]{#1}%
\providecommand \citenamefont [1]{#1}%
\providecommand \href@noop [0]{\@secondoftwo}%
\providecommand \href [0]{\begingroup \@sanitize@url \@href}%
\providecommand \@href[1]{\@@startlink{#1}\@@href}%
\providecommand \@@href[1]{\endgroup#1\@@endlink}%
\providecommand \@sanitize@url [0]{\catcode `\\12\catcode `\$12\catcode
  `\&12\catcode `\#12\catcode `\^12\catcode `\_12\catcode `\%12\relax}%
\providecommand \@@startlink[1]{}%
\providecommand \@@endlink[0]{}%
\providecommand \url  [0]{\begingroup\@sanitize@url \@url }%
\providecommand \@url [1]{\endgroup\@href {#1}{\urlprefix }}%
\providecommand \urlprefix  [0]{URL }%
\providecommand \Eprint [0]{\href }%
\providecommand \doibase [0]{http://dx.doi.org/}%
\providecommand \selectlanguage [0]{\@gobble}%
\providecommand \bibinfo  [0]{\@secondoftwo}%
\providecommand \bibfield  [0]{\@secondoftwo}%
\providecommand \translation [1]{[#1]}%
\providecommand \BibitemOpen [0]{}%
\providecommand \bibitemStop [0]{}%
\providecommand \bibitemNoStop [0]{.\EOS\space}%
\providecommand \EOS [0]{\spacefactor3000\relax}%
\providecommand \BibitemShut  [1]{\csname bibitem#1\endcsname}%
\let\auto@bib@innerbib\@empty
\bibitem [{\citenamefont {Hacker}\ \emph {et~al.}(2019)\citenamefont {Hacker},
  \citenamefont {Welte}, \citenamefont {Daiss}, \citenamefont {Shaukat},
  \citenamefont {Ritter}, \citenamefont {Li},\ and\ \citenamefont
  {Rempe}}]{Hacker2019}%
  \BibitemOpen
  \bibfield  {author} {\bibinfo {author} {\bibfnamefont {Bastian}\ \bibnamefont
  {Hacker}}, \bibinfo {author} {\bibfnamefont {Stephan}\ \bibnamefont {Welte}},
  \bibinfo {author} {\bibfnamefont {Severin}\ \bibnamefont {Daiss}}, \bibinfo
  {author} {\bibfnamefont {Armin}\ \bibnamefont {Shaukat}}, \bibinfo {author}
  {\bibfnamefont {Stephan}\ \bibnamefont {Ritter}}, \bibinfo {author}
  {\bibfnamefont {Lin}\ \bibnamefont {Li}}, \ and\ \bibinfo {author}
  {\bibfnamefont {Gerhard}\ \bibnamefont {Rempe}},\ }\bibfield  {title}
  {\enquote {\bibinfo {title} {Deterministic creation of entangled
  atom\textendash light {{Schr\"odinger}}-cat states},}\ }\href {\doibase
  10.1038/s41566-018-0339-5} {\bibfield  {journal} {\bibinfo  {journal} {Nat.
  Photonics}\ }\textbf {\bibinfo {volume} {13}},\ \bibinfo {pages} {110--115}
  (\bibinfo {year} {2019})}\BibitemShut {NoStop}%
\bibitem [{\citenamefont {Gilchrist}\ \emph {et~al.}(2004)\citenamefont
  {Gilchrist}, \citenamefont {Nemoto}, \citenamefont {Munro}, \citenamefont
  {Ralph}, \citenamefont {Glancy}, \citenamefont {Braunstein},\ and\
  \citenamefont {Milburn}}]{Gilchrist2004}%
  \BibitemOpen
  \bibfield  {author} {\bibinfo {author} {\bibfnamefont {A.}~\bibnamefont
  {Gilchrist}}, \bibinfo {author} {\bibfnamefont {Kae}\ \bibnamefont {Nemoto}},
  \bibinfo {author} {\bibfnamefont {W.~J.}\ \bibnamefont {Munro}}, \bibinfo
  {author} {\bibfnamefont {T.~C.}\ \bibnamefont {Ralph}}, \bibinfo {author}
  {\bibfnamefont {S.}~\bibnamefont {Glancy}}, \bibinfo {author} {\bibfnamefont
  {Samuel~L.}\ \bibnamefont {Braunstein}}, \ and\ \bibinfo {author}
  {\bibfnamefont {G.~J.}\ \bibnamefont {Milburn}},\ }\bibfield  {title}
  {\enquote {\bibinfo {title} {Schr\"odinger cats and their power for quantum
  information processing},}\ }\href {\doibase 10.1088/1464-4266/6/8/032}
  {\bibfield  {journal} {\bibinfo  {journal} {J. Opt. B: Quantum Semiclass.
  Opt.}\ }\textbf {\bibinfo {volume} {6}},\ \bibinfo {pages} {S828--S833}
  (\bibinfo {year} {2004})}\BibitemShut {NoStop}%
\bibitem [{\citenamefont {Giovannetti}\ \emph {et~al.}(2011)\citenamefont
  {Giovannetti}, \citenamefont {Lloyd},\ and\ \citenamefont
  {Maccone}}]{Giovannetti2011}%
  \BibitemOpen
  \bibfield  {author} {\bibinfo {author} {\bibfnamefont {Vittorio}\
  \bibnamefont {Giovannetti}}, \bibinfo {author} {\bibfnamefont {Seth}\
  \bibnamefont {Lloyd}}, \ and\ \bibinfo {author} {\bibfnamefont {Lorenzo}\
  \bibnamefont {Maccone}},\ }\bibfield  {title} {\enquote {\bibinfo {title}
  {Advances in quantum metrology},}\ }\href {\doibase 10.1038/nphoton.2011.35}
  {\bibfield  {journal} {\bibinfo  {journal} {Nat. Photonics}\ }\textbf
  {\bibinfo {volume} {5}},\ \bibinfo {pages} {222--229} (\bibinfo {year}
  {2011})}\BibitemShut {NoStop}%
\bibitem [{\citenamefont {Pezz{\`e}}\ \emph {et~al.}(2018)\citenamefont
  {Pezz{\`e}}, \citenamefont {Smerzi}, \citenamefont {Oberthaler},
  \citenamefont {Schmied},\ and\ \citenamefont {Treutlein}}]{Pezze2018}%
  \BibitemOpen
  \bibfield  {author} {\bibinfo {author} {\bibfnamefont {Luca}\ \bibnamefont
  {Pezz{\`e}}}, \bibinfo {author} {\bibfnamefont {Augusto}\ \bibnamefont
  {Smerzi}}, \bibinfo {author} {\bibfnamefont {Markus~K.}\ \bibnamefont
  {Oberthaler}}, \bibinfo {author} {\bibfnamefont {Roman}\ \bibnamefont
  {Schmied}}, \ and\ \bibinfo {author} {\bibfnamefont {Philipp}\ \bibnamefont
  {Treutlein}},\ }\bibfield  {title} {\enquote {\bibinfo {title} {Quantum
  metrology with nonclassical states of atomic ensembles},}\ }\href {\doibase
  10.1103/RevModPhys.90.035005} {\bibfield  {journal} {\bibinfo  {journal}
  {Rev. Mod. Phys.}\ }\textbf {\bibinfo {volume} {90}},\ \bibinfo {pages}
  {035005} (\bibinfo {year} {2018})}\BibitemShut {NoStop}%
\bibitem [{\citenamefont {Vlastakis}\ \emph {et~al.}(2013)\citenamefont
  {Vlastakis}, \citenamefont {Kirchmair}, \citenamefont {Leghtas},
  \citenamefont {Nigg}, \citenamefont {Frunzio}, \citenamefont {Girvin},
  \citenamefont {Mirrahimi}, \citenamefont {Devoret},\ and\ \citenamefont
  {Schoelkopf}}]{vlastakis2013}%
  \BibitemOpen
  \bibfield  {author} {\bibinfo {author} {\bibfnamefont {Brian}\ \bibnamefont
  {Vlastakis}}, \bibinfo {author} {\bibfnamefont {Gerhard}\ \bibnamefont
  {Kirchmair}}, \bibinfo {author} {\bibfnamefont {Zaki}\ \bibnamefont
  {Leghtas}}, \bibinfo {author} {\bibfnamefont {Simon~E.}\ \bibnamefont
  {Nigg}}, \bibinfo {author} {\bibfnamefont {Luigi}\ \bibnamefont {Frunzio}},
  \bibinfo {author} {\bibfnamefont {S.~M.}\ \bibnamefont {Girvin}}, \bibinfo
  {author} {\bibfnamefont {Mazyar}\ \bibnamefont {Mirrahimi}}, \bibinfo
  {author} {\bibfnamefont {M.~H.}\ \bibnamefont {Devoret}}, \ and\ \bibinfo
  {author} {\bibfnamefont {R.~J.}\ \bibnamefont {Schoelkopf}},\ }\bibfield
  {title} {\enquote {\bibinfo {title} {Deterministically {{Encoding Quantum
  Information Using}} 100-{{Photon Schr\"odinger Cat States}}},}\ }\href
  {\doibase 10.1126/science.1243289} {\bibfield  {journal} {\bibinfo  {journal}
  {Science}\ }\textbf {\bibinfo {volume} {342}},\ \bibinfo {pages} {607--610}
  (\bibinfo {year} {2013})}\BibitemShut {NoStop}%
\bibitem [{\citenamefont {Chamberland}\ and\ \citenamefont
  {others}(2020)\citenamefont {Chamberland} \emph {et~al.}}]{chamberland2020}%
  \BibitemOpen
  \bibfield  {author} {\bibinfo {author} {\bibfnamefont {Christopher}\
  \bibnamefont {Chamberland}} \emph {et~al.},\ }\bibfield  {title} {\enquote
  {\bibinfo {title} {Building a fault-tolerant quantum computer using
  concatenated cat codes},}\ }\href@noop {} {\bibfield  {journal} {\bibinfo
  {journal} {arXiv:2012.04108}\ } (\bibinfo {year} {2020})}\BibitemShut
  {NoStop}%
\bibitem [{\citenamefont {Grimm}\ \emph {et~al.}(2020)\citenamefont {Grimm},
  \citenamefont {Frattini}, \citenamefont {Puri}, \citenamefont {Mundhada},
  \citenamefont {Touzard}, \citenamefont {Mirrahimi}, \citenamefont {Girvin},
  \citenamefont {Shankar},\ and\ \citenamefont {Devoret}}]{grimm2020}%
  \BibitemOpen
  \bibfield  {author} {\bibinfo {author} {\bibfnamefont {A.}~\bibnamefont
  {Grimm}}, \bibinfo {author} {\bibfnamefont {N.~E.}\ \bibnamefont {Frattini}},
  \bibinfo {author} {\bibfnamefont {S.}~\bibnamefont {Puri}}, \bibinfo {author}
  {\bibfnamefont {S.~O.}\ \bibnamefont {Mundhada}}, \bibinfo {author}
  {\bibfnamefont {S.}~\bibnamefont {Touzard}}, \bibinfo {author} {\bibfnamefont
  {M.}~\bibnamefont {Mirrahimi}}, \bibinfo {author} {\bibfnamefont {S.~M.}\
  \bibnamefont {Girvin}}, \bibinfo {author} {\bibfnamefont {S.}~\bibnamefont
  {Shankar}}, \ and\ \bibinfo {author} {\bibfnamefont {M.~H.}\ \bibnamefont
  {Devoret}},\ }\bibfield  {title} {\enquote {\bibinfo {title} {Stabilization
  and operation of a {{Kerr}}-cat qubit},}\ }\href {\doibase
  10.1038/s41586-020-2587-z} {\bibfield  {journal} {\bibinfo  {journal}
  {Nature}\ }\textbf {\bibinfo {volume} {584}},\ \bibinfo {pages} {205--209}
  (\bibinfo {year} {2020})}\BibitemShut {NoStop}%
\bibitem [{\citenamefont {Leghtas}\ \emph {et~al.}(2013)\citenamefont
  {Leghtas}, \citenamefont {Kirchmair}, \citenamefont {Vlastakis},
  \citenamefont {Schoelkopf}, \citenamefont {Devoret},\ and\ \citenamefont
  {Mirrahimi}}]{leghtas2013}%
  \BibitemOpen
  \bibfield  {author} {\bibinfo {author} {\bibfnamefont {Zaki}\ \bibnamefont
  {Leghtas}}, \bibinfo {author} {\bibfnamefont {Gerhard}\ \bibnamefont
  {Kirchmair}}, \bibinfo {author} {\bibfnamefont {Brian}\ \bibnamefont
  {Vlastakis}}, \bibinfo {author} {\bibfnamefont {Robert~J.}\ \bibnamefont
  {Schoelkopf}}, \bibinfo {author} {\bibfnamefont {Michel~H.}\ \bibnamefont
  {Devoret}}, \ and\ \bibinfo {author} {\bibfnamefont {Mazyar}\ \bibnamefont
  {Mirrahimi}},\ }\bibfield  {title} {\enquote {\bibinfo {title}
  {Hardware-{{Efficient Autonomous Quantum Memory Protection}}},}\ }\href
  {\doibase 10.1103/PhysRevLett.111.120501} {\bibfield  {journal} {\bibinfo
  {journal} {Phys. Rev. Lett.}\ }\textbf {\bibinfo {volume} {111}},\ \bibinfo
  {pages} {120501} (\bibinfo {year} {2013})}\BibitemShut {NoStop}%
\bibitem [{\citenamefont {Mirrahimi}\ \emph {et~al.}(2014)\citenamefont
  {Mirrahimi}, \citenamefont {Leghtas}, \citenamefont {Albert}, \citenamefont
  {Touzard}, \citenamefont {Schoelkopf}, \citenamefont {Jiang},\ and\
  \citenamefont {Devoret}}]{mirrahimi2014}%
  \BibitemOpen
  \bibfield  {author} {\bibinfo {author} {\bibfnamefont {Mazyar}\ \bibnamefont
  {Mirrahimi}}, \bibinfo {author} {\bibfnamefont {Zaki}\ \bibnamefont
  {Leghtas}}, \bibinfo {author} {\bibfnamefont {Victor~V.}\ \bibnamefont
  {Albert}}, \bibinfo {author} {\bibfnamefont {Steven}\ \bibnamefont
  {Touzard}}, \bibinfo {author} {\bibfnamefont {Robert~J.}\ \bibnamefont
  {Schoelkopf}}, \bibinfo {author} {\bibfnamefont {Liang}\ \bibnamefont
  {Jiang}}, \ and\ \bibinfo {author} {\bibfnamefont {Michel~H.}\ \bibnamefont
  {Devoret}},\ }\bibfield  {title} {\enquote {\bibinfo {title} {Dynamically
  protected cat-qubits: A new paradigm for universal quantum computation},}\
  }\href {\doibase 10.1088/1367-2630/16/4/045014} {\bibfield  {journal}
  {\bibinfo  {journal} {New J. Phys.}\ }\textbf {\bibinfo {volume} {16}},\
  \bibinfo {pages} {045014} (\bibinfo {year} {2014})}\BibitemShut {NoStop}%
\bibitem [{\citenamefont {Leghtas}\ \emph {et~al.}(2015)\citenamefont
  {Leghtas}, \citenamefont {Touzard}, \citenamefont {Pop}, \citenamefont {Kou},
  \citenamefont {Vlastakis}, \citenamefont {Petrenko}, \citenamefont {Sliwa},
  \citenamefont {Narla}, \citenamefont {Shankar}, \citenamefont {Hatridge},
  \citenamefont {Reagor}, \citenamefont {Frunzio}, \citenamefont {Schoelkopf},
  \citenamefont {Mirrahimi},\ and\ \citenamefont {Devoret}}]{leghtas2015}%
  \BibitemOpen
  \bibfield  {author} {\bibinfo {author} {\bibfnamefont {Z.}~\bibnamefont
  {Leghtas}}, \bibinfo {author} {\bibfnamefont {S.}~\bibnamefont {Touzard}},
  \bibinfo {author} {\bibfnamefont {I.~M.}\ \bibnamefont {Pop}}, \bibinfo
  {author} {\bibfnamefont {A.}~\bibnamefont {Kou}}, \bibinfo {author}
  {\bibfnamefont {B.}~\bibnamefont {Vlastakis}}, \bibinfo {author}
  {\bibfnamefont {A.}~\bibnamefont {Petrenko}}, \bibinfo {author}
  {\bibfnamefont {K.~M.}\ \bibnamefont {Sliwa}}, \bibinfo {author}
  {\bibfnamefont {A.}~\bibnamefont {Narla}}, \bibinfo {author} {\bibfnamefont
  {S.}~\bibnamefont {Shankar}}, \bibinfo {author} {\bibfnamefont {M.~J.}\
  \bibnamefont {Hatridge}}, \bibinfo {author} {\bibfnamefont {M.}~\bibnamefont
  {Reagor}}, \bibinfo {author} {\bibfnamefont {L.}~\bibnamefont {Frunzio}},
  \bibinfo {author} {\bibfnamefont {R.~J.}\ \bibnamefont {Schoelkopf}},
  \bibinfo {author} {\bibfnamefont {M.}~\bibnamefont {Mirrahimi}}, \ and\
  \bibinfo {author} {\bibfnamefont {M.~H.}\ \bibnamefont {Devoret}},\
  }\bibfield  {title} {\enquote {\bibinfo {title} {Confining the state of light
  to a quantum manifold by engineered two-photon loss},}\ }\href {\doibase
  10.1126/science.aaa2085} {\bibfield  {journal} {\bibinfo  {journal}
  {Science}\ }\textbf {\bibinfo {volume} {347}},\ \bibinfo {pages} {853--857}
  (\bibinfo {year} {2015})}\BibitemShut {NoStop}%
\bibitem [{\citenamefont {Ofek}\ \emph {et~al.}(2016)\citenamefont {Ofek},
  \citenamefont {Petrenko}, \citenamefont {Heeres}, \citenamefont {Reinhold},
  \citenamefont {Leghtas}, \citenamefont {Vlastakis}, \citenamefont {Liu},
  \citenamefont {Frunzio}, \citenamefont {Girvin}, \citenamefont {Jiang},
  \citenamefont {Mirrahimi}, \citenamefont {Devoret},\ and\ \citenamefont
  {Schoelkopf}}]{ofek2016}%
  \BibitemOpen
  \bibfield  {author} {\bibinfo {author} {\bibfnamefont {Nissim}\ \bibnamefont
  {Ofek}}, \bibinfo {author} {\bibfnamefont {Andrei}\ \bibnamefont {Petrenko}},
  \bibinfo {author} {\bibfnamefont {Reinier}\ \bibnamefont {Heeres}}, \bibinfo
  {author} {\bibfnamefont {Philip}\ \bibnamefont {Reinhold}}, \bibinfo {author}
  {\bibfnamefont {Zaki}\ \bibnamefont {Leghtas}}, \bibinfo {author}
  {\bibfnamefont {Brian}\ \bibnamefont {Vlastakis}}, \bibinfo {author}
  {\bibfnamefont {Yehan}\ \bibnamefont {Liu}}, \bibinfo {author} {\bibfnamefont
  {Luigi}\ \bibnamefont {Frunzio}}, \bibinfo {author} {\bibfnamefont {S.~M.}\
  \bibnamefont {Girvin}}, \bibinfo {author} {\bibfnamefont {L.}~\bibnamefont
  {Jiang}}, \bibinfo {author} {\bibfnamefont {Mazyar}\ \bibnamefont
  {Mirrahimi}}, \bibinfo {author} {\bibfnamefont {M.~H.}\ \bibnamefont
  {Devoret}}, \ and\ \bibinfo {author} {\bibfnamefont {R.~J.}\ \bibnamefont
  {Schoelkopf}},\ }\bibfield  {title} {\enquote {\bibinfo {title} {Extending
  the lifetime of a quantum bit with error correction in superconducting
  circuits},}\ }\href {\doibase 10.1038/nature18949} {\bibfield  {journal}
  {\bibinfo  {journal} {Nature}\ }\textbf {\bibinfo {volume} {536}},\ \bibinfo
  {pages} {441--445} (\bibinfo {year} {2016})}\BibitemShut {NoStop}%
\bibitem [{\citenamefont {Puri}\ \emph
  {et~al.}(2017{\natexlab{a}})\citenamefont {Puri}, \citenamefont {Boutin},\
  and\ \citenamefont {Blais}}]{puri2017}%
  \BibitemOpen
  \bibfield  {author} {\bibinfo {author} {\bibfnamefont {Shruti}\ \bibnamefont
  {Puri}}, \bibinfo {author} {\bibfnamefont {Samuel}\ \bibnamefont {Boutin}}, \
  and\ \bibinfo {author} {\bibfnamefont {Alexandre}\ \bibnamefont {Blais}},\
  }\bibfield  {title} {\enquote {\bibinfo {title} {Engineering the quantum
  states of light in a {{Kerr}}-nonlinear resonator by two-photon driving},}\
  }\href {\doibase 10.1038/s41534-017-0019-1} {\bibfield  {journal} {\bibinfo
  {journal} {Quantum Inf.}\ }\textbf {\bibinfo {volume} {3}},\ \bibinfo {pages}
  {18} (\bibinfo {year} {2017}{\natexlab{a}})}\BibitemShut {NoStop}%
\bibitem [{\citenamefont {Puri}\ \emph {et~al.}(2019)\citenamefont {Puri},
  \citenamefont {Grimm}, \citenamefont {{Campagne-Ibarcq}}, \citenamefont
  {Eickbusch}, \citenamefont {Noh}, \citenamefont {Roberts}, \citenamefont
  {Jiang}, \citenamefont {Mirrahimi}, \citenamefont {Devoret},\ and\
  \citenamefont {Girvin}}]{puri2019}%
  \BibitemOpen
  \bibfield  {author} {\bibinfo {author} {\bibfnamefont {Shruti}\ \bibnamefont
  {Puri}}, \bibinfo {author} {\bibfnamefont {Alexander}\ \bibnamefont {Grimm}},
  \bibinfo {author} {\bibfnamefont {Philippe}\ \bibnamefont
  {{Campagne-Ibarcq}}}, \bibinfo {author} {\bibfnamefont {Alec}\ \bibnamefont
  {Eickbusch}}, \bibinfo {author} {\bibfnamefont {Kyungjoo}\ \bibnamefont
  {Noh}}, \bibinfo {author} {\bibfnamefont {Gabrielle}\ \bibnamefont
  {Roberts}}, \bibinfo {author} {\bibfnamefont {Liang}\ \bibnamefont {Jiang}},
  \bibinfo {author} {\bibfnamefont {Mazyar}\ \bibnamefont {Mirrahimi}},
  \bibinfo {author} {\bibfnamefont {Michel~H.}\ \bibnamefont {Devoret}}, \ and\
  \bibinfo {author} {\bibfnamefont {S.~M.}\ \bibnamefont {Girvin}},\ }\bibfield
   {title} {\enquote {\bibinfo {title} {Stabilized {{Cat}} in a {{Driven
  Nonlinear Cavity}}: A {{Fault}}-{{Tolerant Error Syndrome Detector}}},}\
  }\href {\doibase 10.1103/PhysRevX.9.041009} {\bibfield  {journal} {\bibinfo
  {journal} {Phys. Rev. X}\ }\textbf {\bibinfo {volume} {9}},\ \bibinfo {pages}
  {041009} (\bibinfo {year} {2019})}\BibitemShut {NoStop}%
\bibitem [{\citenamefont {Touzard}\ \emph {et~al.}(2018)\citenamefont
  {Touzard}, \citenamefont {Grimm}, \citenamefont {Leghtas}, \citenamefont
  {Mundhada}, \citenamefont {Reinhold}, \citenamefont {Axline}, \citenamefont
  {Reagor}, \citenamefont {Chou}, \citenamefont {Blumoff}, \citenamefont
  {Sliwa}, \citenamefont {Shankar}, \citenamefont {Frunzio}, \citenamefont
  {Schoelkopf}, \citenamefont {Mirrahimi},\ and\ \citenamefont
  {Devoret}}]{touzard2018}%
  \BibitemOpen
  \bibfield  {author} {\bibinfo {author} {\bibfnamefont {S.}~\bibnamefont
  {Touzard}}, \bibinfo {author} {\bibfnamefont {A.}~\bibnamefont {Grimm}},
  \bibinfo {author} {\bibfnamefont {Z.}~\bibnamefont {Leghtas}}, \bibinfo
  {author} {\bibfnamefont {S.~O.}\ \bibnamefont {Mundhada}}, \bibinfo {author}
  {\bibfnamefont {P.}~\bibnamefont {Reinhold}}, \bibinfo {author}
  {\bibfnamefont {C.}~\bibnamefont {Axline}}, \bibinfo {author} {\bibfnamefont
  {M.}~\bibnamefont {Reagor}}, \bibinfo {author} {\bibfnamefont
  {K.}~\bibnamefont {Chou}}, \bibinfo {author} {\bibfnamefont {J.}~\bibnamefont
  {Blumoff}}, \bibinfo {author} {\bibfnamefont {K.~M.}\ \bibnamefont {Sliwa}},
  \bibinfo {author} {\bibfnamefont {S.}~\bibnamefont {Shankar}}, \bibinfo
  {author} {\bibfnamefont {L.}~\bibnamefont {Frunzio}}, \bibinfo {author}
  {\bibfnamefont {R.~J.}\ \bibnamefont {Schoelkopf}}, \bibinfo {author}
  {\bibfnamefont {M.}~\bibnamefont {Mirrahimi}}, \ and\ \bibinfo {author}
  {\bibfnamefont {M.~H.}\ \bibnamefont {Devoret}},\ }\bibfield  {title}
  {\enquote {\bibinfo {title} {Coherent {{Oscillations}} inside a {{Quantum
  Manifold Stabilized}} by {{Dissipation}}},}\ }\href {\doibase
  10.1103/PhysRevX.8.021005} {\bibfield  {journal} {\bibinfo  {journal} {Phys.
  Rev. X}\ }\textbf {\bibinfo {volume} {8}},\ \bibinfo {pages} {021005}
  (\bibinfo {year} {2018})}\BibitemShut {NoStop}%
\bibitem [{\citenamefont {Guillaud}\ and\ \citenamefont
  {Mirrahimi}(2019)}]{guillaud2019}%
  \BibitemOpen
  \bibfield  {author} {\bibinfo {author} {\bibfnamefont {J{\'e}r{\'e}mie}\
  \bibnamefont {Guillaud}}\ and\ \bibinfo {author} {\bibfnamefont {Mazyar}\
  \bibnamefont {Mirrahimi}},\ }\bibfield  {title} {\enquote {\bibinfo {title}
  {Repetition {{Cat Qubits}} for {{Fault}}-{{Tolerant Quantum Computation}}},}\
  }\href {\doibase 10.1103/PhysRevX.9.041053} {\bibfield  {journal} {\bibinfo
  {journal} {Phys. Rev. X}\ }\textbf {\bibinfo {volume} {9}},\ \bibinfo {pages}
  {041053} (\bibinfo {year} {2019})}\BibitemShut {NoStop}%
\bibitem [{\citenamefont {Puri}\ \emph {et~al.}(2020)\citenamefont {Puri},
  \citenamefont {{St-Jean}}, \citenamefont {Gross}, \citenamefont {Grimm},
  \citenamefont {Frattini}, \citenamefont {Iyer}, \citenamefont {Krishna},
  \citenamefont {Touzard}, \citenamefont {Jiang}, \citenamefont {Blais},
  \citenamefont {Flammia},\ and\ \citenamefont {Girvin}}]{Puri2020}%
  \BibitemOpen
  \bibfield  {author} {\bibinfo {author} {\bibfnamefont {Shruti}\ \bibnamefont
  {Puri}}, \bibinfo {author} {\bibfnamefont {Lucas}\ \bibnamefont {{St-Jean}}},
  \bibinfo {author} {\bibfnamefont {Jonathan~A.}\ \bibnamefont {Gross}},
  \bibinfo {author} {\bibfnamefont {Alexander}\ \bibnamefont {Grimm}}, \bibinfo
  {author} {\bibfnamefont {Nicholas~E.}\ \bibnamefont {Frattini}}, \bibinfo
  {author} {\bibfnamefont {Pavithran~S.}\ \bibnamefont {Iyer}}, \bibinfo
  {author} {\bibfnamefont {Anirudh}\ \bibnamefont {Krishna}}, \bibinfo {author}
  {\bibfnamefont {Steven}\ \bibnamefont {Touzard}}, \bibinfo {author}
  {\bibfnamefont {Liang}\ \bibnamefont {Jiang}}, \bibinfo {author}
  {\bibfnamefont {Alexandre}\ \bibnamefont {Blais}}, \bibinfo {author}
  {\bibfnamefont {Steven~T.}\ \bibnamefont {Flammia}}, \ and\ \bibinfo {author}
  {\bibfnamefont {S.~M.}\ \bibnamefont {Girvin}},\ }\bibfield  {title}
  {\enquote {\bibinfo {title} {Bias-preserving gates with stabilized cat
  qubits},}\ }\href {\doibase 10.1126/sciadv.aay5901} {\bibfield  {journal}
  {\bibinfo  {journal} {Science Advances}\ }\textbf {\bibinfo {volume} {6}},\
  \bibinfo {pages} {eaay5901} (\bibinfo {year} {2020})}\BibitemShut {NoStop}%
\bibitem [{\citenamefont {Lescanne}\ \emph {et~al.}(2020)\citenamefont
  {Lescanne}, \citenamefont {Villiers}, \citenamefont {Peronnin}, \citenamefont
  {Sarlette}, \citenamefont {Delbecq}, \citenamefont {Huard}, \citenamefont
  {Kontos}, \citenamefont {Mirrahimi},\ and\ \citenamefont
  {Leghtas}}]{Lescanne2020}%
  \BibitemOpen
  \bibfield  {author} {\bibinfo {author} {\bibfnamefont {Rapha{\"e}l}\
  \bibnamefont {Lescanne}}, \bibinfo {author} {\bibfnamefont {Marius}\
  \bibnamefont {Villiers}}, \bibinfo {author} {\bibfnamefont {Th{\'e}au}\
  \bibnamefont {Peronnin}}, \bibinfo {author} {\bibfnamefont {Alain}\
  \bibnamefont {Sarlette}}, \bibinfo {author} {\bibfnamefont {Matthieu}\
  \bibnamefont {Delbecq}}, \bibinfo {author} {\bibfnamefont {Benjamin}\
  \bibnamefont {Huard}}, \bibinfo {author} {\bibfnamefont {Takis}\ \bibnamefont
  {Kontos}}, \bibinfo {author} {\bibfnamefont {Mazyar}\ \bibnamefont
  {Mirrahimi}}, \ and\ \bibinfo {author} {\bibfnamefont {Zaki}\ \bibnamefont
  {Leghtas}},\ }\bibfield  {title} {\enquote {\bibinfo {title} {Exponential
  suppression of bit-flips in a qubit encoded in an oscillator},}\ }\href
  {\doibase 10.1038/s41567-020-0824-x} {\bibfield  {journal} {\bibinfo
  {journal} {Nat. Phys.}\ }\textbf {\bibinfo {volume} {16}},\ \bibinfo {pages}
  {509--513} (\bibinfo {year} {2020})}\BibitemShut {NoStop}%
\bibitem [{\citenamefont {Puri}\ \emph
  {et~al.}(2017{\natexlab{b}})\citenamefont {Puri}, \citenamefont {Andersen},
  \citenamefont {Grimsmo},\ and\ \citenamefont {Blais}}]{Puri2017b}%
  \BibitemOpen
  \bibfield  {author} {\bibinfo {author} {\bibfnamefont {Shruti}\ \bibnamefont
  {Puri}}, \bibinfo {author} {\bibfnamefont {Christian~Kraglund}\ \bibnamefont
  {Andersen}}, \bibinfo {author} {\bibfnamefont {Arne~L.}\ \bibnamefont
  {Grimsmo}}, \ and\ \bibinfo {author} {\bibfnamefont {Alexandre}\ \bibnamefont
  {Blais}},\ }\bibfield  {title} {\enquote {\bibinfo {title} {Quantum annealing
  with a network of all-to-all connected, two-photon driven {{Kerr}} nonlinear
  oscillators},}\ }\href {\doibase 10.1038/ncomms15785} {\bibfield  {journal}
  {\bibinfo  {journal} {Nat. Commun.}\ }\textbf {\bibinfo {volume} {8}},\
  \bibinfo {pages} {15785} (\bibinfo {year} {2017}{\natexlab{b}})}\BibitemShut
  {NoStop}%
\bibitem [{\citenamefont {Sanders}(1992)}]{sanders1992}%
  \BibitemOpen
  \bibfield  {author} {\bibinfo {author} {\bibfnamefont {Barry~C.}\
  \bibnamefont {Sanders}},\ }\bibfield  {title} {\enquote {\bibinfo {title}
  {Entangled coherent states},}\ }\href {\doibase 10.1103/PhysRevA.45.6811}
  {\bibfield  {journal} {\bibinfo  {journal} {Phys. Rev. A}\ }\textbf {\bibinfo
  {volume} {45}},\ \bibinfo {pages} {6811--6815} (\bibinfo {year}
  {1992})}\BibitemShut {NoStop}%
\bibitem [{\citenamefont {{van Enk}}\ and\ \citenamefont
  {Hirota}(2001)}]{vanenk2001}%
  \BibitemOpen
  \bibfield  {author} {\bibinfo {author} {\bibfnamefont {S.~J.}\ \bibnamefont
  {{van Enk}}}\ and\ \bibinfo {author} {\bibfnamefont {O.}~\bibnamefont
  {Hirota}},\ }\bibfield  {title} {\enquote {\bibinfo {title} {Entangled
  coherent states: Teleportation and decoherence},}\ }\href {\doibase
  10.1103/PhysRevA.64.022313} {\bibfield  {journal} {\bibinfo  {journal} {Phys.
  Rev. A}\ }\textbf {\bibinfo {volume} {64}},\ \bibinfo {pages} {022313}
  (\bibinfo {year} {2001})}\BibitemShut {NoStop}%
\bibitem [{\citenamefont {Sanders}(2012)}]{sanders2012}%
  \BibitemOpen
  \bibfield  {author} {\bibinfo {author} {\bibfnamefont {Barry~C.}\
  \bibnamefont {Sanders}},\ }\bibfield  {title} {\enquote {\bibinfo {title}
  {Review of entangled coherent states},}\ }\href {\doibase
  10.1088/1751-8113/45/24/244002} {\bibfield  {journal} {\bibinfo  {journal}
  {Journal of Physics A: Mathematical and Theoretical}\ }\textbf {\bibinfo
  {volume} {45}},\ \bibinfo {pages} {244002} (\bibinfo {year}
  {2012})}\BibitemShut {NoStop}%
\bibitem [{\citenamefont {Joo}\ \emph {et~al.}(2011)\citenamefont {Joo},
  \citenamefont {Munro},\ and\ \citenamefont {Spiller}}]{joo2011}%
  \BibitemOpen
  \bibfield  {author} {\bibinfo {author} {\bibfnamefont {Jaewoo}\ \bibnamefont
  {Joo}}, \bibinfo {author} {\bibfnamefont {William~J.}\ \bibnamefont {Munro}},
  \ and\ \bibinfo {author} {\bibfnamefont {Timothy~P.}\ \bibnamefont
  {Spiller}},\ }\bibfield  {title} {\enquote {\bibinfo {title} {Quantum
  {{Metrology}} with {{Entangled Coherent States}}},}\ }\href {\doibase
  10.1103/PhysRevLett.107.083601} {\bibfield  {journal} {\bibinfo  {journal}
  {Phys. Rev. Lett.}\ }\textbf {\bibinfo {volume} {107}},\ \bibinfo {pages}
  {083601} (\bibinfo {year} {2011})}\BibitemShut {NoStop}%
\bibitem [{\citenamefont {Zhang}\ \emph {et~al.}(2013)\citenamefont {Zhang},
  \citenamefont {Li}, \citenamefont {Yang},\ and\ \citenamefont
  {Jin}}]{zhang2013}%
  \BibitemOpen
  \bibfield  {author} {\bibinfo {author} {\bibfnamefont {Y.~M.}\ \bibnamefont
  {Zhang}}, \bibinfo {author} {\bibfnamefont {X.~W.}\ \bibnamefont {Li}},
  \bibinfo {author} {\bibfnamefont {W.}~\bibnamefont {Yang}}, \ and\ \bibinfo
  {author} {\bibfnamefont {G.~R.}\ \bibnamefont {Jin}},\ }\bibfield  {title}
  {\enquote {\bibinfo {title} {Quantum {{Fisher}} information of entangled
  coherent states in the presence of photon loss},}\ }\href {\doibase
  10.1103/PhysRevA.88.043832} {\bibfield  {journal} {\bibinfo  {journal} {Phys.
  Rev. A}\ }\textbf {\bibinfo {volume} {88}},\ \bibinfo {pages} {043832}
  (\bibinfo {year} {2013})}\BibitemShut {NoStop}%
\bibitem [{\citenamefont {Wang}\ \emph {et~al.}(2016)\citenamefont {Wang},
  \citenamefont {Gao}, \citenamefont {Reinhold}, \citenamefont {Heeres},
  \citenamefont {Ofek}, \citenamefont {Chou}, \citenamefont {Axline},
  \citenamefont {Reagor}, \citenamefont {Blumoff}, \citenamefont {Sliwa},
  \citenamefont {Frunzio}, \citenamefont {Girvin}, \citenamefont {Jiang},
  \citenamefont {Mirrahimi}, \citenamefont {Devoret},\ and\ \citenamefont
  {Schoelkopf}}]{wang2016}%
  \BibitemOpen
  \bibfield  {author} {\bibinfo {author} {\bibfnamefont {Chen}\ \bibnamefont
  {Wang}}, \bibinfo {author} {\bibfnamefont {Yvonne~Y.}\ \bibnamefont {Gao}},
  \bibinfo {author} {\bibfnamefont {Philip}\ \bibnamefont {Reinhold}}, \bibinfo
  {author} {\bibfnamefont {R.~W.}\ \bibnamefont {Heeres}}, \bibinfo {author}
  {\bibfnamefont {Nissim}\ \bibnamefont {Ofek}}, \bibinfo {author}
  {\bibfnamefont {Kevin}\ \bibnamefont {Chou}}, \bibinfo {author}
  {\bibfnamefont {Christopher}\ \bibnamefont {Axline}}, \bibinfo {author}
  {\bibfnamefont {Matthew}\ \bibnamefont {Reagor}}, \bibinfo {author}
  {\bibfnamefont {Jacob}\ \bibnamefont {Blumoff}}, \bibinfo {author}
  {\bibfnamefont {K.~M.}\ \bibnamefont {Sliwa}}, \bibinfo {author}
  {\bibfnamefont {L.}~\bibnamefont {Frunzio}}, \bibinfo {author} {\bibfnamefont
  {S.~M.}\ \bibnamefont {Girvin}}, \bibinfo {author} {\bibfnamefont {Liang}\
  \bibnamefont {Jiang}}, \bibinfo {author} {\bibfnamefont {M.}~\bibnamefont
  {Mirrahimi}}, \bibinfo {author} {\bibfnamefont {M.~H.}\ \bibnamefont
  {Devoret}}, \ and\ \bibinfo {author} {\bibfnamefont {R.~J.}\ \bibnamefont
  {Schoelkopf}},\ }\bibfield  {title} {\enquote {\bibinfo {title} {A
  {{Schr\"odinger}} cat living in two boxes},}\ }\href {\doibase
  10.1126/science.aaf2941} {\bibfield  {journal} {\bibinfo  {journal}
  {Science}\ }\textbf {\bibinfo {volume} {352}},\ \bibinfo {pages} {1087--1091}
  (\bibinfo {year} {2016})}\BibitemShut {NoStop}%
\bibitem [{\citenamefont {Braunstein}\ and\ \citenamefont {{van
  Loock}}(2005)}]{braunstein2005}%
  \BibitemOpen
  \bibfield  {author} {\bibinfo {author} {\bibfnamefont {Samuel~L.}\
  \bibnamefont {Braunstein}}\ and\ \bibinfo {author} {\bibfnamefont {Peter}\
  \bibnamefont {{van Loock}}},\ }\bibfield  {title} {\enquote {\bibinfo {title}
  {Quantum information with continuous variables},}\ }\href {\doibase
  10.1103/RevModPhys.77.513} {\bibfield  {journal} {\bibinfo  {journal} {Rev.
  Mod. Phys.}\ }\textbf {\bibinfo {volume} {77}},\ \bibinfo {pages} {513--577}
  (\bibinfo {year} {2005})}\BibitemShut {NoStop}%
\bibitem [{\citenamefont {Wang}\ and\ \citenamefont
  {Sanders}(2001)}]{wang2001}%
  \BibitemOpen
  \bibfield  {author} {\bibinfo {author} {\bibfnamefont {Xiaoguang}\
  \bibnamefont {Wang}}\ and\ \bibinfo {author} {\bibfnamefont {Barry~C.}\
  \bibnamefont {Sanders}},\ }\bibfield  {title} {\enquote {\bibinfo {title}
  {Multipartite entangled coherent states},}\ }\href {\doibase
  10.1103/PhysRevA.65.012303} {\bibfield  {journal} {\bibinfo  {journal} {Phys.
  Rev. A}\ }\textbf {\bibinfo {volume} {65}},\ \bibinfo {pages} {012303}
  (\bibinfo {year} {2001})}\BibitemShut {NoStop}%
\bibitem [{\citenamefont {Blais}\ \emph
  {et~al.}(2020{\natexlab{a}})\citenamefont {Blais}, \citenamefont {Girvin},\
  and\ \citenamefont {Oliver}}]{Blais2020b}%
  \BibitemOpen
  \bibfield  {author} {\bibinfo {author} {\bibfnamefont {Alexandre}\
  \bibnamefont {Blais}}, \bibinfo {author} {\bibfnamefont {Steven~M.}\
  \bibnamefont {Girvin}}, \ and\ \bibinfo {author} {\bibfnamefont {William~D.}\
  \bibnamefont {Oliver}},\ }\bibfield  {title} {\enquote {\bibinfo {title}
  {Quantum information processing and quantum optics with circuit quantum
  electrodynamics},}\ }\href {\doibase 10.1038/s41567-020-0806-z} {\bibfield
  {journal} {\bibinfo  {journal} {Nat. Phys.}\ }\textbf {\bibinfo {volume}
  {16}},\ \bibinfo {pages} {247--256} (\bibinfo {year}
  {2020}{\natexlab{a}})}\BibitemShut {NoStop}%
\bibitem [{\citenamefont {Blais}\ \emph
  {et~al.}(2020{\natexlab{b}})\citenamefont {Blais}, \citenamefont {Grimsmo},
  \citenamefont {Girvin},\ and\ \citenamefont {Wallraff}}]{blais2020}%
  \BibitemOpen
  \bibfield  {author} {\bibinfo {author} {\bibfnamefont {Alexandre}\
  \bibnamefont {Blais}}, \bibinfo {author} {\bibfnamefont {Arne~L.}\
  \bibnamefont {Grimsmo}}, \bibinfo {author} {\bibfnamefont {S.~M.}\
  \bibnamefont {Girvin}}, \ and\ \bibinfo {author} {\bibfnamefont {Andreas}\
  \bibnamefont {Wallraff}},\ }\bibfield  {title} {\enquote {\bibinfo {title}
  {Circuit {{Quantum Electrodynamics}}},}\ }\href@noop {} {\bibfield  {journal}
  {\bibinfo  {journal} {arXiv:2005.12667}\ } (\bibinfo {year}
  {2020}{\natexlab{b}})}\BibitemShut {NoStop}%
\bibitem [{\citenamefont {Ma}\ \emph {et~al.}(2021)\citenamefont {Ma},
  \citenamefont {Puri}, \citenamefont {Schoelkopf}, \citenamefont {Devoret},
  \citenamefont {Girvin},\ and\ \citenamefont {Jiang}}]{ma2021}%
  \BibitemOpen
  \bibfield  {author} {\bibinfo {author} {\bibfnamefont {Wen-Long}\
  \bibnamefont {Ma}}, \bibinfo {author} {\bibfnamefont {Shruti}\ \bibnamefont
  {Puri}}, \bibinfo {author} {\bibfnamefont {Robert~J.}\ \bibnamefont
  {Schoelkopf}}, \bibinfo {author} {\bibfnamefont {Michel~H.}\ \bibnamefont
  {Devoret}}, \bibinfo {author} {\bibfnamefont {S.~M.}\ \bibnamefont {Girvin}},
  \ and\ \bibinfo {author} {\bibfnamefont {Liang}\ \bibnamefont {Jiang}},\
  }\bibfield  {title} {\enquote {\bibinfo {title} {Quantum control of bosonic
  modes with superconducting circuits},}\ }\href@noop {} {\bibfield  {journal}
  {\bibinfo  {journal} {arXiv:2102.09668}\ } (\bibinfo {year}
  {2021})}\BibitemShut {NoStop}%
\bibitem [{\citenamefont {Poyatos}\ \emph {et~al.}(1996)\citenamefont
  {Poyatos}, \citenamefont {Cirac},\ and\ \citenamefont
  {Zoller}}]{poyatos1996}%
  \BibitemOpen
  \bibfield  {author} {\bibinfo {author} {\bibfnamefont {J.~F.}\ \bibnamefont
  {Poyatos}}, \bibinfo {author} {\bibfnamefont {J.~I.}\ \bibnamefont {Cirac}},
  \ and\ \bibinfo {author} {\bibfnamefont {P.}~\bibnamefont {Zoller}},\
  }\bibfield  {title} {\enquote {\bibinfo {title} {Quantum {{Reservoir
  Engineering}} with {{Laser Cooled Trapped Ions}}},}\ }\href {\doibase
  10.1103/PhysRevLett.77.4728} {\bibfield  {journal} {\bibinfo  {journal}
  {Phys. Rev. Lett.}\ }\textbf {\bibinfo {volume} {77}},\ \bibinfo {pages}
  {4728--4731} (\bibinfo {year} {1996})}\BibitemShut {NoStop}%
\bibitem [{\citenamefont {Metelmann}\ and\ \citenamefont
  {Clerk}(2015)}]{metelmann2015}%
  \BibitemOpen
  \bibfield  {author} {\bibinfo {author} {\bibfnamefont {A.}~\bibnamefont
  {Metelmann}}\ and\ \bibinfo {author} {\bibfnamefont {A.~A.}\ \bibnamefont
  {Clerk}},\ }\bibfield  {title} {\enquote {\bibinfo {title} {Nonreciprocal
  {{Photon Transmission}} and {{Amplification}} via {{Reservoir
  Engineering}}},}\ }\href {\doibase 10.1103/PhysRevX.5.021025} {\bibfield
  {journal} {\bibinfo  {journal} {Phys. Rev. X}\ }\textbf {\bibinfo {volume}
  {5}},\ \bibinfo {pages} {021025} (\bibinfo {year} {2015})}\BibitemShut
  {NoStop}%
\bibitem [{\citenamefont {Gilles}\ \emph {et~al.}(1994)\citenamefont {Gilles},
  \citenamefont {Garraway},\ and\ \citenamefont {Knight}}]{gilles1994}%
  \BibitemOpen
  \bibfield  {author} {\bibinfo {author} {\bibfnamefont {L.}~\bibnamefont
  {Gilles}}, \bibinfo {author} {\bibfnamefont {B.~M.}\ \bibnamefont
  {Garraway}}, \ and\ \bibinfo {author} {\bibfnamefont {P.~L.}\ \bibnamefont
  {Knight}},\ }\bibfield  {title} {\enquote {\bibinfo {title} {Generation of
  nonclassical light by dissipative two-photon processes},}\ }\href {\doibase
  10.1103/PhysRevA.49.2785} {\bibfield  {journal} {\bibinfo  {journal} {Phys.
  Rev. A}\ }\textbf {\bibinfo {volume} {49}},\ \bibinfo {pages} {2785--2799}
  (\bibinfo {year} {1994})}\BibitemShut {NoStop}%
\bibitem [{\citenamefont {Mamaev}\ \emph {et~al.}(2018)\citenamefont {Mamaev},
  \citenamefont {Govia},\ and\ \citenamefont {Clerk}}]{mamaev2018}%
  \BibitemOpen
  \bibfield  {author} {\bibinfo {author} {\bibfnamefont {M.}~\bibnamefont
  {Mamaev}}, \bibinfo {author} {\bibfnamefont {L.~C.~G.}\ \bibnamefont
  {Govia}}, \ and\ \bibinfo {author} {\bibfnamefont {A.~A.}\ \bibnamefont
  {Clerk}},\ }\bibfield  {title} {\enquote {\bibinfo {title} {Dissipative
  stabilization of entangled cat states using a driven {{Bose}}-{{Hubbard}}
  dimer},}\ }\href {\doibase 10.22331/q-2018-03-27-58} {\bibfield  {journal}
  {\bibinfo  {journal} {Quantum}\ }\textbf {\bibinfo {volume} {2}},\ \bibinfo
  {pages} {58} (\bibinfo {year} {2018})}\BibitemShut {NoStop}%
\bibitem [{\citenamefont {Porras}\ and\ \citenamefont
  {{Fern{\'a}ndez-Lorenzo}}(2019)}]{Porras2019}%
  \BibitemOpen
  \bibfield  {author} {\bibinfo {author} {\bibfnamefont {Diego}\ \bibnamefont
  {Porras}}\ and\ \bibinfo {author} {\bibfnamefont {Samuel}\ \bibnamefont
  {{Fern{\'a}ndez-Lorenzo}}},\ }\bibfield  {title} {\enquote {\bibinfo {title}
  {Topological {{Amplification}} in {{Photonic Lattices}}},}\ }\href {\doibase
  10.1103/PhysRevLett.122.143901} {\bibfield  {journal} {\bibinfo  {journal}
  {Phys. Rev. Lett.}\ }\textbf {\bibinfo {volume} {122}},\ \bibinfo {pages}
  {143901} (\bibinfo {year} {2019})}\BibitemShut {NoStop}%
\bibitem [{\citenamefont {Wanjura}\ \emph {et~al.}(2020)\citenamefont
  {Wanjura}, \citenamefont {Brunelli},\ and\ \citenamefont
  {Nunnenkamp}}]{Wanjura2020}%
  \BibitemOpen
  \bibfield  {author} {\bibinfo {author} {\bibfnamefont {Clara~C.}\
  \bibnamefont {Wanjura}}, \bibinfo {author} {\bibfnamefont {Matteo}\
  \bibnamefont {Brunelli}}, \ and\ \bibinfo {author} {\bibfnamefont {Andreas}\
  \bibnamefont {Nunnenkamp}},\ }\bibfield  {title} {\enquote {\bibinfo {title}
  {Topological framework for directional amplification in driven-dissipative
  cavity arrays},}\ }\href {\doibase 10.1038/s41467-020-16863-9} {\bibfield
  {journal} {\bibinfo  {journal} {Nat. Commun.}\ }\textbf {\bibinfo {volume}
  {11}},\ \bibinfo {pages} {3149} (\bibinfo {year} {2020})}\BibitemShut
  {NoStop}%
\bibitem [{\citenamefont {Savona}(2017)}]{Savona2017}%
  \BibitemOpen
  \bibfield  {author} {\bibinfo {author} {\bibfnamefont {Vincenzo}\
  \bibnamefont {Savona}},\ }\bibfield  {title} {\enquote {\bibinfo {title}
  {Spontaneous symmetry breaking in a quadratically driven nonlinear photonic
  lattice},}\ }\href {\doibase 10.1103/PhysRevA.96.033826} {\bibfield
  {journal} {\bibinfo  {journal} {Phys. Rev. A}\ }\textbf {\bibinfo {volume}
  {96}},\ \bibinfo {pages} {033826} (\bibinfo {year} {2017})}\BibitemShut
  {NoStop}%
\bibitem [{\citenamefont {Rota}\ \emph {et~al.}(2019)\citenamefont {Rota},
  \citenamefont {Minganti}, \citenamefont {Ciuti},\ and\ \citenamefont
  {Savona}}]{Rota2019}%
  \BibitemOpen
  \bibfield  {author} {\bibinfo {author} {\bibfnamefont {Riccardo}\
  \bibnamefont {Rota}}, \bibinfo {author} {\bibfnamefont {Fabrizio}\
  \bibnamefont {Minganti}}, \bibinfo {author} {\bibfnamefont {Cristiano}\
  \bibnamefont {Ciuti}}, \ and\ \bibinfo {author} {\bibfnamefont {Vincenzo}\
  \bibnamefont {Savona}},\ }\bibfield  {title} {\enquote {\bibinfo {title}
  {Quantum {{Critical Regime}} in a {{Quadratically Driven Nonlinear Photonic
  Lattice}}},}\ }\href {\doibase 10.1103/PhysRevLett.122.110405} {\bibfield
  {journal} {\bibinfo  {journal} {Phys. Rev. Lett.}\ }\textbf {\bibinfo
  {volume} {122}},\ \bibinfo {pages} {110405} (\bibinfo {year}
  {2019})}\BibitemShut {NoStop}%
\bibitem [{\citenamefont {Goto}(2016)}]{Goto2016}%
  \BibitemOpen
  \bibfield  {author} {\bibinfo {author} {\bibfnamefont {Hayato}\ \bibnamefont
  {Goto}},\ }\bibfield  {title} {\enquote {\bibinfo {title} {Bifurcation-based
  adiabatic quantum computation with a nonlinear oscillator network},}\ }\href
  {\doibase 10.1038/srep21686} {\bibfield  {journal} {\bibinfo  {journal} {Sci.
  Rep.}\ }\textbf {\bibinfo {volume} {6}},\ \bibinfo {pages} {21686} (\bibinfo
  {year} {2016})}\BibitemShut {NoStop}%
\bibitem [{\citenamefont {Nigg}\ \emph {et~al.}(2017)\citenamefont {Nigg},
  \citenamefont {L{\"o}rch},\ and\ \citenamefont {Tiwari}}]{Nigg2017}%
  \BibitemOpen
  \bibfield  {author} {\bibinfo {author} {\bibfnamefont {Simon~E.}\
  \bibnamefont {Nigg}}, \bibinfo {author} {\bibfnamefont {Niels}\ \bibnamefont
  {L{\"o}rch}}, \ and\ \bibinfo {author} {\bibfnamefont {Rakesh~P.}\
  \bibnamefont {Tiwari}},\ }\bibfield  {title} {\enquote {\bibinfo {title}
  {Robust quantum optimizer with full connectivity},}\ }\href {\doibase
  10.1126/sciadv.1602273} {\bibfield  {journal} {\bibinfo  {journal} {Sci.
  Adv.}\ }\textbf {\bibinfo {volume} {3}},\ \bibinfo {pages} {e1602273}
  (\bibinfo {year} {2017})}\BibitemShut {NoStop}%
\bibitem [{\citenamefont {Kraus}\ \emph {et~al.}(2008)\citenamefont {Kraus},
  \citenamefont {B{\"u}chler}, \citenamefont {Diehl}, \citenamefont {Kantian},
  \citenamefont {Micheli},\ and\ \citenamefont {Zoller}}]{kraus2008}%
  \BibitemOpen
  \bibfield  {author} {\bibinfo {author} {\bibfnamefont {B.}~\bibnamefont
  {Kraus}}, \bibinfo {author} {\bibfnamefont {H.~P.}\ \bibnamefont
  {B{\"u}chler}}, \bibinfo {author} {\bibfnamefont {S.}~\bibnamefont {Diehl}},
  \bibinfo {author} {\bibfnamefont {A.}~\bibnamefont {Kantian}}, \bibinfo
  {author} {\bibfnamefont {A.}~\bibnamefont {Micheli}}, \ and\ \bibinfo
  {author} {\bibfnamefont {P.}~\bibnamefont {Zoller}},\ }\bibfield  {title}
  {\enquote {\bibinfo {title} {Preparation of entangled states by quantum
  {{Markov}} processes},}\ }\href {\doibase 10.1103/PhysRevA.78.042307}
  {\bibfield  {journal} {\bibinfo  {journal} {Phys. Rev. A}\ }\textbf {\bibinfo
  {volume} {78}},\ \bibinfo {pages} {042307} (\bibinfo {year}
  {2008})}\BibitemShut {NoStop}%
\bibitem [{Note1()}]{Note1}%
  \BibitemOpen
  \bibinfo {note} {Phases $\phi $ and $\theta $ are connected by the local
  gauge transformation $\protect \mathaccentV {hat}05E{a}_j\rightarrow \protect
  \mathaccentV {hat}05E{a}_je^{ij\nu }$ yielding $\phi \rightarrow \phi +\nu $
  and $\theta \rightarrow \theta + 2\nu $. While these phases are gauge
  dependent, the phase difference $2\phi - \theta $ is gauge invariant. For
  $2\phi - \theta = 0$ (corresponding to $\theta = 2\phi $ set in the main
  text), we are able to express the dark state condition in the form of
  Eq.~\protect \textup {\hbox {\mathsurround \z@ \protect \normalfont
  (\ignorespaces \ref {cond zeta}\unskip \@@italiccorr )}} allowing us to write
  exact closed-form solutions for the dark states.}\BibitemShut {Stop}%
\bibitem [{\citenamefont {Wang}\ \emph {et~al.}(2019)\citenamefont {Wang},
  \citenamefont {Pechal}, \citenamefont {Wollack}, \citenamefont
  {{Arrangoiz-Arriola}}, \citenamefont {Gao}, \citenamefont {Lee},\ and\
  \citenamefont {{Safavi-Naeini}}}]{Wang2019}%
  \BibitemOpen
  \bibfield  {author} {\bibinfo {author} {\bibfnamefont {Zhaoyou}\ \bibnamefont
  {Wang}}, \bibinfo {author} {\bibfnamefont {Marek}\ \bibnamefont {Pechal}},
  \bibinfo {author} {\bibfnamefont {E.~Alex}\ \bibnamefont {Wollack}}, \bibinfo
  {author} {\bibfnamefont {Patricio}\ \bibnamefont {{Arrangoiz-Arriola}}},
  \bibinfo {author} {\bibfnamefont {Maodong}\ \bibnamefont {Gao}}, \bibinfo
  {author} {\bibfnamefont {Nathan~R.}\ \bibnamefont {Lee}}, \ and\ \bibinfo
  {author} {\bibfnamefont {Amir~H.}\ \bibnamefont {{Safavi-Naeini}}},\
  }\bibfield  {title} {\enquote {\bibinfo {title} {Quantum {{Dynamics}} of a
  {{Few}}-{{Photon Parametric Oscillator}}},}\ }\href {\doibase
  10.1103/PhysRevX.9.021049} {\bibfield  {journal} {\bibinfo  {journal} {Phys.
  Rev. X}\ }\textbf {\bibinfo {volume} {9}},\ \bibinfo {pages} {021049}
  (\bibinfo {year} {2019})}\BibitemShut {NoStop}%
\bibitem [{\citenamefont {Wielinga}\ and\ \citenamefont
  {Milburn}(1993)}]{Wielinga1993}%
  \BibitemOpen
  \bibfield  {author} {\bibinfo {author} {\bibfnamefont {B.}~\bibnamefont
  {Wielinga}}\ and\ \bibinfo {author} {\bibfnamefont {G.~J.}\ \bibnamefont
  {Milburn}},\ }\bibfield  {title} {\enquote {\bibinfo {title} {Quantum
  tunneling in a {{Kerr}} medium with parametric pumping},}\ }\href {\doibase
  10.1103/PhysRevA.48.2494} {\bibfield  {journal} {\bibinfo  {journal} {Phys.
  Rev. A}\ }\textbf {\bibinfo {volume} {48}},\ \bibinfo {pages} {2494--2496}
  (\bibinfo {year} {1993})}\BibitemShut {NoStop}%
\bibitem [{Note2()}]{Note2}%
  \BibitemOpen
  \bibinfo {note} {Eq.~\protect \textup {\hbox {\mathsurround \z@ \protect
  \normalfont (\ignorespaces \ref {many cats}\unskip \@@italiccorr )}} can be
  derived using $|\pm \zeta \rangle = \mathcal {D}_{\hat {b}_{\phi }}(\pm \zeta
  )|0\rangle $ and $ \mathcal {D}_{\hat {b}_{\phi }}(\pm \zeta ) = \prod
  _{j=1}^N\mathcal {D}_{\hat {a}_{j}}(\pm \zeta _j)$ which follows from $\hat
  {b}_k = \frac {1}{\sqrt {N}}\sum _{j=1}^N e^{ijk}\hat {a}_j$ and $[\hat
  {a}_j,\hat {a}^{\dagger }_m] = \delta _{jm}$, where $\mathcal {D}_{\hat
  {c}}(\zeta )=\exp (\zeta \hat {c}^{\dagger } - \zeta ^*\hat {c})$ is the
  displacement operator and $|0\rangle = \bigotimes _{j=1}^N|0\rangle _j=
  \bigotimes _{k}|0\rangle _k$ is the $N$-mode vacuum state.}\BibitemShut
  {Stop}%
\bibitem [{\citenamefont {Ansari}\ and\ \citenamefont
  {Man'ko}(1994)}]{Ansari1994}%
  \BibitemOpen
  \bibfield  {author} {\bibinfo {author} {\bibfnamefont {Nadeem~A.}\
  \bibnamefont {Ansari}}\ and\ \bibinfo {author} {\bibfnamefont {V.~I.}\
  \bibnamefont {Man'ko}},\ }\bibfield  {title} {\enquote {\bibinfo {title}
  {Squeezing in multimode {{Schr\"odinger}} cat states},}\ }\href {\doibase
  10.1007/BF02580951} {\bibfield  {journal} {\bibinfo  {journal} {J Russ Laser
  Res}\ }\textbf {\bibinfo {volume} {15}},\ \bibinfo {pages} {377--390}
  (\bibinfo {year} {1994})}\BibitemShut {NoStop}%
\bibitem [{\citenamefont {Vogel}\ and\ \citenamefont
  {Sperling}(2014)}]{Vogel2014}%
  \BibitemOpen
  \bibfield  {author} {\bibinfo {author} {\bibfnamefont {W.}~\bibnamefont
  {Vogel}}\ and\ \bibinfo {author} {\bibfnamefont {J.}~\bibnamefont
  {Sperling}},\ }\bibfield  {title} {\enquote {\bibinfo {title} {Unified
  quantification of nonclassicality and entanglement},}\ }\href {\doibase
  10.1103/PhysRevA.89.052302} {\bibfield  {journal} {\bibinfo  {journal} {Phys.
  Rev. A}\ }\textbf {\bibinfo {volume} {89}},\ \bibinfo {pages} {052302}
  (\bibinfo {year} {2014})}\BibitemShut {NoStop}%
\bibitem [{\citenamefont {Jeong}\ and\ \citenamefont {Kim}(2002)}]{Jeong2002}%
  \BibitemOpen
  \bibfield  {author} {\bibinfo {author} {\bibfnamefont {H.}~\bibnamefont
  {Jeong}}\ and\ \bibinfo {author} {\bibfnamefont {M.~S.}\ \bibnamefont
  {Kim}},\ }\bibfield  {title} {\enquote {\bibinfo {title} {Efficient quantum
  computation using coherent states},}\ }\href {\doibase
  10.1103/PhysRevA.65.042305} {\bibfield  {journal} {\bibinfo  {journal} {Phys.
  Rev. A}\ }\textbf {\bibinfo {volume} {65}},\ \bibinfo {pages} {042305}
  (\bibinfo {year} {2002})}\BibitemShut {NoStop}%
\bibitem [{Note3()}]{Note3}%
  \BibitemOpen
  \bibinfo {note} {The joint Wigner function of the $N$-mode state $\protect
  \mathaccentV {hat}05E{\rho }$ is $W(\vec {x},\vec {p})=\pi ^{-N}\int
  _{\mathbb {R}^{N}}\textrm {d}\vec {y}\,\exp (-2\,i\,\vec {y}\cdot \vec
  {p})\langle \vec {x}+\vec {y}|\hat {\rho }|\vec {x}-\vec {y}\rangle $, where
  $\protect \mathaccentV {vec}17E{x}$, $\protect \mathaccentV {vec}17E{p}$ and
  $\protect \mathaccentV {vec}17E{y}$ are real-valued $N$-dimensional vectors,
  and $x_j$ and $p_j $ are quadratures of resonator $j$. For cat states
  stabilized in mode $\phi \not =0$, the amplitudes $\zeta _j$ and $\zeta
  _{j+1}$ at neighboring resonators have a relative phase difference $\phi $
  [see Eq.~\protect \textup {\hbox {\mathsurround \z@ \protect \normalfont
  (\ignorespaces \ref {many cats}\unskip \@@italiccorr )}}]. This results in a
  relative rotation between the local coherent states $|\zeta _j\rangle _j$ and
  $|\zeta _{j+1}\rangle _{j+1}$ in phase space by an angle $\phi
  $.}\BibitemShut {Stop}%
\bibitem [{Note4()}]{Note4}%
  \BibitemOpen
  \bibinfo {note} {Note that the dynamics of an array with only two resonators,
  i.e. $N=2$, is fundamentally different from that of larger arrays with $N\geq
  3$ as the photon parity $\protect \mathaccentV {hat}05E{\protect \mathcal
  {P}}_{\phi } = \protect \qopname \relax o{exp}{\mathopen {\setbox \z@ \hbox
  {\frozen@everymath \@emptytoks \mathsurround \z@ $\nulldelimiterspace \z@
  \left (\vcenter to\@ne \big@size {}\right .$}\box \z@ }i\pi \protect
  \mathaccentV {hat}05E{b}_{\phi }^{\dagger }\protect \mathaccentV
  {hat}05E{b}_{\phi }\mathclose {\setbox \z@ \hbox {\frozen@everymath
  \@emptytoks \mathsurround \z@ $\nulldelimiterspace \z@ \left )\vcenter to\@ne
  \big@size {}\right .$}\box \z@ }}$ of mode $\phi $ is conserved, where the
  quasi-momentum can take only values $\phi = 0,\pi $. As a result, the
  even-parity cat state $|\mathcal {C}^{+}\rangle $ is deterministically
  approached from the initial vacuum state for any $\gamma $. However, in this
  manuscript we focus on arrays $N\geq 3$ for which the photon parities
  $\protect \mathaccentV {hat}05E{\protect \mathcal {P}}$ and $\protect
  \mathaccentV {hat}05E{\protect \mathcal {P}}_{\phi }$ are not
  conserved.}\BibitemShut {Stop}%
\bibitem [{\citenamefont {Albert}\ and\ \citenamefont
  {Jiang}(2014)}]{albert2014}%
  \BibitemOpen
  \bibfield  {author} {\bibinfo {author} {\bibfnamefont {Victor~V.}\
  \bibnamefont {Albert}}\ and\ \bibinfo {author} {\bibfnamefont {Liang}\
  \bibnamefont {Jiang}},\ }\bibfield  {title} {\enquote {\bibinfo {title}
  {Symmetries and conserved quantities in {{Lindblad}} master equations},}\
  }\href {\doibase 10.1103/PhysRevA.89.022118} {\bibfield  {journal} {\bibinfo
  {journal} {Phys. Rev. A}\ }\textbf {\bibinfo {volume} {89}},\ \bibinfo
  {pages} {022118} (\bibinfo {year} {2014})}\BibitemShut {NoStop}%
\bibitem [{\citenamefont {Misra}\ and\ \citenamefont
  {Sudarshan}(1977)}]{misra1977}%
  \BibitemOpen
  \bibfield  {author} {\bibinfo {author} {\bibfnamefont {B.}~\bibnamefont
  {Misra}}\ and\ \bibinfo {author} {\bibfnamefont {E.~C.~G.}\ \bibnamefont
  {Sudarshan}},\ }\bibfield  {title} {\enquote {\bibinfo {title} {The
  {{Zeno}}'s paradox in quantum theory},}\ }\href {\doibase 10.1063/1.523304}
  {\bibfield  {journal} {\bibinfo  {journal} {J. of Math. Phys.}\ }\textbf
  {\bibinfo {volume} {18}},\ \bibinfo {pages} {756--763} (\bibinfo {year}
  {1977})}\BibitemShut {NoStop}%
\bibitem [{\citenamefont {Koshino}\ and\ \citenamefont
  {Shimizu}(2005)}]{koshino2005}%
  \BibitemOpen
  \bibfield  {author} {\bibinfo {author} {\bibfnamefont {Kazuki}\ \bibnamefont
  {Koshino}}\ and\ \bibinfo {author} {\bibfnamefont {Akira}\ \bibnamefont
  {Shimizu}},\ }\bibfield  {title} {\enquote {\bibinfo {title} {Quantum
  {{Zeno}} effect by general measurements},}\ }\href {\doibase
  10.1016/j.physrep.2005.03.001} {\bibfield  {journal} {\bibinfo  {journal}
  {Phys. Rep.}\ }\textbf {\bibinfo {volume} {412}},\ \bibinfo {pages}
  {191--275} (\bibinfo {year} {2005})}\BibitemShut {NoStop}%
\bibitem [{\citenamefont {Popkov}\ \emph {et~al.}(2018)\citenamefont {Popkov},
  \citenamefont {Essink}, \citenamefont {Presilla},\ and\ \citenamefont
  {Sch{\"u}tz}}]{popkov2018}%
  \BibitemOpen
  \bibfield  {author} {\bibinfo {author} {\bibfnamefont {Vladislav}\
  \bibnamefont {Popkov}}, \bibinfo {author} {\bibfnamefont {Simon}\
  \bibnamefont {Essink}}, \bibinfo {author} {\bibfnamefont {Carlo}\
  \bibnamefont {Presilla}}, \ and\ \bibinfo {author} {\bibfnamefont {Gunter}\
  \bibnamefont {Sch{\"u}tz}},\ }\bibfield  {title} {\enquote {\bibinfo {title}
  {Effective quantum {{Zeno}} dynamics in dissipative quantum systems},}\
  }\href {\doibase 10.1103/PhysRevA.98.052110} {\bibfield  {journal} {\bibinfo
  {journal} {Phys. Rev. A}\ }\textbf {\bibinfo {volume} {98}},\ \bibinfo
  {pages} {052110} (\bibinfo {year} {2018})}\BibitemShut {NoStop}%
\bibitem [{\citenamefont {Kirchmair}\ \emph {et~al.}(2013)\citenamefont
  {Kirchmair}, \citenamefont {Vlastakis}, \citenamefont {Leghtas},
  \citenamefont {Nigg}, \citenamefont {Paik}, \citenamefont {Ginossar},
  \citenamefont {Mirrahimi}, \citenamefont {Frunzio}, \citenamefont {Girvin},\
  and\ \citenamefont {Schoelkopf}}]{kirchmair2013}%
  \BibitemOpen
  \bibfield  {author} {\bibinfo {author} {\bibfnamefont {Gerhard}\ \bibnamefont
  {Kirchmair}}, \bibinfo {author} {\bibfnamefont {Brian}\ \bibnamefont
  {Vlastakis}}, \bibinfo {author} {\bibfnamefont {Zaki}\ \bibnamefont
  {Leghtas}}, \bibinfo {author} {\bibfnamefont {Simon~E.}\ \bibnamefont
  {Nigg}}, \bibinfo {author} {\bibfnamefont {Hanhee}\ \bibnamefont {Paik}},
  \bibinfo {author} {\bibfnamefont {Eran}\ \bibnamefont {Ginossar}}, \bibinfo
  {author} {\bibfnamefont {Mazyar}\ \bibnamefont {Mirrahimi}}, \bibinfo
  {author} {\bibfnamefont {Luigi}\ \bibnamefont {Frunzio}}, \bibinfo {author}
  {\bibfnamefont {S.~M.}\ \bibnamefont {Girvin}}, \ and\ \bibinfo {author}
  {\bibfnamefont {R.~J.}\ \bibnamefont {Schoelkopf}},\ }\bibfield  {title}
  {\enquote {\bibinfo {title} {Observation of quantum state collapse and
  revival due to the single-photon {{Kerr}} effect},}\ }\href {\doibase
  10.1038/nature11902} {\bibfield  {journal} {\bibinfo  {journal} {Nature}\
  }\textbf {\bibinfo {volume} {495}},\ \bibinfo {pages} {205--209} (\bibinfo
  {year} {2013})}\BibitemShut {NoStop}%
\bibitem [{Note5()}]{Note5}%
  \BibitemOpen
  \bibinfo {note} {The dissipator $\protect \mathcal {L}_{\gamma }\protect
  \mathaccentV {hat}05E{\rho }= \DOTSB \sum@ \slimits@ _{j,l=1}^N D_{jl} \left
  [\protect \mathaccentV {hat}05E{a}_j\protect \mathaccentV {hat}05E{\rho
  }\protect \mathaccentV {hat}05E{a}_l^{\dagger } -\protect \frac {1}{2}\left
  \protect \{\protect \mathaccentV {hat}05E{a}_l^{\dagger }\protect
  \mathaccentV {hat}05E{a}_j,\protect \mathaccentV {hat}05E{\rho } \right
  \protect \}\right ] = \DOTSB \sum@ \slimits@ _{q=1}^N \protect \mathaccentV
  {bar}016{\gamma }_q \left [\protect \mathaccentV {hat}05E{b}_q\protect
  \mathaccentV {hat}05E{\rho }\protect \mathaccentV {hat}05E{b}_q^{\dagger }
  -\protect \frac {1}{2}\left \protect \{\protect \mathaccentV
  {hat}05E{b}_q^{\dagger }\protect \mathaccentV {hat}05E{b}_q,\protect
  \mathaccentV {hat}05E{\rho } \right \protect \}\right ]$ can be always
  diagonalized by the linear transformation $T_{jq}$ of annihilation operators,
  where columns of the transformation matrix $T_{jq}$ and single-photon loss
  rates $\protect \mathaccentV {bar}016{\gamma }_q$ are eigenvectors and
  eigenvalues, respectively, of the dynamical matrix $D_{jl}$.}\BibitemShut
  {Stop}%
\bibitem [{Note6()}]{Note6}%
  \BibitemOpen
  \bibinfo {note} {Jump-operator phases $\phi _j\rightarrow \phi _j+\nu _j-\nu
  _{j+1}$ and two-photon pump phases $\theta _j\rightarrow \theta _j+2\nu _j$
  transform under the local gauge transformation $\protect \mathaccentV
  {hat}05E{a}_j\rightarrow \protect \mathaccentV {hat}05E{a}_j e^{i\nu _j}$.
  The master equation \protect \textup {\hbox {\mathsurround \z@ \protect
  \normalfont (\ignorespaces \ref {master loss}\unskip \@@italiccorr )}} with
  vanishing two-photon pump phases $\theta _j=0$, for all $j$, describes a
  general situation with any two-photon pump phases $\theta _j\not =0$ (that
  can also vary at different resonators $j$) for the particular gauge choice
  $\nu _j = - \theta _j/2$.}\BibitemShut {Stop}%
\bibitem [{\citenamefont {Reagor}\ \emph {et~al.}(2013)\citenamefont {Reagor},
  \citenamefont {Paik}, \citenamefont {Catelani}, \citenamefont {Sun},
  \citenamefont {Axline}, \citenamefont {Holland}, \citenamefont {Pop},
  \citenamefont {Masluk}, \citenamefont {Brecht}, \citenamefont {Frunzio},
  \citenamefont {Devoret}, \citenamefont {Glazman},\ and\ \citenamefont
  {Schoelkopf}}]{reagor2013}%
  \BibitemOpen
  \bibfield  {author} {\bibinfo {author} {\bibfnamefont {Matthew}\ \bibnamefont
  {Reagor}}, \bibinfo {author} {\bibfnamefont {Hanhee}\ \bibnamefont {Paik}},
  \bibinfo {author} {\bibfnamefont {Gianluigi}\ \bibnamefont {Catelani}},
  \bibinfo {author} {\bibfnamefont {Luyan}\ \bibnamefont {Sun}}, \bibinfo
  {author} {\bibfnamefont {Christopher}\ \bibnamefont {Axline}}, \bibinfo
  {author} {\bibfnamefont {Eric}\ \bibnamefont {Holland}}, \bibinfo {author}
  {\bibfnamefont {Ioan~M.}\ \bibnamefont {Pop}}, \bibinfo {author}
  {\bibfnamefont {Nicholas~A.}\ \bibnamefont {Masluk}}, \bibinfo {author}
  {\bibfnamefont {Teresa}\ \bibnamefont {Brecht}}, \bibinfo {author}
  {\bibfnamefont {Luigi}\ \bibnamefont {Frunzio}}, \bibinfo {author}
  {\bibfnamefont {Michel~H.}\ \bibnamefont {Devoret}}, \bibinfo {author}
  {\bibfnamefont {Leonid}\ \bibnamefont {Glazman}}, \ and\ \bibinfo {author}
  {\bibfnamefont {Robert~J.}\ \bibnamefont {Schoelkopf}},\ }\bibfield  {title}
  {\enquote {\bibinfo {title} {Reaching 10\,ms single photon lifetimes for
  superconducting aluminum cavities},}\ }\href {\doibase 10.1063/1.4807015}
  {\bibfield  {journal} {\bibinfo  {journal} {Appl. Phys. Lett.}\ }\textbf
  {\bibinfo {volume} {102}},\ \bibinfo {pages} {192604} (\bibinfo {year}
  {2013})}\BibitemShut {NoStop}%
\bibitem [{\citenamefont {Reagor}\ \emph {et~al.}(2016)\citenamefont {Reagor},
  \citenamefont {Pfaff}, \citenamefont {Axline}, \citenamefont {Heeres},
  \citenamefont {Ofek}, \citenamefont {Sliwa}, \citenamefont {Holland},
  \citenamefont {Wang}, \citenamefont {Blumoff}, \citenamefont {Chou},
  \citenamefont {Hatridge}, \citenamefont {Frunzio}, \citenamefont {Devoret},
  \citenamefont {Jiang},\ and\ \citenamefont {Schoelkopf}}]{reagor2016}%
  \BibitemOpen
  \bibfield  {author} {\bibinfo {author} {\bibfnamefont {Matthew}\ \bibnamefont
  {Reagor}}, \bibinfo {author} {\bibfnamefont {Wolfgang}\ \bibnamefont
  {Pfaff}}, \bibinfo {author} {\bibfnamefont {Christopher}\ \bibnamefont
  {Axline}}, \bibinfo {author} {\bibfnamefont {Reinier~W.}\ \bibnamefont
  {Heeres}}, \bibinfo {author} {\bibfnamefont {Nissim}\ \bibnamefont {Ofek}},
  \bibinfo {author} {\bibfnamefont {Katrina}\ \bibnamefont {Sliwa}}, \bibinfo
  {author} {\bibfnamefont {Eric}\ \bibnamefont {Holland}}, \bibinfo {author}
  {\bibfnamefont {Chen}\ \bibnamefont {Wang}}, \bibinfo {author} {\bibfnamefont
  {Jacob}\ \bibnamefont {Blumoff}}, \bibinfo {author} {\bibfnamefont {Kevin}\
  \bibnamefont {Chou}}, \bibinfo {author} {\bibfnamefont {Michael~J.}\
  \bibnamefont {Hatridge}}, \bibinfo {author} {\bibfnamefont {Luigi}\
  \bibnamefont {Frunzio}}, \bibinfo {author} {\bibfnamefont {Michel~H.}\
  \bibnamefont {Devoret}}, \bibinfo {author} {\bibfnamefont {Liang}\
  \bibnamefont {Jiang}}, \ and\ \bibinfo {author} {\bibfnamefont {Robert~J.}\
  \bibnamefont {Schoelkopf}},\ }\bibfield  {title} {\enquote {\bibinfo {title}
  {Quantum memory with millisecond coherence in circuit {{QED}}},}\ }\href
  {\doibase 10.1103/PhysRevB.94.014506} {\bibfield  {journal} {\bibinfo
  {journal} {Phys. Rev. B}\ }\textbf {\bibinfo {volume} {94}},\ \bibinfo
  {pages} {014506} (\bibinfo {year} {2016})}\BibitemShut {NoStop}%
\bibitem [{\citenamefont {Sliwa}\ \emph {et~al.}(2015)\citenamefont {Sliwa},
  \citenamefont {Hatridge}, \citenamefont {Narla}, \citenamefont {Shankar},
  \citenamefont {Frunzio}, \citenamefont {Schoelkopf},\ and\ \citenamefont
  {Devoret}}]{sliwa2015}%
  \BibitemOpen
  \bibfield  {author} {\bibinfo {author} {\bibfnamefont {K.~M.}\ \bibnamefont
  {Sliwa}}, \bibinfo {author} {\bibfnamefont {M.}~\bibnamefont {Hatridge}},
  \bibinfo {author} {\bibfnamefont {A.}~\bibnamefont {Narla}}, \bibinfo
  {author} {\bibfnamefont {S.}~\bibnamefont {Shankar}}, \bibinfo {author}
  {\bibfnamefont {L.}~\bibnamefont {Frunzio}}, \bibinfo {author} {\bibfnamefont
  {R.~J.}\ \bibnamefont {Schoelkopf}}, \ and\ \bibinfo {author} {\bibfnamefont
  {M.~H.}\ \bibnamefont {Devoret}},\ }\bibfield  {title} {\enquote {\bibinfo
  {title} {Reconfigurable {{Josephson Circulator}}/{{Directional
  Amplifier}}},}\ }\href {\doibase 10.1103/PhysRevX.5.041020} {\bibfield
  {journal} {\bibinfo  {journal} {Phys. Rev. X}\ }\textbf {\bibinfo {volume}
  {5}},\ \bibinfo {pages} {041020} (\bibinfo {year} {2015})}\BibitemShut
  {NoStop}%
\bibitem [{\citenamefont {Lecocq}\ \emph {et~al.}(2017)\citenamefont {Lecocq},
  \citenamefont {Ranzani}, \citenamefont {Peterson}, \citenamefont {Cicak},
  \citenamefont {Simmonds}, \citenamefont {Teufel},\ and\ \citenamefont
  {Aumentado}}]{lecocq2017}%
  \BibitemOpen
  \bibfield  {author} {\bibinfo {author} {\bibfnamefont {F.}~\bibnamefont
  {Lecocq}}, \bibinfo {author} {\bibfnamefont {L.}~\bibnamefont {Ranzani}},
  \bibinfo {author} {\bibfnamefont {G.~A.}\ \bibnamefont {Peterson}}, \bibinfo
  {author} {\bibfnamefont {K.}~\bibnamefont {Cicak}}, \bibinfo {author}
  {\bibfnamefont {R.~W.}\ \bibnamefont {Simmonds}}, \bibinfo {author}
  {\bibfnamefont {J.~D.}\ \bibnamefont {Teufel}}, \ and\ \bibinfo {author}
  {\bibfnamefont {J.}~\bibnamefont {Aumentado}},\ }\bibfield  {title} {\enquote
  {\bibinfo {title} {Nonreciprocal {{Microwave Signal Processing}} with a
  {{Field}}-{{Programmable Josephson Amplifier}}},}\ }\href {\doibase
  10.1103/PhysRevApplied.7.024028} {\bibfield  {journal} {\bibinfo  {journal}
  {Phys. Rev. Applied}\ }\textbf {\bibinfo {volume} {7}},\ \bibinfo {pages}
  {024028} (\bibinfo {year} {2017})}\BibitemShut {NoStop}%
\bibitem [{\citenamefont {Ra}\ \emph {et~al.}(2020)\citenamefont {Ra},
  \citenamefont {Dufour}, \citenamefont {Walschaers}, \citenamefont {Jacquard},
  \citenamefont {Michel}, \citenamefont {Fabre},\ and\ \citenamefont
  {Treps}}]{Ra2020}%
  \BibitemOpen
  \bibfield  {author} {\bibinfo {author} {\bibfnamefont {Young-Sik}\
  \bibnamefont {Ra}}, \bibinfo {author} {\bibfnamefont {Adrien}\ \bibnamefont
  {Dufour}}, \bibinfo {author} {\bibfnamefont {Mattia}\ \bibnamefont
  {Walschaers}}, \bibinfo {author} {\bibfnamefont {Cl{\'e}ment}\ \bibnamefont
  {Jacquard}}, \bibinfo {author} {\bibfnamefont {Thibault}\ \bibnamefont
  {Michel}}, \bibinfo {author} {\bibfnamefont {Claude}\ \bibnamefont {Fabre}},
  \ and\ \bibinfo {author} {\bibfnamefont {Nicolas}\ \bibnamefont {Treps}},\
  }\bibfield  {title} {\enquote {\bibinfo {title} {Non-{{Gaussian}} quantum
  states of a multimode light field},}\ }\href {\doibase
  10.1038/s41567-019-0726-y} {\bibfield  {journal} {\bibinfo  {journal} {Nat.
  Phys.}\ }\textbf {\bibinfo {volume} {16}},\ \bibinfo {pages} {144--147}
  (\bibinfo {year} {2020})}\BibitemShut {NoStop}%
\bibitem [{\citenamefont {Zurek}(2003)}]{Zurek2003}%
  \BibitemOpen
  \bibfield  {author} {\bibinfo {author} {\bibfnamefont {Wojciech~Hubert}\
  \bibnamefont {Zurek}},\ }\bibfield  {title} {\enquote {\bibinfo {title}
  {Decoherence, einselection, and the quantum origins of the classical},}\
  }\href {\doibase 10.1103/RevModPhys.75.715} {\bibfield  {journal} {\bibinfo
  {journal} {Rev. Mod. Phys.}\ }\textbf {\bibinfo {volume} {75}},\ \bibinfo
  {pages} {715--775} (\bibinfo {year} {2003})}\BibitemShut {NoStop}%
\bibitem [{\citenamefont {Li}\ \emph {et~al.}(2021)\citenamefont {Li},
  \citenamefont {Soret},\ and\ \citenamefont {Ciuti}}]{li2020}%
  \BibitemOpen
  \bibfield  {author} {\bibinfo {author} {\bibfnamefont {Zejian}\ \bibnamefont
  {Li}}, \bibinfo {author} {\bibfnamefont {Ariane}\ \bibnamefont {Soret}}, \
  and\ \bibinfo {author} {\bibfnamefont {Cristiano}\ \bibnamefont {Ciuti}},\
  }\bibfield  {title} {\enquote {\bibinfo {title} {Dissipation-induced
  antiferromagneticlike frustration in coupled photonic resonators},}\ }\href
  {\doibase 10.1103/PhysRevA.103.022616} {\bibfield  {journal} {\bibinfo
  {journal} {Phys. Rev. A}\ }\textbf {\bibinfo {volume} {103}},\ \bibinfo
  {pages} {022616} (\bibinfo {year} {2021})}\BibitemShut {NoStop}%
\bibitem [{\citenamefont {Zhou}\ \emph {et~al.}(2021)\citenamefont {Zhou},
  \citenamefont {Gneiting}, \citenamefont {You},\ and\ \citenamefont
  {Nori}}]{zhou2021a}%
  \BibitemOpen
  \bibfield  {author} {\bibinfo {author} {\bibfnamefont {Zheng-Yang}\
  \bibnamefont {Zhou}}, \bibinfo {author} {\bibfnamefont {Clemens}\
  \bibnamefont {Gneiting}}, \bibinfo {author} {\bibfnamefont {J.~Q.}\
  \bibnamefont {You}}, \ and\ \bibinfo {author} {\bibfnamefont {Franco}\
  \bibnamefont {Nori}},\ }\bibfield  {title} {\enquote {\bibinfo {title}
  {Generating and detecting entangled cat states in dissipatively coupled
  degenerate optical parametric oscillators},}\ }\href {\doibase
  10.1103/PhysRevA.104.013715} {\bibfield  {journal} {\bibinfo  {journal}
  {Phys. Rev. A}\ }\textbf {\bibinfo {volume} {104}},\ \bibinfo {pages}
  {013715} (\bibinfo {year} {2021})}\BibitemShut {NoStop}%
\end{thebibliography}%

\FloatBarrier
\newpage

\onecolumngrid

\clearpage\onecolumngrid

\setcounter{equation}{0}
\setcounter{figure}{0}
\setcounter{table}{0}
\setcounter{page}{1}
\makeatletter
\renewcommand{\theequation}{S\arabic{equation}}
\renewcommand{\thefigure}{S\arabic{figure}}
\renewcommand{\bibnumfmt}[1]{[S#1]}

\begin{center}
\textbf{\large{}Stabilization of multi-mode Schr\" odinger cat states via normal-mode dissipation engineering – Supplementary material}{\large\par}
\par\end{center}

\section*{Second-order term in the Dyson series (\ref{dyson proj})}\label{app:Zeno term}

We now explicitly evaluate the term $\mathcal{P}_d \,\mathcal{K} \int_{0}^{\tau}{\rm d}\tau_1\int_{0}^{\tau_1}{\rm d}\tau_2 \,e^{\bar{\mathcal{L}}_d \left( \tau_1 - \tau_2\right)}\,\mathcal{K}\,\mathcal{P}_d $ of the second order in $\mathcal{K}$ in the Dyson series [Eq.~(\ref{dyson proj}) of Appendix~\ref{app:Zeno}]. We start by evaluating 
\begin{align}
\mathcal{K}\mathcal{P}_d \hat{\rho} =-\frac{i}{\gamma}\left[\hat{H},\hat{\rho}_{d\phi}\right] 
=-\frac{i}{\gamma}\left[\hat{H}_{\phi},\hat{\rho}_{d\phi}\right] -\frac{i}{\gamma} \frac{U}{N}\sum_{{k\neq\phi}}\left\{
\hat{B}_k^{\dagger}\hat{\rho}_{d\phi}-\hat{\rho}_{d\phi}\hat{B}_k
\right\},
\end{align}
where $\hat{\rho}_{d\phi} = \hat{\rho}_d \otimes\hat{\rho}_{\phi}$, $\hat{B}_k = \left(\hat{b}_{\phi}^{\dagger\,2} - \zeta^{*\,2} \right)\hat{b}_{k}\hat{b}_{2\phi - k}$ and we used that $\hat{b}_k\,\hat{\rho}_{d\phi}  = \hat{\rho}_{d\phi}  \,\hat{b}_k^{\dagger} =0$ for all $k\neq\phi$. The next important observation is that 
\begin{equation}
\bar{\mathcal{L}}_d \hat{\rho}_{d\phi} \hat{B}_k =  \sum_{l\neq\phi}\frac{\gamma_l}{\gamma}\mathcal{D}[ \hat{b}_l] \, \hat{\rho}_{d\phi}\hat{B}_{k} = - \frac{1}{2}\sum_{l\neq\phi}\frac{\gamma_l}{\gamma}\hat{\rho}_{d\phi}\hat{B}_{k} \hat{b}_l^{\dagger} \hat{b}_l =-\frac{1}{2}\sum_{l\neq\phi}\frac{\gamma_l}{\gamma}\left(\delta_{kl} + \delta_{(2\phi-k)l}\right) \hat{\rho}_{d\phi}\hat{B}_{k} = -\frac{\gamma_k}{\gamma} \hat{\rho}_{d\phi} \hat{B}_k,\label{emat1}
\end{equation}
where we used that $\hat{b}_k\,\hat{\rho}_{d\phi}  = \hat{\rho}_{d\phi}  \,\hat{b}_k^{\dagger} =0$ in the second equality and in the third equality as well as that $\gamma_k = \gamma_{2\phi - k}$ in the last equality. Analogically, we obtain
\begin{equation}
\bar{\mathcal{L}}_d \hat{B}_k^{\dagger} \hat{\rho}_{d\phi}= -\frac{\gamma_k}{\gamma} \hat{B}_k^{\dagger}\hat{\rho}_{d\phi}.\label{emat2}
\end{equation}
Using $e^{\bar{\mathcal{L}}_d \, \tau}\hat{\rho}_{d\phi} = \hat{\rho}_{d\phi} $, $\left[\bar{\mathcal{L}}_d,\hat{H}_{\phi}\right]=0$ as well as equations  (\ref{emat1}) and (\ref{emat2}), we can evaluate the integrals
\begin{equation}
 \int_{0}^{\tau}{\rm d}\tau_1\int_{0}^{\tau_1}{\rm d}\tau_2 \,e^{\bar{\mathcal{L}}_d \left( \tau_1 - \tau_2\right)}\mathcal{K}\mathcal{P}_d \hat{\rho}\nonumber\\ 
 = -\frac{i}{\gamma}\frac{\tau^2}{2}\left[\hat{H}_{\phi},\hat{\rho}_{d\phi}\right] 
- i \sum_{{k\neq\phi}} C_k \left\{
\hat{B}_{k}^{\dagger} \hat{\rho}_{d\phi} - \textrm{h.c.}
\right\},
\end{equation}
where
\begin{equation}
C_k = \frac{1}{\gamma}\frac{U}{N} \int_{0}^{\tau}{\rm d}\tau_1\int_{0}^{\tau_1}{\rm d}\tau_2\, e^{-\frac{\gamma_k}{\gamma}\left(\tau_1-\tau_2\right)} = \frac{U}{N} \left\{\frac{\tau}{\gamma_k} + \frac{\gamma}{\gamma_k^2} \left(e^{-\frac{\gamma_k}{\gamma}\,\tau} - 1\right)\right\}.
\end{equation}
Finally we can evaluate the term of the second order in $\mathcal{K}$ in the Dyson series [Eq.~(\ref{dyson proj}) of Appendix~\ref{app:Zeno}]
\begin{align}
 \mathcal{P}_d\,\mathcal{K}\,\int_{0}^{\tau}{\rm d}\tau_1\int_{0}^{\tau_1}{\rm d}\tau_2 \,e^{\bar{\mathcal{L}}_d \left( \tau_1 - \tau_2\right)}\mathcal{K}\mathcal{P}_d \hat{\rho} 
&= \left(-\frac{i}{\gamma}\right)^2\frac{\tau^2}{2}\mathcal{P}_d\left[\hat{H},\left[\hat{H}_{\phi},\hat{\rho}_{d\phi}\right]\right] 
-  \frac{1}{\gamma}\sum_{{k\neq\phi}} C_k \left\{
\mathcal{P}_d\left[\hat{H},\hat{B}_{k}^{\dagger} \hat{\rho}_{d\phi}\right]  
+ \textrm{h.c.}
 \right\}\label{dt2 1}\\
&= \left(-\frac{i}{\gamma}\right)^2\frac{\tau^2}{2}\left[\hat{H}_{\phi},\left[\hat{H}_{\phi},\hat{\rho}_{d\phi} \right]\right] + \tau\frac{\Gamma}{\gamma}\mathcal{D}[\hat{Z}_{\phi}]\hat{\rho}_{d\phi},\label{dt2 2}
\end{align}
where $\hat{Z}_{\phi} = \left(\hat{b}_{\phi}^2 - \zeta^2 \right)$ and $\Gamma = 4\frac{U}{\tau N}\sum_{{k\neq\phi}} C_k$. To simplify the first term on the right hand side of Eq.~(\ref{dt2 1}) we used that $ {\rm Tr}_{d}\left[ \left[ \hat{H}, \left(\hat{\rho}_d \otimes\,\cdot\,\right) \right]\right] =  \left[\hat{H}_{\phi}, \,\cdot\,\right]$.
The terms in the curly brackets on the right hand side of Eq.~(\ref{dt2 1}) are simplified as follows
\begin{equation}
\mathcal{P}_d\left[\hat{H},\hat{B}_{k}^{\dagger} \hat{\rho}_{d\phi}\right] + \textrm{h.c.} = 2\frac{U}{N}\left[\hat{Z}_{\phi}^{\dagger},\hat{Z}_{\phi}\hat{\rho}_{d\phi}\right] + \textrm{h.c.} =  - 4\frac{U}{N}\mathcal{D}[\hat{Z}_{\phi}]\hat{\rho}_{d\phi},\label{comm 2p 1}
\end{equation}
where, in the first equality, we used that ${\rm Tr}_{d}\left[\hat{b}_{l}^{\dagger}\hat{b}^{\dagger}_{2\phi - l}\hat{B}_{k}^{\dagger}\hat{\rho}_{d\phi} \right] = 0$ and ${\rm Tr}_{d}\left[\hat{b}_{l}\hat{b}_{2\phi - l}\hat{B}_{k}^{\dagger}\hat{\rho}_{d\phi} \right] = \left(\delta_{kl} + \delta_{k(2\phi-l)}\right)\hat{Z}_{\phi}\hat{\rho}_{\phi}$ for all $l$ as well as $ {\rm Tr}_{d}\left[\left[\hat{b}_{k_1}^{\dagger}\hat{b}^{\dagger}_{k_2}\hat{b}_{k_3}\hat{b}_{k_4},\hat{B}_{k}^{\dagger}\hat{\rho}_{d\phi} \right] \right] = \delta_{k_1\phi} \delta_{k_2\phi}\left(\delta_{kk_3} + \delta_{k(2\phi-k_3)}\right)\left[\hat{b}_{\phi}^{\dagger\,2},\hat{Z}_{\phi}\hat{\rho}_{\phi}\right]$ for all $k_1$, $k_2$, $k_3$ and $k_4$ satisfying $k_1+k_2-k_3-k_4 \mod 2\pi= 0$. Inspecting the time dependent constant
\begin{equation}
\tau\frac{\Gamma}{\gamma}= \frac{4}{N^2}\sum_{{k\neq\phi}} \left\{ \frac{U^2}{\gamma\gamma_k}\,\tau + \frac{U^2}{\gamma_k^2}\left(e^{-\frac{\gamma_k}{\gamma} \tau} - 1 \right) \right\}\approx 4\,\frac{\tau}{\gamma} \frac{U^2}{N^2}\sum_{{k\neq\phi}}  \frac{1}{\gamma_k}
\end{equation}
 in front of the Dissipator in Eq.~(\ref{dt2 2}), we see that the second term in the curly brackets is bounded and it is of order $\mathcal{O}\left( \frac{U^2}{\gamma_k^2}\right)$ for all $k\neq\phi$. As a result, this term can be neglected \cite{popkov2018}. Neglecting this term in equation (\ref{dt2 2}), we can simplify the term of the second order in $\mathcal{K}$ in the Dyson series
\begin{align}
 \mathcal{P}_d\,\mathcal{K}\,\int_{0}^{\tau}{\rm d}\tau_1\int_{0}^{\tau_1}{\rm d}\tau_2 \,e^{\bar{\mathcal{L}}_d \left( \tau_1 - \tau_2\right)}\mathcal{K}\mathcal{P}_d = \frac{\tau^2}{2}\, \left(\mathcal{P}_d \,\mathcal{K}\,\mathcal{P}_d \right)^2  + \tau\,\frac{\Gamma}{\gamma} \mathcal{D}[\hat{Z}_{\phi}]\mathcal{P}_d,
\end{align}
where we also used that $\mathcal{P}_d \,\mathcal{K}\,\mathcal{P}_d \left(\hat{\rho}_d\otimes\,\cdot\,\right) =-\frac{i}{\gamma} \left[ \hat{H}_{\phi},\left(\hat{\rho}_d\otimes\,\cdot\,\right)\right].$ Substituting in the term of the second order in $\mathcal{K}$ in the Dyson series [Eq.~(\ref{dyson proj}) of Appendix~\ref{app:Zeno}] we obtain the explicit form of the Dyson series [Eq.~(\ref{dyson proj 2}) of Appendix~\ref{app:Zeno}].

\end{document}